\documentclass[reqno]{amsart}

\usepackage{amscd,amsfonts,amssymb}
\usepackage{graphicx}
\usepackage{color}
\usepackage[english]{babel}

\copyrightinfo{2000}{American Mathematical Society}

\numberwithin{equation}{section}

\newtheorem{theorem}{Theorem}[section]

\theoremstyle{definition}
\newtheorem{definition}{Definition}[section]
\theoremstyle{remark}
\newtheorem{remark}{Remark}[section]

\newtheorem{assum}{Assumption}[section]
\newtheorem{utv}{Proposition}[section]

\author{G.~S.~Kambarbaeva}
\address{Elbrus Technologies, Vavilova 24, Moscow, Russia.} \email{kambarg@mail.ru}

\author{O.~S.~Rozanova}
\address{Mechanics and Mathematics Faculty, Moscow State University, Leninskie Gory 1, Moscow, Russia.} \email{ rozanova@mech.math.msu.su}

\title[Optimal strategies of investment in a linear model of market]{Optimal strategies of investment in a linear stochastic
model of market}

\thanks{}



\date{ }

\begin{document}

\begin{abstract}
We study the continuous time portfolio optimization model on the
market where the mean returns of individual securities or asset
categories are linearly dependent on underlying economic factors. We
introduce the functional $Q_\gamma$ featuring the expected earnings
yield of portfolio minus a penalty term proportional with a
coefficient $\gamma$  to the variance when we keep the value of the
factor levels fixed. The coefficient $\gamma$ plays the role of a
risk-aversion parameter. We find the optimal trading positions that
can be obtained as the solution to a maximization problem for
$Q_\gamma$ at any moment of time. The single-factor case is analyzed
in more details. We present a simple asset allocation example
featuring an   interest rate which affects a stock index and also
serves as a second investment opportunity. We consider two
possibilities: the interest rate for the bank account is governed by
Vasicek-type and Cox-Ingersoll-Ross dynamics, respectively.  Then we
compare our results with the theory of Bielecki and Pliska where the
authors employ the methods of the risk-sensitive control theory
thereby using an infinite horizon objective featuring the long run
expected growth rate, the asymptotic variance, and a risk-aversion
parameter similar to $\gamma$.

\end{abstract}

\maketitle

\section{Introduction}

The art of making  decisions about investment mix in order to meet
specified investment goals for the benefit of the investors  and
balancing risk against performance is called the portfolio
 management (portfolio is  a collection of investments all owned by the same individual or
 organization).

 A modern study of portfolio selection begins in works by
Markowitz \cite{markowitz1952}, \cite{markowitz}. He showed how to
formulate the problem of minimizing a portfolio's variance subject
to the constraint that its expected return equals a prescribed level
as a quadratic program. Such an optimal portfolio is said to be
variance minimizing and if it also achieves the maximum expected
return among all portfolios having the same variance of return then
it is said to be efficient \cite{sharpe}.

There has been considerable research involving stochastic processes
models of assets taking into account optimal investment decisions.
 Several researchers (e.g.
Merton \cite{merton1971}, Karatzas \cite{karatzas}) used stochastic
control theory to develop continuous time portfolio management model
where the assets are modeled bas stochastic processes but financial
and economic factors are ignored. At the same time,  a number of
empirical studies (e.g.\cite{pestim}, \cite{patelis},
\cite{ilmanen}) provided evidences that macroeconomic factors such
as unemployment rate, inflation rate, dividend yield, change of
industrial production, an interest rate, etc, influence on the stock
return.

Lucas  \cite{lucas} introduced  a discrete time model  including
stochastic process models as factors.   Brennan,  Schwartz and
Lagnado \cite{brenschwlagn} considered the factors as diffusion
processes and assets as correlated Brownian motions, with the drift
and diffusion coefficients for the asset pricesses taken to be
deterministic functions of the factor levels. Their objective was to
maximize expected utility of wealth at a terminal date.  A key
limitation of the Brennan-Schwartz-Lagnado approach is that one is
unlikely to obtain tractable formulas for the optimal strategies.

In the past decades, the applications of risk-sensitive control to
asset management is very popular. The risk-sensitive control differs
from traditional stochastic control in that it explicitly models the
risk-aversion of the decision maker as an integral part of the
control framework, rather than importing it in the problem via an
externally defined utility function \cite{whittle}.
 Risk-sensitive control was first
applied to solve financial problems by Lefebvre and Montulet
\cite{lefemont} in a corporate finance context and by Fleming
\cite{fleming1995} in a portfolio selection context.

Bielecki and Pliska \cite{bielplis} were the first to apply the
continuous time risk-sensitive control as a practical tool that
could be used to solve "real world" portfolio selection problems. In
the series of works by T.Bielecki, S.Pliska et al.(\cite{bielplis},
\cite{biplsh}, etc) was devoted to a infinite-horizon
continuous-time risk sensitive portfolio optimization problem.
 The authors considered a model of
market analogous to the Brennan-Schwartz-Lagnado one
\cite{brenschwlagn} where the mean returns of individual securities
are explicitly affected by underlying economic factors such as
dividend yields, a firm's return on equity, an interest rate, and
unemployment rate. The factors are random processes,  and the drift
coefficients for the securities are linear functions of these
factors. The main result of the theory by Bielecki and Pliska  is a
construction of admissible trading strategies, which   have a simple
characterization in terms of the factor levels. The results are
illustrated on a simple but important example of two asset
allocation, having  independent interest to financial economists.
Here one of assets is a bank account and the unique factor is  a
Vasicek-type interest rate. The Vasicek model of the interest rate
is linear and this gives a possibility to obtain an explicit
formulas for the optimal  strategy.

The strategy proposed in the works of Bielecki and Pliska refers to
the strategic asset allocation. According \cite{brenschwlagn} the
primary goal of a strategic asset allocation is to create an asset
mix that will provide the optimal balance between expected risk and
return for a long-term investment horizon.

In the present paper we propose an alternative method of capital
allocation in which an investor takes a more active approach that
tries to position a portfolio into those assets, sectors, or
individual stocks that show the most potential for gains. The model
refers to the tactical asset allocation (e.g. \cite{rey2003}). Our
strategy also  has a simple characterization in terms of the factor
levels, but  it is more flexible comparing with the Bielecki-Pliska
model and can be actualized within all the time of investment.
Moreover, we get  explicit formulae not only for the lineal model of
factor (in particular, for the Vasicek model of the interest rate),
but for more complicated one such that the Cox-Ingersoll-Ross model.
This paper summarizes and extends the results of
\cite{Kambarbaeva_2010}, \cite{Kambarbaeva_2011},
\cite{Kambarbaeva_2012}.

This paper is organized as follows. In Sec.\ref{market} we
 describe the model of linear market where we are going to
consider the allocation of capital. In Sec.\ref{BPstrategy} we give
an outline of the Bielecki and Pliska theory and write their
explicit optimal strategy. Sec.\ref{algorithm} contains  auxiliary
results: an algorithm of finding the conditional expectation and
conditional variance for a couple of stochastic differential
equations. To find these values  we have to solve a Fokker-Planck
equation for the joint probability density of two respective random
values. We show that there exist two approaches to solving this
problem. The first approach  uses an ansatz for the solution, such
that the problem is reduced to solving a nonlinear system of
ordinary differential equations. The second approach refers to the
application of the Fourier analysis. In
Sec.\ref{portfolio_selection} we formulate the problem of the fixed
time portfolio selection and give a general algorithm for its
solution.

In Sec.\ref{Vasicek} we consider the case of linear interest rate
(the Vasicek model). First we solve the fixed time optimization
problem for the the example of portfolio consisting of two assets
mentioned before. Then we find asymptotics of the proportions of
capital invested in securities  as the time tends to infinity for
different initial distribution of factor. We dwell on the case of
two and three risky assets and discuss the influence of parameters
of the model on the strategy. At last we compare our asset
allocation with its analogous for the long-run Bielecki and Pliska
strategy.

Sec.\ref{CIR} deals with the Cox-Ingersoll-Ross model of the
interest rate, we consider again the case of  portfolio consisting
of two assets.

In Sec.\ref{comparing_V_CIR} we compare the fixed time optimal
strategies for a portfolio consisting of  two assets for both model
of interest rate and conclude that the Cox-Ingersoll-Ross model is
in some sense preferable.

The formulae appearing in this work are sometime very cumbersome and
we have no possibility to write them out. To obtain them we used the
computer algebra system MAPLE.

\section{A stochastic market model}\label{market}

We study a portfolio optimization problem in the frame of the market
model of $\, m\geq2 $ assets and $n\geq1$ factors used by T.Bielecki
and S.Pliska (e.g. \cite{bielplis},\cite{biplsh}). Below we describe
this model.

Let $(\Omega, \{\mathcal{F}_{t}\}_{t\geq0}, \mathcal{F},
\mathbf{P})$ be the underlying probability space. Denoting by
$S_{i}$, $i=1,...,m$ the price of the $i$-th security and by
$X_{j},$ $j=1,..., n$ the level of the $j$-th factor at time t, we
consider the following market model for the dynamics of the security
prices and factors:
\begin{equation}
\begin{array}{ll}
\label{genAssets} \displaystyle \frac{dS_{i}(t)}{S_{i}(t)} =
\displaystyle (A_{i} + \sum_{p=1}^{n} \alpha_{ip} X_{p}(t))dt +
\sum_{k=1}^{m+n}\sigma_{ik} dW_{k}(t),\\
S_{i}(0)=s_{i}>0, i=1,..., m,
\end{array}
\end{equation}
\begin{equation}
\begin{array}{ll}
\label{genFactors} dX_{j}(t) = \displaystyle (B_{j} +
\sum_{p=1}^{n}\beta_{jp} X_{p}(t))dt +
\sum_{k=1}^{m+n} \lambda_{jk} dW_{k}(t), \\
X_{j}(0)=x_{j}, j=1,..., n,
\end{array}
\end{equation}
where $W(t)$ is a $\mathcal{R}^{m+n}$ -- valued standard Brownian
motion process with components  $W_{k}(t)$; $X(t)$ is the
$\mathcal{R}^{n}$ -- valued factor process with components
$X_{j}(t)$; the market parameters $A:=[A_{i}],$ $B:= [B_{j}],$
$\alpha:=[\alpha_{ip}],$ $\beta:=[\beta_{jp}],$
$\Sigma:=[\sigma_{ik}],$ $\Lambda:=[\lambda_{jk}]$ matrices of
appropriate dimensions. According to \cite{karashre} (chapter 5) a
unique, strong solution exists for \eqref{genAssets},
\eqref{genFactors}, and the processes $S_{i}(t)$ are positive with
probability~1.

Let $\mathcal{G}_{t}:=\sigma((S(s),X(s)),0\leq s\leq t)$, where
$S(t)=(S_{1}(t), ..., S_{m}(t))$ is the security price process. Let
$h(t)=(h_{1}(t), ..., h_{m}(t))$ denote an $\mathcal{R}^{m}$ valued
investment process or strategy whose component $h_{i}(t)$ represents
the proportion of capital that is invested in security $i$ at time
$t$. We define the admissible investment strategy according to
\cite{bielplis}.
\begin{definition}\label{defBP}
An investment process $h(t)$ is
\emph{admissible} if the following conditions are
{satisfied\footnote[1]{$(\cdot)^{T}$ stands for a transposition
operator}}:
$$
\begin{array}{ll}
(i) \displaystyle \quad \sum_{i=1}^{m}h_{i}(t)=1;\\

(ii) \quad h(t) $ is measurable, $ \mathcal{G}_{t}-$adapted$;\\

(iii) \displaystyle \quad
\mathbf{P}[\int_{0}^{t}h^{T}(s)h(s)ds<\infty]=1 $ for all finite $
t\geq0.
\end{array}
$$
\end{definition}

The class of admissible investment strategies will be denoted by
$\mathcal{H}$.

Let $h(t)$ be an admissible investment process. Then there exists a
unique, strong and almost surely positive solution $V(t)$ to the
following equation:
\begin{equation}
\begin{array}{ll}
\label{genVCapital} dV(t) = \displaystyle
\sum_{i=1}^{m}h_{i}(t)V(t)\left[(A_{i} + \sum_{p=1}^{n}\alpha_{ip}
X_{p}(t))dt + \sum_{k=1}^{m+n}\sigma_{ik}
dW_{k}(t)\right],\\
V(0)=v>0.
\end{array}
\end{equation}
The process $V(t)$ represents the investor's capital at time $t$
where $h_{i}(t)$ represents the proportion of capital that is
invested in security $i$.

\begin{remark} In \cite{davislleo} the model of the linear market
was extended to the case of  asset prices represented by SDEs driven
by Brownian motion and a Poisson random measure, with drifts that
are functions of an auxiliary diffusion factor process.
\end{remark}

\section{The optimal investment strategy by Bielecki and Pliska \cite{bielplis} }\label{BPstrategy}

A new kind of portfolio optimization model to the type of asset
allocation problems $\eqref{genAssets}$ considering a portfolio of
$m\geq2$ assets affected by $n\geq1$ financial and economic factors
was introduced by Bielecki and Pliska. Namely, they considered the
following {functional\footnote[2] {$\mathbf{E}(\cdot)$ and
$\mathbf{Var}(\cdot)$ are  expectation and variance  in the
probability space $(\Omega, \{\mathcal{F}_{t}\}_{t\geq0},
\mathcal{F}, \mathbf{P})$, respectively}}
$$J_{\theta} := \liminf_{t\rightarrow\infty}\frac{Q_{\theta}(t)}{t},
\mbox{\,уфх\,} Q_{\theta}(t) := \frac{-2}{\theta} \ln
\mathbf{E}(e^{(-\theta/2)\ln V(t)}),\quad \theta>-2, \quad
\theta\neq 0.$$ A Taylor expansion of $Q_{\theta}$ around $\theta=0$
{yields}
\begin{equation}
\label{TailorSeries} Q_{\theta}(t) = \mathbf{E} (\ln V(t)) -
\frac{\theta}{4} \mathbf{Var}(\ln V(t)) + O(\theta^{2}),
\end{equation}
hence $J_{\theta}$ can be interpreted as the long-run expected
growth rate minus a penalty term, with an error that is proportional
to $\theta^{2}$. The penalty term is also proportional to $\theta$,
so $\theta$ was interpreted as a risk sensitivity parameter or risk
aversion parameter, with $\theta>0$ and $\theta<0$ corresponding to
risk averse and risk seeking investors, respectively and $\theta=0$
is the risk null case.

Bielecki and Pliska \cite{bielplis} proposed to solve the following
family of risk sensitive optimal investment problems, labeled as
$(P_{\theta})$:
\begin{quote}
\emph{for $\theta\in(0, \infty)$ maximize the risk sensitized
expected growth rate
$$J_{\theta}(v,x; h(\cdot)) = \displaystyle
\liminf_{t\rightarrow\infty} \frac{-2}{\theta} t^{-1} \ln
\mathbf{E}[e^{(-\theta/2)\ln V(t)}| V(0)=v, X(0)=x]$$ over the class
of all admissible investment processes $h(\cdot)$, subject to
definition 1.3, where $V(t), X(t)$ obey equations
\eqref{genVCapital}, \eqref{genFactors}.}
\end{quote}

Authors noticed that $J_{\theta}$ has a  large-deviations-type
functional for the capital process $V(t)$. Maximizing $J_{\theta}$
for $\theta>0$ protects an investor interested in maximizing the
expected growth rate of the capital against large deviations of the
actually realized rate from the expectations.

\begin{remark} In case of $\theta<0$ the problem $(P_{\theta})$ can be solved
similar to the case of $\theta>0$. The case $\theta=0$ is considered
separately as a limiting case $\theta\rightarrow0$.
\end{remark}

An algorithm to find optimal investment strategy $H_\theta$ together
with corresponding maximum value of $\displaystyle J_{\theta}$
labeled as $\rho (\theta)$ was proposed in \cite{bielplis}. In order
to present the main results pertaining to these investment problems,
authors introduced the following notation for $\theta\geq0$ and
$x\in R^{n}$:
\begin{equation}
\label{Ktheta} \emph{K}_{\theta}(x) := \inf_{h\in\chi, 1^{T}h=1}
\left[\frac{1}{2} \left( \frac{\theta}{2}+1
\right)h^{T}\Sigma\Sigma^{T}h - h^{T}(A + \alpha x) \right],
\end{equation}
where $(A + \alpha x)$ denotes the vector with components $(A +
\alpha x)_{i}=\displaystyle (A_{i} +
\sum_{p=1}^{n}\alpha_{ip}x_{p})$.

Also they made the following assumptions:

\begin{assum}
\label{usl1} $\chi = \mathbb{R}^{n}$
\end{assum}

\begin{assum}
\label{usl2} $\displaystyle \lim_{\|x\|\rightarrow\infty}
\emph{K}_{\theta}(x)=-\infty$ for $\theta>0$. Here $\|\cdot\|$ is
the norm in $\mathbb{R}^{n}$.
\end{assum}

\begin{assum}
\label{usl3} The matrix $\Lambda\Lambda^{T}$ is positive definite.
\end{assum}

\begin{assum}
\label{usl4} The matrix $\Sigma\Lambda^{T}$ is zero.
\end{assum}

\begin{remark} (i) Note that if $\Sigma\Sigma^{T}$ is positive definite,
then assumption \ref{usl2} is implied by assumption \ref{usl1}

(ii) Assumptions  \ref{usl1}-\ref{usl4} are sufficient for the
results below, but Assumption \ref{usl2} is not necessary (see
\cite{biplsh}, Sec. 4).
\end{remark}

Following two theorems contain key results concerning the solution
of the problem   $(P_{\theta})$.

\begin{theorem}\cite{bielplis}
\label{teor1} Assume \ref{usl1}-\ref{usl4}. For a fixed value of
$\theta>0$ let $H_{\theta}(x)$  denote a minimizing selector in
\eqref{Ktheta}, that is
$$\emph{K}_{\theta}(x) := \left[\frac{1}{2} \left( \frac{\theta}{2}+1 \right)H_{\theta}(x)^{T}\Sigma\Sigma^{T}H_{\theta}(x)
- H_{\theta}(x)^{T}(A + \alpha x) \right].$$ Then the investment
process $h_{\theta}$ is optimal for problem $(P_{\theta})$, where
$\forall t\geq 0$
\begin{equation}
\label{Hoptim} h_{\theta}(t) = H_{\theta}(X(t)).
\end{equation}
\end{theorem}

\begin{theorem}\cite{bielplis}
\label{teor2} Let us assume \ref{usl1}-\ref{usl3} and consider
problem $(P_{\theta})$ for a fixed value of $\theta>0$. Let
$h_{\theta}(t)$ satisfy theorem \ref{teor1}. Then
\begin{enumerate}
\item
For all $v>0$ ш $x\in R^{n}$ we have
$$
\begin{array}{rcl}
J_{\theta}(v,x; h_{\theta}(\cdot)) &=& \displaystyle
\lim_{t\rightarrow\infty} \left( \frac{-2}{\theta} \right) t^{-1}
\ln
\mathbf{E}[e^{(-\theta/2)\ln V(t)}| V(0)=v, X(0)=x]\\
&:=& \rho(\theta).
\end{array}
$$
\item
The constant $\rho(\theta)$ is the unique non-negative constant
which is a part of the solution $(\rho(\theta),v(x;\theta))$ to the
following equation:
\begin{equation}
\begin{array}{rcl}
\label{PDErhotheta} \rho &=& \displaystyle (B + \beta x)^{T}grad_{x}
v(x) - \frac{\theta}{4} \sum_{i,j=1}^{n} \frac{\partial
v(x)}{\partial x_{i}} \frac{\partial
v(x)}{\partial x_{j}} \sum_{k=1}^{n+m}\lambda_{ik}\lambda_{jk}+\\
& &  \displaystyle +\frac{1}{2} \sum_{i,j=1}^{n} \frac{\partial^2
v(x)}{\partial x_{i}
\partial x_{j}} \sum_{k=1}^{n+m} \lambda_{ik}\lambda_{jk} -
\emph{K}_{\theta}(x),\\
& & \displaystyle v(x)\in\mathbf{C}^{2}(R^{n}), \quad
\lim_{\|x\|\rightarrow \infty} v(x) = \infty, \quad \rho = const.
\end{array}
\end{equation}
\end{enumerate}
\end{theorem}

It remains to consider the case corresponding to $\theta=0$. This is
the classical problem of maximizing the portfolio's expected growth
rate, that is, the growth rate under the log-utility function (see
e.g. Karatzas \cite{karatzas}). This problem was labeled $(P_{0})$
and formulated as follows:
\begin{quote}
\emph{maximize the functional
$$J_{0}(v,x; h(\cdot)) =
\displaystyle \liminf_{t\rightarrow\infty} t^{-1} \ln \mathbf{E}[\ln
V(t)| V(0)=v, X(0)=x]$$ over the class of all admissible investment
processes $h(\cdot)$ subject to definition 1.3, where $V(t), X(t)$
described by equations \eqref{genVCapital}, \eqref{genFactors}.}
\end{quote}

It turns out that to solve $(P_{0})$ it is necessary to make three
additional assumptions:

\begin{assum}
\label{usl5} For each $\theta \geq 0$ the function
$\emph{K}_{\theta}(x)$ defined in \eqref{Ktheta} is of the quadratic
form
$$\emph{K}_{\theta}(x) = \frac{1}{2}x^{T}K_{1}(\theta)x + K_{2}(\theta)x + K_{3}(\theta),$$
where $K_{1}(\theta),$  $K_{2}(\theta)$ ш  $K_{3}(\theta)$ are
functions of appropriate dimensions depending only on $\theta$.
\end{assum}

\begin{assum}
\label{usl6} For each $\theta \geq 0$ the matrix $K_{1}(\theta)$ is
symmetric and negative definite.
\end{assum}

\begin{assum}
\label{usl7} The $n \times n$ matrix $\beta$ with components
$\beta_{jp}$ in \eqref{genFactors} is stable.
\end{assum}

\begin{remark}
(i) Assumption \ref{usl5} is satisfied if, for example, the matrix
$\Sigma\Sigma^{T}$ is non-singular  and $\chi = R^{n}$.

(ii) According to \cite{bankgkkt}  Assumption \ref{usl5} implies $\,
\displaystyle \lim_{\theta\rightarrow0} K_{i}(\theta) = K_{i}(0)$
for $i=1, 2, 3$.
\end{remark}

In order to establish  relationships between the risk-neutral
problem $(P_{0})$ and the risk-sensitive problem $(P_{\theta}),$
$\theta>0$ we consider the following equation:
\begin{equation}
\begin{array} {rcl}
& & \label{rho0} \rho(0) = \displaystyle (B + \beta x)^{T} grad_{x}
v_{0}(x) + \frac{1}{2} \sum_{i,j=1}^{n}
\frac{\partial^{2}v_{0}(x)}{\partial x_{i} \partial x_{j}}
\sum_{k=1}^{n+m} \lambda_{ik}\lambda_{jk} -
K_{0}(x),\\
& & \displaystyle v_{0}(x)\in C^{2}(R^{n}), \quad \lim_{\parallel
x\parallel\rightarrow\infty} v_{0}(x) = \infty, \quad \rho(0) =
const.
\end{array}
\end{equation}

The following two results are true:

\begin{theorem}\cite{bielplis}
\label{teor3} Assume \ref{usl3}-\ref{usl7}. Then the optimal
strategy for $(P_{0})$ is as in Theorem \ref{teor1} with $\theta=0,$
and the optimal objective value $\rho(0)=\rho(\theta)$ in Theorem
\ref{teor2} with $\theta=0.$ Moreover, the optimal objective values
$\rho(\theta), \theta>0,$ being the solution of \eqref{PDErhotheta},
converge to the optimal objective value $\rho(0)$ as
$\theta\rightarrow0$.
\end{theorem}

The next result characterizes the portfolio's expected growth rate
under the optimal investment strategy for the risk aversion level
$\theta>0$. We denote this growth rate by $\rho_{\theta}$, which is
to be distinguished from the optimal objective value $\rho(\theta)$,
as in Theorem \ref{teor1}.

\begin{theorem}\cite{bielplis}
\label{teor4} Assume \ref{usl3}-\ref{usl7}. Fix $\theta>0$, let
$H_{\theta}(x)$ be as in theorem \ref{teor1} and suppose that
$H_{\theta}(x)$ is an affine function and that
$$\lim_{\|x\|\rightarrow\infty} [\frac{1}{2}
H_{\theta}(x)^{T}\Sigma\Sigma^{T}H_{\theta}(x) - H_{\theta}(x)^{T}(A
+ \alpha x)] = - \infty.$$ Consider the equation
\begin{equation}
\begin{array} {rcl}
\label{rhotheta} \rho_{\theta} = \displaystyle (B + \beta x)^{T}
grad_{x} v_{\theta, 0}(x) + \frac{1}{2} \sum_{i,j=1}^{n}
\frac{\partial^{2}v_{\theta, 0}(x)}{\partial x_{i} \partial x_{j}}
\sum_{k=1}^{n+m} \lambda_{ik}\lambda_{jk} -\\
\displaystyle - [\frac{1}{2} H_{\theta}(x)^{T}\Sigma\Sigma^{T}H_{\theta}(x) - H_{\theta}(x)^{T}(A + \alpha x)],\\
\displaystyle v_{\theta, 0}(x)\in C^{2}(R^{n}), \quad
\lim_{\parallel x\parallel\rightarrow\infty} v_{\theta, 0}(x) =
\infty, \quad \rho_{\theta} = const.
\end{array}
\end{equation}
Then there exists a solution $(\rho_{\theta}, v_{\theta,0})$ the
preceding equation, the constant $\rho_{\theta}$ is unique, and we
have
$$J_{0}(v,x;h_{\theta}(\cdot)) = \rho_{\theta}$$
for all $(v,x)\in(0,\infty)\times R^{n},$ where $h_{\theta}(\cdot)$
is defined as in \eqref{Hoptim}.
\end{theorem}

 The
main result of Bielecki and Pliska is that the optimal investment
strategy problem is converted to the problem of solution to PDE
\eqref{PDErhotheta} . Authors solve the problem explicitly for the
classical example of  portfolio consisting of two assets, where one
of them is a bank account, and a linear interest rate as a factor.

Namely, they consider a  single  risky   asset,  say a   stock
index, that   is governed by  a stochastic  differential equation
$$\frac{d S_{1}(t)}{S_{1}(t)} = (A_{1} + \alpha_{1} R(t))dt + \sigma_{1} d W_{1}(t), \quad S_{1}(0) = s > 0,$$
where the spot interest rate $R(t)$ is satisfies the classical
Vasicek dynamics:
$$dR(t) = (B + \beta R(t))dt + \lambda dW_{2}(t), \quad R(0) = r. $$
Here $A_{1}, \alpha_{1}, B, \beta, \sigma_{1}, \lambda$ are fixed,
scalar parameters to be estimated, while $W_{1}, W_{2}$ are two
independent Brownian motions. Hereafter we assume $B>0, \beta<0$. in
all that follows.

The investor can take a long or short position in the stock index as
well as borrow or lend money, with continuous compounding, at the
prevailing interest rate. It is therefore convenient to follow the
common approach and introduce the "bank account" process $S_{2}$,
where
$$\frac{dS_{2}(t)}{S_{2}(t)} = R(t) dt.$$

Thus $S_{2}(t)$ represents the time $t$ value of a savings account
when $S_{2}(0)=1$ dollar is deposited at the zero time.

With only two assets it is convenient to describe the investor's
trading strategy in terms of the scalar valued function $H_{\theta}$
which is interpreted  as the proportion of capital invested in the
stock index, leaving  the proportion $1-H_{\theta}$ invested    in
the bank account.

This enables us to formulate the investor's problem as
$(P_{\theta})$ in the market model \eqref{genAssets},
\eqref{genFactors} for there are $m=2,$ $n=1,$ and we can set
$(h_{1},h_{2})=(H_{\theta},1-H_{\theta}),$ $X(t)=R(t),$ $B=B,$
$\beta=\beta,$ $\Lambda=(0,0,\lambda)^{T},$ $A=(A_{1},0)^{T},$
$\alpha=(\alpha_{1},1)^{T},$
$$
\Sigma = \left(
\begin{array} {ccc}
\sigma_{1} & 0 & 0 \\
         0 & 0 & 0
\end{array}
\right).
$$

According to Theorem \ref{teor1}
$$K_{\theta}(R) = \inf_{h\in \mathbb{R}} [(1/2)(\theta/2 + 1)(h,1-h)\Sigma\Sigma^{T}(h,1-h)^{T} - (h,1-h)(A + \alpha R)],$$
in which case
\begin{equation}
\label{exBPH} H_{\theta} := H_{\theta}(R) =
\frac{A_{1}+(\alpha_{1}-1)R}{(1+\frac{\theta}{2})\sigma_1^{2}},
\end{equation}

$$
K_{\theta}(R) = -R
-\frac{(A_{1}+(\alpha_{1}-1))^2R^2}{(\theta+2)\sigma_{1}^2}.
$$
Theorem \ref{teor2} implies that $\rho(\theta)$ is the part fo
solution $(\rho,v)$ of the equation
$$\rho = \frac{1}{2}\lambda^2 v''(R) + (B + \beta R)v'(R) - \frac{\theta}{4}\lambda^2 (v'(R))^2 - K_{\theta}(R),$$
and, according to \cite{biplsh}, equals
\begin{equation}
\label{exBP} \rho(\theta) = \lambda^{2}N_{1} + B
N_{2}-\frac{\lambda^{2} \theta
N_{2}^{2}}{4}+\frac{A_{1}^{2}}{(\theta+2)\sigma_{1}^{2}},
\end{equation}
where
$$\displaystyle
N_{1}=\frac{\beta+\sqrt{\beta^{2}+\frac{\theta\lambda^{2}(\alpha_{1}-1)^{2}}{(\theta+2)\sigma_{1}^{2}}}}{\lambda^{2}\theta},
\displaystyle
N_{2}=\frac{1-\frac{2A_{1}(\alpha_{1}-1)}{(\theta+2)\sigma_{1}^{2}}+2B
N_{1}}{\sqrt{\beta^{2}+\frac{\theta\lambda^{2}(\alpha_{1}-1)^{2}}{(\theta+2)\sigma_{1}^{2}}}}.$$

 Theorem \ref{teor4} results that the solution of equation
$$
\begin{array} {rcl}
\rho &=& \frac{1}{2}\lambda^2 v''(R) + (B + \beta R)
v'(R)-\\
&& -
\Big[\frac{1}{2}(H_{\theta}(R),1-H_{\theta}(R))\Sigma\Sigma^{T}(H_{\theta}(R),1-H_{\theta}(R))^{T}
-\\
&& - (H_{\theta}(R),1-H_{\theta}(R))(A + \alpha R) \Big],
\end{array}
$$
gives
\begin{equation}
\label{exBP0} \rho_{\theta} = -\frac{B}{\beta} + \frac{2(\theta +
1)}{(\theta + 2)^2\sigma^2} \Big[ [A_{1} -
\frac{B}{\beta}(\alpha_{1}-1)]^2 -
\frac{\lambda^2(\alpha_{1}-1)^2}{2\beta} \Big].
\end{equation}

We note that Bielecki and Pliska introduced an optimal investment
strategy to maximize portfolio return to an infinite time horizon.
In this paper we introduce another strategy which can be used by
investor to manage portfolio and maximize return at any fixed time
moment.


\section{Conditional expectation and variance for a couple of SDEs: two
approaches to the solution}\label{algorithm} Let us consider a
system of stochastic differential equations
\begin{equation}
\begin{array}{ll}
\label{genSDE}
dF = A(t, F, X) dt + \sigma(t,F,X) dW_{1}, \\
dX = B(t,F,X)dt + \lambda(t,F,X) dW_{2}, \\
F(0)=f, X(0)=x, t\geq 0, \, f \in \mathbb R, x \in \mathbb R,
\end{array}
\end{equation}
where $W=(W_{1},W_{2}) $ ia a two-dimensional Brownian motion with
independent components, $A,B,\sigma,\lambda$ are given functions.

The joint distribution density $P(t,f,x)$ of stochastic variables
$F$ and $X$ is described by the Fokker-Plank equation (e.g.
\cite{shiryaev}, \cite{risken})
\begin{equation}
\begin{array}{rcl}
\label{genFPK} \displaystyle \frac{\partial P(t,f,x)}{\partial t} &
= & \displaystyle - \frac{\partial A(t,F,X)P(t,f,x)}{\partial
f}+\frac{1}{2}\frac{\partial^2
\sigma^2(t,F,X)P(t,f,x)}{\partial f^2}\\
& &  \displaystyle - \frac{\partial B(t,F,X) P(t,f,x)}{\partial
x}+\frac{1}{2}\frac{\partial^2 \lambda^2(t,F,X)P(t,f,x)}{\partial
x^2}
\end{array}
\end{equation}
with initial data
\begin{equation}
\label{genID} P(0,f,x)=P_{0}(f,x),
\end{equation}
determined by initial distributions of $F$ and $X$.

Provided $P(t,f,x)$ is known, the conditional expectation of $F$
with given value of $X$ at the  moment $t$ can be found by the
following formula (see, e.g. \cite{shiryaev}, \cite{chorin})
\begin{equation}
\label{genCME} \bar {f}(t,x) := \mathbf{E}(F|X=x) = \frac{\int_{
\mathbb R} f P(t,f,x)d f}{\int_{ \mathbb R} P(t,f,x)d f}.
\end{equation}
If we set $P_{0}(f,x)=\delta (f-f_{0})g(x),$ where
$f_{0}\in\mathbb{R}$ and $g(x)$ is an arbitrary function,
$\int_{\mathbb{R}}g(x)dx=1$, then $\bar{f}(0,x) = f_{0}.$ Some
characteristics of \eqref{genCME} are studied in \cite{alroz1},
\cite{alroz2}.

The conditional variance of a stochastic variable $F$ with given
value of $X$ at the moment $t$ is defined as follows:
\begin{equation}
\label{genCVar} \bar {v}(t,x) := \mathbf{Var}(F|X=x) = \frac{\int_{
\mathbb R} f^2 P(t,f,x)d f}{\int_{ \mathbb R} P(t,f,x)d
f}-\bar{f}^2(t,x).
\end{equation}

The fundamental solution of  equation \eqref{genFPK} can be found by
means of the Riccati matrix equations \cite{yau2004},
\cite{corlopsuasus2008}.  For some simple but important for
application choice of initial data  problem \eqref{genFPK},
\eqref{genID} can be solved in terms of elementary functions.

Moreover, sometimes the Fourier transform of $P(t,f,x)$ can be found
easier than this function itself. Further we are providing two
approaches to the problem.

\subsection{Approach 1: a reduction to the system of ODEs} \label{apr1}Let us assume that the
variables $F$ and $X$ obey the following system  of stochastic
differential equations:
\begin{equation}
\begin{array}{ll}
\label{constSDE}
dF(t) = (A + \alpha_{1} X(t) + \alpha_{2} F(t)) dt + \sigma_{1} dW_{1}(t) + \sigma_{2} dW_{2}(t), \\
dX(t) = (B + \beta_{1} X(t) + \beta_{2} F(t))dt + \lambda_{1} dW_{1}(t) + \lambda_{2} dW_{2}(t), \\
F(0)=f, X(0)=x, t\geq0, f, x \in \mathbb{R},
\end{array}
\end{equation}
where $W(t)=(W_{1}(t),W_{2}(t)) $ is a two-dimensional Brownian
motion; $A,B, \alpha_{i}, \beta_{i}, \sigma_{i}, \lambda_{i}$~are
known smooth functions of $t$, $i=1,2$.

Then joint distribution density $P(t,f,x)$ of  $F$ and $X$ solves
the Fokker-Planck equation
\begin{equation}
\begin{array}{rcl}
\label{constFPK} & & \displaystyle \frac{\partial P(t,f,x)}{\partial
t} = -(A(t)+\alpha_{1}(t) x) \displaystyle \frac{\partial
P(t,f,x)}{\partial f} -\\
& & - \displaystyle \alpha_{2}(t) \Big(P(t,f,x)+ f \displaystyle
\frac{\partial P(t,f,x)}{\partial f}\Big)-
(B(t)+\beta_{2}(t) f)\displaystyle \frac{\partial P(t,f,x)}{\partial x} -\\
& & - \displaystyle \beta_{1}(t) \Big(P(t,f,x)+x \displaystyle
\frac{\partial P(t,f,x)}{\partial x}\Big) +
\frac{1}{2}(\sigma_{1}^{2}(t) + \sigma_{2}^{2}(t)) \displaystyle
\frac{\partial^{2} P(t,f,x)}{\partial f^{2}} +\\
& & + \displaystyle(\sigma_{1}(t) \lambda_{1}(t) + \sigma_{2}(t)
\lambda_{2}(t)) \displaystyle \frac{\partial^{2} P(t,f,x)}{\partial
f
\partial x}+ \frac{1}{2}(\lambda_{1}^{2}(t) + \lambda_{2}^{2}(t))
\displaystyle \frac{\partial^{2}P(t,f,x)}{\partial x^{2}}
\end{array}
\end{equation}
subject to  initial data
\begin{equation}
\label{constID} P(0,f,x)=P_{0}(f,x)=\delta (f-f_{0})\,g(x).
\end{equation}

We perform the Fourier transform of $P(t,f,x)$ in $f$ of
$\eqref{constFPK}$ and $\eqref{constID}$ and get:
\begin{equation}
\begin{array}{rcl}
\label{FconstFPK} & & \displaystyle\frac{\partial
\hat{P}(t,\mu,x)}{\partial t} =  -(A(t)+\alpha_{1}(t) x)\mu
\hat{P}(t,\mu,x)i -\\
& & - \alpha_{2}(t)\left(\hat{P}(t,\mu,x) - \mu
\displaystyle\frac{\partial \hat{P}(t,\mu,x)}{\partial \mu}\right) -
(B(t)+\beta_{1}(t) x)\displaystyle\frac{\partial \hat{P}(t,\mu,x)}{\partial x} -\\
& & - \beta_{1}(t)\hat{P}(t,\mu,x) - \beta_{2}(t)
\displaystyle\frac{\partial^{2}\hat{P}(t,\mu,x)}{\partial x \partial
\mu}i -\\
& & -\displaystyle\frac{1}{2}(\sigma_{1}^{2}(t)+\sigma_{2}^{2}(t))
\mu^{2} \hat{P}(t,\mu,x) +
\displaystyle\frac{1}{2}(\lambda_{1}^{2}(t) + \lambda_{2}^{2}(t))
\displaystyle\frac{\partial^{2}\hat{P}(t,\mu,x)}{\partial x^{2}} +\\
& & + (\sigma_{1}(t) \lambda_{1}(t) + \sigma_{2}(t) \lambda_{2}(t))
\mu \displaystyle\frac{\partial \hat{P}(t,\mu,x)}{\partial x}i,
\end{array}
\end{equation}

\begin{equation}
\label{FconstID} \hat{P}(0,\mu,x) = \frac{1}{\sqrt{2 \pi}}e^{-i \mu
f_0}g(x).
\end{equation}

To obtain explicit formulae we restrict ourselves to the case
\begin{equation}
\label{FconstSpecID}
g(x)=\frac{e^{\frac{-(x-x_0)^2}{2s^2}}}{s\sqrt{2\pi}},
\end{equation}
where $x_0\in\mathbb{R},$ $s\in\mathbb{R}_{+}.$ Here  $x_0$ is the
mean of the variable $X$  at the initial time, $s^2$ is the
variance. Thereafter we seek the solution of the problem
\eqref{FconstFPK}, \eqref{FconstID} using the  following ansatz:
\begin{equation}
\label{FurPview} \hat{P}(t,\mu,x) = \frac{e^{\gamma_{1}(t) +
\gamma_{2}(t) \mu + \gamma_{3}(t)x + \gamma_{4}(t) \mu^{2} +
\gamma_{5}(t)x \mu + \gamma_{6}(t)x^{2}}}{s\sqrt{2 \pi}}.
\end{equation}

We substitute  \eqref{FurPview} into  \eqref{FconstFPK} and
\eqref{FconstID}. Equating coefficients at the powers of $\mu$ and
$x$ gives  the   system for $\gamma_{j}(t), j=1,...,6$:
\begin{equation}
\begin{array}{rcl}
\label{ODE} \displaystyle \frac{\partial\gamma_1(t)}{\partial t} &=&
\frac{1}{2} \lambda_1^2(t) \gamma_3^2(t)+\lambda_2^2(t)
\gamma_6(t)+\frac{1}{2} \lambda_2^2(t)
\gamma_3^2(t)-\alpha_2(t)-\beta_1(t)-\\
&& -B(t) \gamma_3(t)-i \beta_2(t) \gamma_5(t)+\lambda_1^2(t)
\gamma_6(t)-i \beta_2(t)
\gamma_2(t) \gamma_3(t),\\

\displaystyle \frac{\partial\gamma_2(t)}{\partial t} &=&
\left(\lambda_1^2(t)+\lambda_2^2(t)\right) \gamma_3(t)
\gamma_5(t)-B(t)
\gamma_5(t) + \alpha_2(t) \gamma_2(t)-\\
&& -2 i \beta_2(t) \gamma_4(t) \gamma_3(t)+\left(\sigma_1(t)
\lambda_1(t) +\sigma_2(t) \lambda_2(t)\right) \gamma_3(t) i -\\
&& - i A(t)-i \beta_2(t) \gamma_2(t) \gamma_5(t),\\

\displaystyle \frac{\partial\gamma_3(t)}{\partial t}  &=&
-\beta_1(t) \gamma_3(t)-i \beta_2(t) \gamma_5(t) \gamma_3(t)-2 B(t)
\gamma_6(t) -\\
&& -2 i \beta_2(t) \gamma_2(t) \gamma_6(t)+2 (\lambda_1^2(t) +
\lambda_2^2(t)) \gamma_3(t) \gamma_6(t),\\

\displaystyle \frac{\partial\gamma_4(t)}{\partial t}  &=&
\frac{1}{2} (\lambda_1^2(t) + \lambda_2^2(t)) \gamma_5^2(t)-2 i
\beta_2(t) \gamma_4(t) \gamma_5(t)+2 \alpha_2(t)
\gamma_4(t) -\\
&& -\frac{1}{2} \sigma_1^2-\frac{1}{2} \sigma_2^2(t)+(\sigma_1(t)
\lambda_1(t) + \sigma_2(t) \lambda_2(t)) \gamma_5(t) i,\\

\displaystyle \frac{\partial\gamma_5(t)}{\partial t}  &=&
-\beta_1(t) \gamma_5(t)-i \beta_2(t) \gamma_5^2(t)+\alpha_2(t)
\gamma_5(t)-i \alpha_1(t)-\\
&& -4 i \beta_2(t) \gamma_4(t) \gamma_6(t) + 2 (\lambda_1^2(t) + \lambda_2^2(t)) \gamma_5(t) \gamma_6(t) +\\
&& + 2 i (\sigma_1(t) \lambda_1(t) + \sigma_2(t)
\lambda_2(t))\gamma_6(t),\\

\displaystyle \frac{\partial\gamma_6(t)}{\partial t}  &=& -2
\beta_1(t) \gamma_6(t)-2 i \beta_2(t) \gamma_5(t) \gamma_6(t) + 2
(\lambda_1^2(t) + \lambda_2^2(t) )\gamma_6^2(t),
\end{array}
\end{equation}
with initial data
\begin{equation}
\begin{array}{rcl}
\label{ODEID} \gamma_{1}(0) &=& \displaystyle -\frac{x_0^2}{2s^2},
\quad
\gamma_{2}(0)=-if_0, \quad \gamma_{3}(0)=\displaystyle \frac{x_0}{s^2},\\
\gamma_{4}(0) &=& 0, \quad \gamma_{5}(0)=0, \quad
\gamma_{6}(0)=\displaystyle -\frac{1}{2s^2}.
\end{array}
\end{equation}

If we succeed to solve the problem \eqref{ODE}, \eqref{ODEID}
explicitly then we find $\hat{P}(t,\mu,x)$ substituting
$\gamma_{j}(t), j=1,..., 6,$ into \eqref{FurPview}. Further we apply
the inverse Fourier transform to find the function $P(t,f,x)$ and
from \eqref{genCME} we obtain  after  integration   $\bar{f}(t,x)$
under  given initial data \eqref{FconstSpecID}.

Let us we assume that in \eqref{constID}
$$g(x)=\frac{1}{2L}\chi_{[-L,L]}(x)=\frac{1}{2L}, \quad x\in[-L,L].$$
This choice corresponds to the initial uniform distribution of the
stochastic variable $X$ on the segment $[-L,L].$ Here $\bar{f}(t,x)$
should be read as:
\begin{equation}
\label{genLCME} \bar{f}(t,x)=\lim_{L\rightarrow+\infty}
\frac{\int_{[-L,L]}fP(t,f,x)df}{\int_{[-L,L]}P(t,f,x)df}.
\end{equation}
Consequently,
\begin{equation}
\label{genLCVar} \bar{v}(t,x)=\lim_{L\rightarrow+\infty}
\frac{\int_{[-L,L]}f^2P(t,f,x)df}{\int_{[-L,L]}P(t,f,x)df} -
\bar{f}^2(t,x).
\end{equation}
Therefore the problem of finding $\hat{P}(t,\mu,x)$ is reduced to
the solution of ODE system \eqref{ODE}   with initial data
$\gamma_{1}(0) = 0, \, \gamma_{2}(0)=-if_0, \, \gamma_{3}(0)=0,
\gamma_{4}(0) = 0, \, \gamma_{5}(0)=0, \, \gamma_{6}(0)=0.$

Below we consider a special cases of system \eqref{constSDE} arising
from economic applications where function $\hat{f}(t,x)$ can be
obtained in an explicit form.

\subsection{Approach 2: a representation in terms of the Fourier
transform}\label{apr2}

We denote by $\hat P (t, \mu, \xi)$ the Fourier transform in
variables $(f, x)$ of the function $P(t,f,x)$ being the solution of
the problem \eqref{genFPK}, \eqref{genID}. Let us assume  that $\hat
P (t, 0, \xi)$ and $\partial_\mu \hat P(t,0,\xi)$  are functions
decreasing at infinity with respect to $\xi$ faster then any power
of it. Then  $\bar{f}(t,x),$ defined in \eqref{genCME} can be
obtained as follows:
\begin{equation}
\label{FurCME} \bar {f}(t,x)=\frac{i F^{-1}_\xi \,  [\partial_\mu
\hat P(t,0,\xi)](t,x) } {F^{-1}_\xi \,  [ \hat P(t,0,\xi)](t,x)
},\quad t\ge 0,\quad x\in \mathbb{R}.
\end{equation}
Hereinafter we denote by $F^{-1}_\mu$ and $F^{-1}_\xi$ the inverse
Fourier transforms in variables $\mu$ and $\xi,$ correspondingly,
let $(\cdot,\cdot)_\mu$ be the action distribution on a trial
function of the variable $\mu.$ Here $(e^{i\mu f}, 1)_f$ means
$\displaystyle \lim_{L\to\infty}(e^{i\mu f},
\omega_\varepsilon(f)\ast \chi_{[-L,L]})_f, $ where $\chi_\Omega$ is
the indicator of the set $\Omega$ and $\omega_\varepsilon(f)$  is
the standard mollifier. The proof of \eqref{FurCME} is an exercise
in the harmonic analysis \cite{martroz}.

Namely, the denominator of \eqref{genCME} is
$$ \int\limits_{\mathbb{R}}
P(t,f,x)\, df = \int\limits_{\mathbb{R}} F^{-1}_\mu [F^{-1}_\xi[\hat
P(t,\mu,\xi)]]\, df =$$
$$=F^{-1}_\xi[\Big( F^{-1}_f[1](\mu),\hat P(t,\mu,\xi)\Big)_\mu]=
\sqrt{2\pi}  \, F^{-1}_\xi[\Big( \delta(\mu),\hat
P(t,\mu,\xi)\Big)_\mu]=$$
$$ = \sqrt{2\pi} F^{-1}_\xi[\hat
P(t,0,\xi)].
$$


Analogously  we compute the numerator:
$$
\int\limits_{\mathbb{R}}f  P(t,f,x)\, df = \int\limits_{\mathbb{R}}f
F^{-1}_\mu [F^{-1}_\xi[\hat P(t,\mu,\xi)]]\, df =$$
$$= F^{-1}_\xi[\Big( F^{-1}_f[f](\mu),\hat
P(t,\mu,\xi)\Big)_\mu] = -\sqrt{2\pi} i\, F^{-1}_\xi[\Big(
\delta'(\mu),\hat P(t,\mu,\xi)\Big)_\mu]=$$
$$=\sqrt{2\pi} i\,
F^{-1}_\xi[\Big( \delta(\mu),\partial_\mu \hat
P(t,\mu,\xi)\Big)_\mu] = i \sqrt{2\pi} F^{-1}_\xi[\partial_\mu \hat
P(t,0,\xi)].
$$


The conditional variance of $F$ at a given value of $X$  defined by
formula \eqref{genCVar} can be represented in terms of the Fourier
transform of the joint distribution density $P(t,f,x)$ as follows:
\begin{equation}
\label{FurCVar} \bar {v}(t,x)=\frac{ (F^{-1}_\xi \, [\partial_\mu
\hat P(t,0,\xi)])^{2} - F^{-1}_\xi \, [\partial_\mu^{2} \hat
P(t,0,\xi)] F^{-1}_\xi \,  [ \hat P(t,0,\xi)]} { (F^{-1}_\xi \,  [
\hat P(t,0,\xi)])^{2} } (t,x).
\end{equation}

We will apply the formulae to the case when the  $X$ factor
volatility is proportional to the square root of the factor. Such
model falls into the category of affine models \cite{duffilsch2003}
therefore the Fokker-Planck equation is integrated in quadratures.

\begin{remark}
Affine models are popular in financial mathematics since many
problems can be solved analytically in their frame. In particular,
affine models include Merton, Vasicek, Cox-Ingersoll-Ross interest
rate models (see \cite{daising2000}, \cite{grassteba2008},
\cite{singleton2001}).
\end{remark}

\section{A problem of the portfolio selection at a fixed
time}\label{portfolio_selection}

\subsection{The problem statement} Let us recall that the
optimal strategy by Bielecki and Pliska corresponds to the infinite
time horizon. we are going to  demonstrate another strategy which
can be used by investor to manage portfolio and maximize return at
any fixed time moment.

We consider the market model of the security prices and factors
$\eqref{genAssets},$ $\eqref{genFactors}$ and investment process
$\eqref{genVCapital}$ defined in the Bielecki and Pliska model.
Denoting $F(t)=\ln V(t)$ and using $\rm It\hat{o}$ formula
\cite{oksendal} we derive following equation from
$\eqref{genVCapital}$:

\begin{equation}
\begin{array}{rcl}
\label{genFCapital} dF(t) &=&  \displaystyle
\left[\sum_{i=1}^{m}(h_{i}A_{i}-\frac{1}{2}h_{i}^{2}
\sum_{k=1}^{m+n}\sigma_{ik}^{2}) + \sum_{i=1}^{m}h_{i}
\sum_{p=1}^{n}\alpha_{ip}X_{p}(t)\right]dt +\\
&&  \displaystyle + \sum_{i=1}^{m}h_{i}
\sum_{k}^{m+n}\sigma_{ik}dW_{k}(t),\\
&& F(0)=\ln V(0)=f.
\end{array}
\end{equation}

We define a functional $\bar{Q}_{\gamma}(t, x; h)$ analogous to the
first two elements of the Taylor series of $Q_{\theta}(t)$ about
$\theta=0$ in the Bielecki and Pliska model, that is
\begin{equation}
\label{FTfunc} \bar{Q}_{\gamma}(t, x; h) = \bar{f}(t,x; h) - \gamma
\bar{v}(t, x; h), \quad x=(x_{1},...,x_{n}),
\end{equation}
where $\gamma=\frac{\theta}{4}\geq 0$ is a risk sensitive parameter
analogous to $\theta$ in the Bielecki and Pliska model, $ \bar{f}(t,
x; h)$ and $ \bar{v}(t, x; h)$ are conditional expectation and
conditional variance of stochastic variable $F(t)$ with given values
of $X_{1}(t)=x_{1},..., X_{n}(t)=x_{n}$. Then we solve the following
problem:
\begin{quote}
\emph{to find  $\displaystyle \max_{h=(h_1,...,h_m)}
\bar{Q}_{\gamma}(t, x; h)$, $x=(x_{1},...,x_{n})$, over the class of
admissible investment strategies ${h}$ (see  Definition \ref{def}),
with given values of the factors $X_{1}(t)=x_{1},...,
X_{n}(t)=x_{n}$ at a given  moment of time $t$}.
\end{quote}
\begin{definition}\label{def}
A strategy $\bar{H}_{\gamma}$ is  called \emph{optimal} strategy if
it gives a maximum of the functional $\bar{Q}_{\gamma}(t,x; h)$ with
given values of the factors $X_1(t)=x_1,..., X_n(t)=x_n$ at a given
time $t$.
\end{definition}

\vspace{0.7cm} Once we find the maximum over the class of denoted
strategies then we find the strategy which provides the maximum
portfolio return with regard to loss of random nature described by
the variance. Changing the value of parameter $\gamma$, we can
overstate or understate the role of randomness, or do not take into
account the randomness at all, setting $\gamma=0$. The model can be
interpreted in the following way. Let us assume the investor be
going to allocate the initial capital between assets $S_{i}$,
$i=1,..., m$, with the prices depending on a set of exogenous
economic factor  $X_{j}$, $j=1,..., n$. The prices of assets and
values of factors obey equations $\eqref{genAssets},$
$\eqref{genFactors}$. The investor solves a dynamic asset management
problem featuring a risk sensitive optimality criterion. Let us
assume that  the investor knows an explicit values of factors in a
fixed moment of time. Thus, the investor has to find the optimal
portfolio  taking into account  a new information on the factors,
such that the model is flexible and can be actualized within all the
time of investment. The model refers to the tactical asset
allocation (e.g. \cite{rey2003}).

\subsection{The algorithm of solution}\label{algorithm_1} Let us give a outline of solution to the optimization problem for a model with one
factor, that is we consider system $\eqref{genFCapital},$
$\eqref{genFactors}$ for $n=1$ (here $X_{1}(t)=X(t)$):
\begin{equation}
\begin{array}{rcl}
\label{1FactorFCapital} dF(t) & = & \displaystyle
\left[\sum_{i=1}^{m}(h_{i}A_{i}-\frac{1}{2}h_{i}^{2}
\sum_{k=1}^{m+1}\sigma_{ik}^{2}) + \sum_{i=1}^{m}h_{i}
\alpha_{i}X(t)\right]dt + \\
& & \displaystyle + \sum_{i=1}^{m}h_{i}
\sum_{k}^{m+1}\sigma_{ik}dW_{k}(t), \quad F(0)=f,
\end{array}
\end{equation}
\begin{equation}
\label{1Factor} dX(t) = \displaystyle (B + \beta X(t))dt +
\sum_{k=1}^{m+1} \lambda_{k} dW_{k}(t), \quad X(0)=x.
\end{equation}

We can use formulae $\eqref{genCME},$ $\eqref{genCVar}$,
$\eqref{genLCME},$ $\eqref{genLCVar}$ or $\eqref{FurCME},$
$\eqref{FurCVar}$ to find    $\bar{f}(t,x;h)$ and $\bar{v}(t,x;h)$.


Then we can write $\bar{Q}_{\gamma}(t, x; h)$ as a quadratic
function with respect to  $h=(h_1,...,h_m)$. Below we will write
this function explicitly for several important cases.

To find a conditional extremum of  $\bar{Q}_{\gamma}(t, x, h)$ with
the constraint $\displaystyle \sum_{i=1}^{m}h_{i}-1=0$ the Lagrange
method can be applied. The Lagrange function is
$$
\begin{array} {rcl}
L (h, \xi) &=& \displaystyle \bar{Q}_{\gamma}(t, x; h) +
\xi(\sum_{i=1}^{m}h_{i}-1)\\
&=& \displaystyle \sum_{i,j=1}^{m}K_{ij}(t,x)h_{i}h_{j} +
\sum_{i=1}^{m} (K_{i}(t,x)+\xi) h_{i} + K_{0}(t,x)-\xi,
\end{array}
$$
where $K_{ij}, K_{i}, K_{0}$ are functions of $t,x$ and coefficients
 $A_{i}$, $\alpha_{i}$, $B$, $\beta$, $\sigma_{ik}$, $\lambda_{k}$,
$i=1,...,m,$ $k=1,...,m+1$.

We get a system of  $m+1$ equations by equating to zero partial
derivatives with respect to $h_{i}, \xi$ of the Lagrange function
$L(h,\xi)$:
$$
\begin{array} {rcl}
\displaystyle \frac{\partial L(h,\xi)}{\partial h_{i}} &=&
\displaystyle
\sum_{j=1}^{m}\left(K_{ij}(t,x)+K_{ji}(t,x)\right)h_{j}+
K_{i}(t,x)+\xi =
0,\\
\displaystyle \frac{\partial L(h,\xi)}{\partial \xi} &=&
\displaystyle \sum_{i=1}^{m}h_{i} - 1 = 0.
\end{array}
$$
This is a  nonhomogeneous system of linear algebraic equations with
respect to variables  $h_{1},..., h_{m},$ $\xi$. The unknown
$h_{1},..., h_{m}$, $\xi$ can be found uniquely provided the
determinant of the system does not vanish. If  $\displaystyle
\lim_{|h|\rightarrow\infty} \bar{Q}_{\gamma}(t, x; h)=-\infty$ and
$\bar{Q}_{\gamma}(t, x; h)$ is continuous function in  $h$, then the
point $h_{1},..., h_{m}$ is a unique maximum.

\begin{remark}
Our considerations hold for $\gamma> \displaystyle -\frac{1}{2}$.
\end{remark}

\begin{remark} We restrict ourselves by consideration of
the  model with one factor  $X(t)$, since our main goal is a study
of influence of such factor as the interest rate. Nevertheless, the
results can be extended to the case of vectorial equation with $n$
components for the factor process $X(t)$. Here $\bar{f}(t,x)$ and
$\bar{v}(t,x)$ are functions of time and $n$ spatial variables. The
factor process can have correlated components.
\end{remark}

Further we find an explicit optimal strategy of investment for the
case of portfolio consisting of two assets, depending on one market
factor, the bank interest rate. For the interest rate we choose
first the Vasicek model and then the Cox-Ingersoll-Ross model.

\section{The linear  interest rate (the Vasicek
model)}\label{Vasicek}

\subsection{An example of portfolio consisting of two assets}
Let us write \eqref{1FactorFCapital}, \eqref{1Factor} in a general
way:
\begin{equation}
\begin{array}{rcl}
\label{1FactorGen} dF\,\, &=& (A + \alpha X )dt + (\sigma, dW),\\
dX\,\,&=& (B + \beta X)dt + (\lambda, dW),\\
 F(0) &=& f, \quad X(0)=x,
\end{array}
\end{equation}
where
\begin{equation}
\begin{array} {rcl}
\label{1FactorCoefs} A &=& \displaystyle
\sum_{i=1}^{m}(h_{i}A_{i}-\frac{1}{2}h_{i}^{2}
\sum_{k=1}^{m+1}\sigma_{ik}^{2}), \qquad \alpha =
\displaystyle\sum_{i=1}^{m}h_{i} \alpha_{i}, \qquad \lambda =
(\lambda_{1},..., \lambda_{m+1}),\\
\sigma &=& \displaystyle (\sum_{i=1}^{m}h_{i}\sigma_{i1},...,
\sum_{i=1}^{m}h_{i}\sigma_{i,m+1}),
\end{array}
\end{equation}
$W=(W_1(t), ..., W_{m+1}(t))$ is a $(m+1)$- dimensional Brownian
motion. Recall that  $\beta<0$.

Equation \eqref{1FactorGen} is a particular case of
\eqref{constSDE}. Therefore we can use the result of Sec.\ref{apr1}.

Equation \eqref{constFPK} is
\begin{equation}
\begin{array}{rcl}
\label{1FactorGenFPK} \displaystyle \frac{\partial
P(t,f,x)}{\partial t} = -(A + \alpha x )\frac{\partial
P(t,f,x)}{\partial f} -\beta P(t,f,x) -(B + \beta x)\frac{\partial
P(t,f,x)}{\partial x}\\
\displaystyle + \frac{1}{2}\Sigma_{1}\frac{\partial^2
P(t,f,x)}{\partial f^2} + \frac{1}{2}\Sigma_{2}\frac{\partial^2
P(t,f,x)}{\partial x^2} + \Sigma_{3}\frac{\partial^2
P(t,f,x)}{\partial f \partial x},
\end{array}
\end{equation}
where
\begin{equation}
\begin{array} {rcl}
\label{1FactorSigma} \Sigma_{1} &=& \sigma\sigma^{T} = \displaystyle
(\sum_{i=1}^{m}h_{i}\sum_{k=1}^{m+1}\sigma_{ik})^2, \qquad
\Sigma_{2} = \lambda\lambda^{T} = \displaystyle
\sum_{k=1}^{m+1}\lambda_{k}^2, \\
\Sigma_{3} &=& \sigma\lambda^{T} = \displaystyle
\sum_{k=1}^{m+1}\lambda_{k} \sum_{i=1}^{m}h_{i}\sigma_{ik}.
\end{array}
\end{equation}
The initial conditions are
$$P(0,f,x)=\delta (f-f_{0})\,g(x).$$

To solve this equation we use the first approach from
Sec.\ref{algorithm}, nevertheless, the second approach can be
applied, too.  We will show how this second approach works in
Sec.\ref{CIR} on the example of the Cox-Ingersoll-Ross interest
rate.

\subsubsection{Gaussian initial distribution of the factor} To get
explicit formulae we consider  Gaussian initial distribution of the
random value  $X$, that is  $g(x)= \displaystyle
\frac{e^{-\frac{(x-x_{0})^2}{2s^2}}}{\sqrt{2\pi}s},$ where
$x_{0}\in\mathbb{R}$ is the mean value of $X$ at initial moment of
time, the constant  $s^2,$ $s\in\mathbb{R}_{+}$ is the variance. The
limit case  $s\rightarrow 0$ corresponds to the factor that equals
$x_0$ initially. Thus,
\begin{equation}
\label{1FactorGenID} P(0,f,x) = \displaystyle \delta (f-f_{0})
\frac{e^{-\frac{(x-x_{0})^2}{2s^2}}}{\sqrt{2\pi}s}.
\end{equation}

The Fourier transform with respect to  $f$ maps
\eqref{1FactorGenFPK}, \eqref{1FactorGenID} into
\begin{equation}
\begin{array}{rcl}
\label{1FactorFPKFur} \displaystyle\frac{\partial
\hat{P}(t,\mu,x)}{\partial t} & = & \displaystyle -i\mu(A + \alpha
x) \hat{P}(t,\mu,x) - (B + \beta x)\displaystyle\frac{\partial
\hat{P}(t,\mu,x)}{\partial x} -
\beta \hat{P}(t,\mu,x) -\\
& & - \displaystyle \frac{1}{2}\Sigma_{1}\mu^{2} \hat{P}(t,\mu,x) +
\frac{1}{2}\Sigma_{2}\frac{\partial^{2}\hat{P}(t,\mu,x)}{\partial
x^{2}}
+ i\mu\Sigma_{3} \frac{\partial \hat{P}(t,\mu,x)}{\partial x},\\
\end{array}
\end{equation}
\begin{equation}
\label{1FactorIDFur} \hat{P}(0,\mu,x) = \frac{e^{-i \mu
f_{0}}e^{-\frac{(x-x_{0})^2}{2s^2}}}{\sqrt{2\pi}s}.
\end{equation}

The anzats for the solution to
\eqref{1FactorFPKFur},\eqref{1FactorIDFur} is
\begin{equation}
\label{1FactorFurPview} \hat{P}(t,\mu,x) = \frac{e^{\gamma_{1}(t) +
\gamma_{2}(t) \mu + \gamma_{3}(t)x + \gamma_{4}(t) \mu^{2} +
\gamma_{5}(t)x \mu + \gamma_{6}(t)x^{2}}}{\sqrt{2 \pi} s}.
\end{equation}

We substitute \eqref{1FactorFurPview} into \eqref{1FactorFPKFur},
\eqref{1FactorIDFur}, then equating the coefficients at the same
powers of $\mu$ and $x$, we get a system for $\gamma_{j}(t),\,$
$j=1,...,6$, a particular case of \eqref{ODE}
\begin{equation}
\begin{array} {rcl}
\label{1FactorODE} \displaystyle
\frac{\partial\gamma_{1}(t)}{\partial t} &=& \Sigma_{2}\gamma_{6}(t)
+
\frac{1}{2}\Sigma_{2}\gamma_{3}^{2}(t) - \beta - B\gamma_{3}(t),\\
\displaystyle \frac{\partial\gamma_{2}(t)}{\partial t} &=&
\Sigma_{2}\gamma_{3}(t)\gamma_{5}(t) - B\gamma_{5}(t)-i A +
i\Sigma_{3}\gamma_{3}(t),\\
\displaystyle \frac{\partial\gamma_{3}(t)}{\partial t} &=&
2\Sigma_{2}\gamma_{3}(t)\gamma_{6}(t) - 2 B\gamma_{6}(t) -
\beta\gamma_{3}(t),\\
\displaystyle \frac{\partial\gamma_{4}(t)}{\partial t} &=&
-\frac{1}{2}\Sigma_{1} +
i\Sigma_{3}\gamma_{5}(t) + \frac{1}{2}\Sigma_{2}\gamma_{5}^{2}(t),\\
\displaystyle \frac{\partial\gamma_{5}(t)}{\partial t} &=&
2i\Sigma_{3}\gamma_{6}(t) - i\alpha +
2\Sigma_{2}\gamma_{5}(t)\gamma_{6}(t) -
\beta\gamma_{5}(t),\\
\displaystyle \frac{\partial\gamma_{6}(t)}{\partial t} &=& -
2\beta\gamma_{6}(t) + 2\Sigma_{2}\gamma_{6}^{2}(t),
\end{array}
\end{equation}
subject to  initial data
\begin{equation}
\label{1FactorODEID}
\begin{array} {rcl}
\gamma_{1}(0) &=& -\frac{x_{0}^2}{2s^2},
\gamma_{2}(0)=-if_{0}, \gamma_{3}(0)=\frac{x_{0}}{s^2},\\
\gamma_{4}(0) &=& 0, \gamma_{5}(0)=0, \gamma_{6}(0)=-\frac{1}{2s^2}.
\end{array}
\end{equation}

Solving problem \eqref{1FactorODE}, \eqref{1FactorODEID}, we get
$\gamma_{j}(t), j=1,..., 6,$  explicitly:

\begin{equation}
\label{1FactorODEsol}
\begin{array} {rcl}
\gamma_{1}(t) &=& \displaystyle -\frac{2B^2 + \beta\Sigma_{2}
\ln\left(\frac{2\beta s^2}{-\Sigma_{2} + 2e^{2\beta t}\beta s^{2} +
e^{2\beta t}\Sigma_{2}}\right) }{2\beta( -\Sigma_{2} + 2e^{2\beta
t}\beta s^{2} + e^{2\beta t}\Sigma_{2})}+\\
&& \displaystyle +\frac{(\Sigma_{2} + 2\beta s^2)\ln\left(\frac{
-\Sigma_{2} + 2e^{2\beta t}\beta s^{2} + e^{2\beta
t}\Sigma_{2}}{2\beta s^2}\right) e^{2\beta t}}{2( -\Sigma_{2} +
2e^{2\beta t}\beta s^{2} + e^{2\beta t}\Sigma_{2})}+\\
&& \displaystyle + \frac{2B(B + x_{0}\beta)e^{\beta t} + (B +
x_{0}\beta)^2 e^{2\beta t}}{\beta(
-\Sigma_{2} + 2e^{2\beta t}\beta s^{2} + e^{2\beta t}\Sigma_{2})},\\

\gamma_{2}(t) &=& \displaystyle -if_{0} -i\frac{(2B\alpha +
x_{0}\beta\alpha + \beta B\alpha t - \beta^{2}At)\Sigma_{2} -
2B\Sigma_{3}\beta}{(2\beta s^{2} + \Sigma_{2})\beta^{2} e^{2\beta t}
- \Sigma_{2}\beta^{2}}-\\
&& \displaystyle -i\frac{e^{\beta t}((-4B\alpha - 2
x_{0}\beta\alpha)\Sigma_{2} + (4B\beta + 2\beta^{2}x_{0})\Sigma_{3}
- 2s^{2}\beta B\alpha)}{(2\beta s^{2} + \Sigma_{2})\beta^{2}
e^{2\beta t} - \Sigma_{2}\beta^{2}}-\\
&& \displaystyle -i\frac{e^{2\beta t}((B\alpha (2 - \beta t) +
\beta^{2}At + x_{0}\beta\alpha )\Sigma_{2} -2\beta (B + \beta
x_{0})\Sigma_{3})}{(2\beta s^{2} + \Sigma_{2})\beta^{2}
e^{2\beta t} - \Sigma_{2}\beta^{2}}-\\
&& \displaystyle -i\frac{e^{2\beta t}(-2B\beta^{2}s^{2}\alpha t +
2s^{2}\beta B\alpha + 2s^{2}\beta^{3}At)}{(2\beta s^{2} +
\Sigma_{2})\beta^{2} e^{2\beta t} - \Sigma_{2}\beta^{2}},\\

\gamma_{3}(t) &=& \displaystyle  \frac{2(-B + e^{\beta t}x_{0}\beta
+ e^{\beta t}B)}{-\Sigma_{2} + 2 e^{2\beta t} \beta s^{2} +
e^{2\beta t}
\Sigma_{2}},\\

\gamma_{4}(t) &=& \displaystyle \frac{\Sigma_{1}t}{2} +
\frac{-2(\Sigma_{2}\alpha - \Sigma_{3}\beta)^2 +
\Sigma_{2}\alpha\beta(2\Sigma_{3}\beta t - \alpha s^{2} -
\Sigma_{2}\alpha t)}{(2\beta s^{2} + \Sigma_{2})\beta^{3} e^{2\beta
t} - \Sigma_{2}\beta^{3}}+\\
&& \displaystyle +\frac{4e^{\beta t} (\Sigma_{2}\alpha -
\Sigma_{3}\beta)(\Sigma_{2}\alpha - \Sigma_{3}\beta + \alpha\beta
s^{2})}{(2\beta s^{2} + \Sigma_{2})\beta^{3} e^{2\beta t} -
\Sigma_{2}\beta^{3}}+\\
&& \displaystyle + \frac{e^{2\beta t} \left((2s^{2}\beta^{2}t +
(\Sigma_{2}t - 3s^{2})\beta -
2\Sigma_{2})\Sigma_{2}\alpha^{2}\right)}{(2\beta s^{2} +
\Sigma_{2})\beta^{3} e^{2\beta t} - \Sigma_{2}\beta^{3}}+\\
&& \displaystyle + \frac{e^{2\beta t} 2\left((-2s^{2}\beta^{2}t +
(2s^{2} - \Sigma_{2}t)\beta + 2\Sigma_{2})\Sigma_{3}\alpha -
2\Sigma_{3}^{2}\beta \right)}{(2\beta s^{2} + \Sigma_{2})\beta^{2}
e^{2\beta t} - \Sigma_{2}\beta^{2}},\\

\gamma_{5}(t) &=& \displaystyle \frac{2i(\alpha\beta s^2 +
\alpha\Sigma_{2} - \Sigma_{3}\beta)e^{\beta t} -i \alpha(\Sigma_{2}
+ 2\beta s^2)e^{2\beta t} - 2\Sigma_{3}\beta +
\alpha\Sigma_{2}}{((2\beta s^2 + \Sigma_{2})e^{2\beta t} -
\Sigma_{2})\beta},\\

\gamma_{6}(t) &=& \displaystyle - \frac{\beta}{-\Sigma_{2} +
2e^{2\beta t}\beta s^{2} + e^{2\beta t}\Sigma_{2}}.
\end{array}
\end{equation}

We substitute the expressions for $\gamma_{j}(t), j=1,..., 6$, in
\eqref{1FactorFurPview} and find  $\hat{P}(t,\mu,x)$ explicitly.
This expression is cumbersome, we do not write this.
The inverse Fourier transform gives
$$P(t,f,x) = \displaystyle \frac{1}{\sqrt{2\pi
}}\int_{-\infty}^{\infty} \hat{P}(t,\mu,x)e^{if\mu}d\mu= $$
$$
\displaystyle \frac{\beta^2 e^{\mathcal{C}_{1}(t,f,x)}}{\sqrt{\pi
s}\mathcal{C}_{2}(t)},
$$
with
$$
\begin{array} {rcl}
\mathcal{C}_{2} (t) &=& \Big(2\beta^4\Sigma_1 t s^2 +
(8\Sigma_3\alpha t s^2 +
\Sigma_1 t \Sigma_2)\beta^3+\\
&& + (-4 s^2\Sigma_2\alpha^2 t + 4\Sigma_3^2 - 8s^2\alpha\Sigma_3 +
4\alpha\Sigma_3
t\Sigma_2)\beta^2 + \\
&& + (-8\alpha\Sigma_2\Sigma_3 -2\alpha^2\Sigma_2^2 t + 6\alpha^2
s^2\Sigma_2)\beta + 4\alpha^2\Sigma_2^2 \Big)e^{2\beta t} +\\
&& + \Big((-8\Sigma_3^2 + 8 s^2\alpha\Sigma_3)\beta^2 +
(16\alpha\Sigma_2\Sigma_3 - 8\alpha^2 s^2\Sigma_2)\beta -
8\alpha^2\Sigma2^2 \Big)e^{\beta t}-\\
&& - \Sigma_1 t\Sigma_2\beta^3 + (-4\alpha\Sigma_3 t\Sigma_2 +
4\Sigma_3^2)\beta^2 +\\
&& + (2\alpha^2\Sigma_2^2 t - 8\alpha\Sigma_2\Sigma_3 + 2\alpha^2
s^2\Sigma_2)\beta + 4\alpha^2\Sigma_2^2,
\end{array}
$$
$\mathcal{C}_{1}(t,f,x)$ is a function of variables $t,f,x$ and
parameters $f_{0},$ $x_{0},$ $s,$ $A,$ $\alpha,$  $B,$ $\beta,$
$\Sigma_{1},$ $\Sigma_{2},$ $\Sigma_{3}$. The multiple integration
by parts was performed by means of the computer algebra system
MAPLE.

The integral converges under  condition
\begin{equation}
\begin{array} {rcl}
\label{1FactorUsl} & & U := \Big(\Big[- 4\alpha^2\Sigma_2^2 -
4\Sigma_3^2\beta^2 + 8 s^2\alpha\Sigma_3\beta^2 - 6\alpha^2
s^2\Sigma_2\beta +
8\alpha\Sigma_2\Sigma_3\beta  +\\
&&\qquad +(\Sigma_2+2\beta s^2)(2\alpha^2\Sigma_2-4\alpha\beta\Sigma_3-\beta^2\Sigma_1)\beta t \Big]e^{2\beta t} +\\
&&\qquad +8 \Big[(\Sigma_3\beta-\Sigma_2\alpha)^2-s^2\alpha\beta(\Sigma_3\beta-\Sigma_2\alpha)\Big]e^{\beta t} +\\
&&\qquad + (\Sigma_1\beta^2 - 2\alpha^2\Sigma_2 +
4\alpha\Sigma_3\beta)\Sigma_2\beta t - 4\Sigma_3^2\beta^2 -
4(\Sigma_3\beta-\Sigma_2\alpha)^2 - \\
&&\qquad - 2\alpha^2 s^2\Sigma_2\beta \Big) \Big((2\beta s^2 +
\Sigma_2)e^{2\beta t} - \Sigma_2 \Big)^{-1}>0.
\end{array}
\end{equation}
Since  $U_{t}'|_{t=0} = -\Sigma_1 \beta^3 >0,$ then there exists $\,
t_*>0$, such that  for all $t\in(0,t_*)$ this condition holds.
Nevertheless, $\displaystyle \lim_{t\rightarrow\infty}U=\infty$, the
sign of the infinity is equal to the sign of   $(-\Sigma_1\beta^3 -
4\alpha\Sigma_3\beta^2 + 2\Sigma_2\alpha^2\beta),$ therefore the
condition \eqref{1FactorUsl} can be not satisfied for large $t$.

Let us denote
$$
\begin{array} {rcl}
U_{s_{\infty}} &:=& \displaystyle \lim_{s\rightarrow\infty} U =\\
&& = 4\alpha \Sigma_3 \beta - 3\Sigma_2 \alpha^2 + (2 \beta \Sigma_2
\alpha^2-\Sigma_1 \beta^3-4 \beta^2 \Sigma_3 \alpha) t + \\
&& + (-4 \alpha \Sigma_3 \beta+4 \Sigma_2 \alpha^2)e^{-\beta t} -
\alpha^2 \Sigma_2 e^{-2 \beta t},
\\
U_{s_{0}} &:=& \displaystyle \lim_{s\rightarrow0} U = \\
&& \displaystyle = \frac{-4 \Sigma_3^2 \beta^2+8 \Sigma_2 \alpha
\beta \Sigma_3-4 \alpha^2 \Sigma_2^2}{(e^{2\beta t}-1)\Sigma_2} +
\frac{(\Sigma_1 \Sigma_2 \beta^3+4 \alpha \Sigma_3 \Sigma_2
\beta^2-2 \alpha^2 \Sigma_2^2 \beta) t}{(e^{2\beta t}-1)\Sigma_2} +\\
&& \displaystyle - \frac{4(\Sigma_3^2 \beta^2 - 2\Sigma_2 \alpha
\beta \Sigma_3 + \alpha^2 \Sigma_2^2) e^{2\beta t}}{(e^{2\beta
t}-1)\Sigma_2} +  \frac{(2 \alpha^2 \Sigma_2 \beta -\Sigma_1
\beta^3-4 \alpha \Sigma_3 \beta^2) e^{2\beta t} t}{(e^{2\beta
t}-1)} + \\
&& \displaystyle + \frac{(8 \Sigma_3^2 \beta^2-16 \Sigma_2 \alpha
\beta \Sigma_3+8 \alpha^2 \Sigma_2^2) e^{\beta t}}{(-1+e^{2\beta
t})\Sigma_2}.
\end{array}
$$
Therefore for large  $s$ condition  \eqref{1FactorUsl} is not
satisfied as  $t\rightarrow \infty$, since  $\displaystyle
\lim_{t\rightarrow\infty} U_{s_{\infty}} = -\infty.$ For small $s$
the condition  \eqref{1FactorUsl} is not satisfied for all $t>0$ if
$-\Sigma_1\beta^3 - 4\beta^2\Sigma_3\alpha +
2\alpha^2\Sigma_2\beta<0.$

Then we substitute  $P(t,f,x)$ in  \eqref{genCME},\eqref{genCVar},
integrate and get the expectation and dispersion:
\begin{equation}
\begin{array} {rcl}
\label{1FactorCME} && \bar{f}(t,x) = -\Big( \Big[ (2\beta^2 x
\alpha + 2 B \alpha \beta ) e^{\beta t}+\\
&&\quad + (-2\beta^2 x \alpha + (-2 A t - 2 f_{0}) \beta^3 +
2\beta^2 B \alpha t - 2B \alpha \beta ) e^{2\beta
t} \Big]s^2 + \\
&&\quad + 2 \Big[ (\Sigma_{2}\alpha - \Sigma_{3}\beta)\beta x -
\beta^2 x_{0}\Sigma_{3} + (\Sigma_{2}x_{0}\alpha -
2B\Sigma_{3})\beta +
2\Sigma_{2}B\alpha \Big] e^{\beta t} + \\
&&\quad \Big[ -\beta x\alpha\Sigma_{2} + (2x_{0}\Sigma_{3} -
At\Sigma_{2}
- f_{0}\Sigma_{2})\beta^2 +\\
&&\quad +((Bt\Sigma_{2} - \Sigma_{2}x_{0})\alpha +
2B\Sigma_{3})\beta -
2\Sigma_{2}B\alpha \Big]e^{2\beta t}+\\
&&\quad + (-\Sigma_{2}\beta\alpha + 2\Sigma_{3}\beta^2)x +
(At\Sigma_{2}
+ f_{0}\Sigma_{2})\beta^2+\\
&&\quad + ((-\Sigma_{2}x_{0} - Bt\Sigma_{2})\alpha +
2B\Sigma_{3})\beta -
2\Sigma_{2}B\alpha \Big) \times \\
&&\quad \times \Big( \beta^2(-\Sigma_{2} + 2e^{2\beta t}\beta s^2 +
e^{2\beta t}\Sigma_{2})\Big)^{-1},
\end{array}
\end{equation}

\begin{equation}
\begin{array} {rcl}
\label{1FactorCVar} && \bar{v}(t,x) = \Big(
\Big[8(\beta^2\Sigma_{3}\alpha -
\Sigma_{2}\alpha^2\beta)e^{\beta t} + 2\Sigma_{2}\alpha^2\beta +\\
&&\quad +(2\Sigma_{1}t\beta^4 + 8\Sigma_{3}\beta^3\alpha t
-4 (2\Sigma_{3}\alpha + \Sigma_{2}\alpha^2 t)\beta^2 + 6\Sigma_{2}\alpha^2\beta)e^{2\beta t} \Big]s^2+\\
&&\quad + 8\Big[2\Sigma_{2}\alpha\Sigma_{3}\beta -
\Sigma_{2}^2\alpha^2 - \Sigma_{3}^2\beta^2 \Big]e^{\beta t}+
\Big[4(\Sigma_3\alpha
t\Sigma_2 + \Sigma_3^2)\beta^2 -\\
&&\quad -2 (4\alpha\Sigma_3\Sigma_2 + \Sigma_2^2\alpha^2 t)\beta + 4\Sigma_2^2\alpha^2 + \Sigma_{1}t\beta^3\Sigma_{2} \Big]e^{2\beta t}+\\
&&\quad -4(\Sigma_3\alpha t\Sigma_2 - \Sigma_3^2)\beta^2 -2
(4\alpha\Sigma_3\Sigma_2
- \Sigma_2^2\alpha^2 t)\beta +\\
&&\quad + 4\Sigma_2^2\alpha^2 - \Sigma_1 t\beta^3\Sigma_2 \Big)
\Big( \beta^3(-\Sigma_{2} + 2e^{2\beta t}\beta s^2 + e^{2\beta
t}\Sigma_{2})\Big)^{-1}.
\end{array}
\end{equation}

From \eqref{1FactorCME}, \eqref{1FactorCVar}, \eqref{1FactorCoefs},
\eqref{1FactorSigma} we get an explicit formula for
$\bar{Q}_{\gamma}(t,x; h)$.

Let us consider the optimal strategy on the example  of two assets,
depending on one market factor, the linear interest rate. It this
case it is convenient to write the strategy in the following
way:$(h_{1},h_{2}) = (h,1-h)$. Then from  \eqref{1FactorCME},
\eqref{1FactorCVar}, \eqref{1FactorCoefs} and \eqref{1FactorSigma},
where $m=2,$ $h_{1}=h,$ $h_{2}=1-h,$ we get:
\begin{equation}
\label{1FactorQ} \bar{Q}_{\gamma}(t,x; h) = K_{2}h^2 + K_{1}h +
K_{0},
\end{equation}
where
\begin{equation}
\begin{array} {rcl}
\label{1FactorK2} && K_2 = \Big( 8 \Big[(\beta \Sigma_2 s^2 +
\Sigma_2^2) (\alpha_1-\alpha_2)^2 - (\beta^2 s^2+2 \beta \Sigma_2)
S_2 (\alpha_1-\alpha_2) +  S_2^2 \beta^2 \Big] \gamma e^{\beta t}
+\\
&&\qquad + \Big[((-4+2 \beta t) \Sigma_2^2+(-6 \beta+4 \beta^2 t)
s^2 \Sigma_2) (\alpha_1-\alpha_2)^2 - \beta^3 t(\Sigma_2 + 2 \beta
s^2)S_3 +\\
&&\qquad +4\left((2 - \beta t)\beta \Sigma_2+2(1-\beta t)\beta^2
s^2\right) S_2 (\alpha_1-\alpha_2)  - 4 S_2^2 \beta^2 \Big]\gamma
e^{2 \beta t} +\\
&&\qquad + \Big[ \beta^3 t( \Sigma_2 + 2 \beta s^2)S_1 \Big] e^{2
\beta t} + \Big[-2((2 + \beta t) \Sigma_2^2 + \beta \Sigma_2 s^2)
(\alpha_1-\alpha_2)^2+\\
&&\qquad + 4( \beta^2 t + 2 \beta) S_2 \Sigma_2
(\alpha_1-\alpha_2)-4 S_2^2 \beta^2+\beta^3 t S_3 \Sigma_2 \Big]
\gamma - \beta^3 t S_1
\Sigma_2\Big)\times \\
&&\qquad \times \Big( \beta^3(2e^{2\beta t}\beta s^2+(-1+e^{2\beta
t})\Sigma_2)
\Big)^{-1}, \\
\end{array}
\end{equation}
\begin{equation}
\begin{array} {rcl}
\label{1FactorK1}
&& K_1 = \Big(-8\beta^2(1-e^{\beta t})^2\gamma S_5 S_2 + \Big[-8\beta\alpha_2(2 \Sigma_2 + \beta s^2) e^{\beta t} +\\
&&\qquad+ \left(4\beta \alpha_2(2 -\beta t)\Sigma_2 + 8\beta^2
\alpha_2(1
-\beta t)s^2 \right) e^{2 \beta t} + 4 \beta \alpha_2(2+ \beta t)\Sigma_2 \Big]\gamma S_2+\\
&&\qquad+ \Big[2((x_0+x)\beta+2B)\beta^2 e^{\beta t}-2(\beta x_0 +
B)\beta^2 e^{2 \beta t} - 2 \beta^2(\beta x+B)\Big] S_2 +\\
&&\qquad+ \Big[(2 \beta^4 s^2 t+\beta^3 t \Sigma_2) e^{2 \beta
t}-\beta^3 t \Sigma_2 \Big] S_4 + \Big[-8(\beta s^2 +
2\Sigma_2)\beta e^{\beta
t} +\\
&&\qquad + (4\beta(-\beta t+2) \Sigma_2 + 8\beta^2(1 - \beta t)s^2)
e^{2 \beta t} + 4(
\beta t+2)\beta \Sigma_2 \Big](\alpha_1-\alpha_2)\gamma S_5+\\
&&\qquad + \Big[(-2 \beta^3 t \Sigma_2-4 \beta^4 s^2 t) e^{2 \beta
t} + 2 \beta^3 t \Sigma_2 \Big] \gamma S_6 +
4\alpha_2\Sigma_2\Big[4(\Sigma_2 + \beta s^2)e^{\beta t} +\\
&&\qquad + ((\beta t-2)\Sigma_2 + \beta(2 \beta t -3) s^2) e^{2
\beta t} -
\Sigma_2(2+\beta t) - \beta s^2\Big](\alpha_1-\alpha_2) \gamma -\\
&&\qquad -2 \left((\beta(x+x_0)+ 2B)\beta \Sigma_2 + (\beta x +
B)\beta^2
s^2\right)(\alpha_1-\alpha_2) e^{\beta t} +\\
&&\qquad + \left((2B - B\beta t+(x+x_0)\beta)\beta \Sigma_2 + 2(B -
B \beta t +
\beta x)\beta^2 s^2 \right)(\alpha_1-\alpha_2)e^{2 \beta t} +\\
&&\qquad + (B \beta t+(x+x_0) \beta + 2B)\beta \Sigma_2(\alpha_1-\alpha_2) + (A_1-A_2)(\Sigma_2(e^{2\beta t}-1)+ 2\beta s^2) \Big)\times\\
&&\qquad \times\Big(\beta^3 ((\Sigma_2 (e^{2 \beta t}-1)+2
\beta s^2) \Big)^{-1},\\
\end{array}
\end{equation}

\begin{equation}
\begin{array} {rcl}
&&K_{0} = \Big(2 \Sigma_2 B \alpha_2 + (-2 S_5 B + \alpha_2 \Sigma_2
(x_0 + x
+ B t )) \beta + \\
&&\quad + (-2 S_5 x_0 + \Sigma_2( f_0 -\frac{t}{2} S_4 + \frac{A_2
t}{2}) + 2 s^2 \alpha_2(x - B t)) \beta^2 e^{2\beta t} - \\
&&\quad -2 (\alpha_2 \Sigma_2 (x_0 + x) + B \alpha_2 s^2
-2 B S_5) \beta e^{\beta t} + \\
&&\quad + 2(S_5 (x_0+x) - s^2 x \alpha_2) \beta^2 e^{\beta t} + 2
s^2( f_0 + \frac{A_2 t}{2}-\frac{t}{2}
S_4) e^{2\beta t} \beta^3 + \\
&&\quad + (\Sigma_2(\frac{t}{2} \sum_{k=1}^3 \sigma_{2k}^2
-\frac{A_2
t}{2} - f_0) - 2 x S_5) \beta^2 + \\
&&\quad + (\alpha_2 \Sigma_2(-B  t +x_0 + x) + 2 B \alpha_2 s^2 - 2
B S_5) e^{2\beta t}
\beta + \\
&&\quad + 2 B \alpha_2 \Sigma_2 (e^{2\beta t} - e^{\beta t}) \Big)
\Big((2 e^{2\beta t} \beta s^2+\Sigma_2 (e^{2\beta t}-1))\beta^2
\Big)^{-1},
\end{array}
\end{equation}
$$
\begin{array} {rcl}
&& S_1=-\frac{1}{2}\sum_{i=1}^3 \sigma_{1i}^2+\sigma_{2i}^2, \quad
S_2=\sum_{i=1}^3(\sigma_{1i}-\sigma_{2i})\lambda_i, \quad
S_3=\sum_{i=1}^3(\sigma_{1i}-\sigma_{2i})^2,\\
&& S_4=\sum_{i=1}^3 \sigma_{2i}^2, \quad
S_5=\sum_{i=1}^3\sigma_{2i}\lambda_{i}, \quad
S_6=\sum_{i=1}^3\sigma_{2i}(\sigma_{1i}-\sigma_{2i}).
\end{array}
$$
Since $K_2=0$ at $t=0$ and $ \frac{\partial K_2}{\partial t}|_{t=0}
= -\gamma\sum_{i=1}^3
(\sigma_{1i}-\sigma_{2i})^2-\frac{1}{2}\sum_{i=1}^3
(\sigma_{1i}^2+\sigma_{2i}^2)<0$, then there exists $ \, t_*>0$ such
that for $\forall t\in(0,t_*)$ a unique point of extremum
$\bar{H}_{\gamma}=\frac{-K_{1}}{2K_{2}}$ is a maximum point of
$\bar{Q}_{\gamma}(t,x;h)$ with respect to $h$.

 Thus, knowing the
value of the interest rate at any moment of time $t\in(0,t_*)$ we
can maximize the income investing a part $\bar{H}_{\gamma}$ in the
first asset and the rest $(1-\bar{H}_{\gamma})$ in the second one.

In the limit cases $s\rightarrow0$ and $s\rightarrow\infty$ the
expression  $\bar{Q}_{\gamma}(t,x;h)$ becomes simpler:
$$
\begin{array} {rcl}
\displaystyle \lim_{s\rightarrow \infty} \bar{Q}_{\gamma}(t,x; h)
&=& \displaystyle f_0 + t A -\gamma t \Sigma_1 -\frac{(3 +
e^{-2\beta t}-2 \beta t-4 e^{-\beta
t})\Sigma_2 \gamma \alpha^2}{\beta^3} -\\
&&\displaystyle  -\frac{4(\beta t - 1 + e^{-\beta t})\gamma \Sigma_3
\alpha}{\beta^2} - \frac{((\beta x+B)(e^{-\beta t}-1)+ \beta B
t)\alpha}{\beta^2},
\end{array}
$$
$$
\begin{array} {rcl}
\displaystyle \lim_{s\rightarrow 0} \bar{Q}_{\gamma}(t,x; h) &=&
\displaystyle At + f_0 - \gamma t \Sigma_1 + \frac{(2 (\beta t-2)
e^{\beta t}+\beta t+2) \Sigma_2 \gamma
\alpha^2}{\beta^3(e^{\beta t}+1)} - \\
&& \displaystyle -4 \frac{((\beta t-2) e^{\beta t}+\beta t+2)
\Sigma_3 \gamma
\alpha}{\beta^2(e^{\beta t}+1)} + \\
&& \displaystyle + \frac{(((-B t+x_0+x) \beta+2 B) e^{\beta
t}-(x_0+B t+x) \beta-2
B) \alpha}{\beta^2(e^{\beta t}+1)} - \\
&& \displaystyle + \frac{4(1-e^{\beta t}) \gamma
\Sigma_3^2}{\beta(e^{\beta t}+1)\Sigma_2} + \frac{2(-(B+\beta x_0)
e^{\beta t}+\beta x+B) \Sigma_3}{\beta(e^{\beta t}+1)\Sigma_2},
\end{array}
$$
where $A, \alpha$ and $\Sigma_1,\Sigma_2, \Sigma_3$ are given by
\eqref{1FactorCoefs} and \eqref{1FactorSigma}, respectively.

\subsubsection{Uniform initial distribution of
factor}\label{uniform}

We consider a random value $X$ distributed uniformly on the segment
 $[-L,L]$, that is $g(x)=\frac{1}{2L}\chi_{[-L,L]}(x)$. Then
\begin{equation}
\label{1FactorRavnID} P(0,f,x)=
\frac{\delta(f-f_0)\chi_{[-L,L]}(x)}{2L} = \frac{\delta(f-f_0)}{2L},
\quad x\in[-L,L].
\end{equation}

After the Fourier transform with respect to  $f$ the equation
\eqref{1FactorRavnID} takes the form
\begin{equation}
\label{1FactorRavnFurID} \hat{P}(0,\mu,x)= \frac{e^{-i\mu
f_0}}{2L\sqrt{2\pi}}, \quad x\in[-L,L].
\end{equation}
For the solution of problem  \eqref{1FactorGenFPK},
\eqref{1FactorRavnFurID} we use the anzats:
\begin{equation}
\label{1FactorRavnFurPview} \hat{P}(t,\mu,x) =
\frac{e^{\gamma_{1}(t) + \gamma_{2}(t) \mu + \gamma_{3}(t)x +
\gamma_{4}(t) \mu^{2} + \gamma_{5}(t)x \mu +
\gamma_{6}(t)x^{2}}}{2L\sqrt{2 \pi}}.
\end{equation}
We have to solve  \eqref{1FactorODE} with initial conditions
$\gamma_1(0)=0, \gamma_2(0)=-i f_0, \gamma_3(0)=0, \gamma_4(0)=0,
\gamma_5(0)=0, \gamma_6(0)=0.$ Thus, $\gamma_j(t), j=1,...,6,$ can
be found explicitly:
$$
\gamma_1(t) = -\beta t,\quad \gamma_2(t) = \frac{\alpha B i
t}{\beta} + \frac{\alpha B i e^{-\beta t}}{\beta^2} - \frac{\alpha B
i}{\beta^2} - A t i -f_0 i, \quad \gamma_3(t)=0,
$$
$$
\begin{array} {rcl}
\gamma_4(t) = \frac{(-4 \Sigma_2 \alpha^2+4 \alpha \Sigma_3
\beta)}{4\beta^3 exp(-\beta t)} + \frac{\alpha^2 \Sigma_2 exp(-2
\beta t)}{4\beta^3} + \frac{(4 \alpha t \Sigma_3 \beta^2-2 \Sigma_1
t \beta^3+3 \Sigma_2 \alpha^2-2 \alpha^2 t \Sigma_2 \beta-4 \alpha
\Sigma_3 \beta)}{4\beta^3},
\end{array}
$$
$$
\gamma_5(t) = \frac{\alpha i}{\beta}(e^{-\beta t}-1), \quad
\gamma_6(t)=0.
$$
We substitute these expressions to \eqref{1FactorRavnFurPview} and
find  $\hat{P}(t,\mu,x)$.
The inverse Fourier transform gives us $P(t,f,x)$.

The following restrictions have to be imposed to guarantee the
convergence of integrals for  $t>0$:
\begin{equation}
\label{1FactorRavnUsl1}
\begin{array} {rcl}
&& -\frac{1}{4\beta^{3}}\,\Big(\alpha^{2} \Sigma_2 e^{-2\beta t} -
4(\alpha^{2}\Sigma_2-\alpha\beta\Sigma_3)e^{-\beta
t}-\\
&& -2\beta^{3}\Sigma_1 t- \alpha^{2}\Sigma_2(2\beta t-3)
-4\alpha\beta \Sigma_3(1-\beta t)\Big)>0,
\end{array}
\end{equation}
\begin{equation}
\label{1FactorRavnUsl2}
\begin{array} {rcl}
&&-\Big((2\beta^{3}\Sigma_1 - 4\alpha \beta^{2}\Sigma_3 +
2\alpha^{2}\beta \Sigma_2) t +
4\alpha\Sigma_3\beta-\\
&&-3\alpha^{2}\Sigma_2 \Big)e^{2\beta t} - (-4\alpha\beta\Sigma_3 +
4\alpha^{2}\Sigma_2)e^{\beta t} + \alpha^{2}\Sigma_2\ge 0.
\end{array}
\end{equation}
One can show that  inequality  \eqref{1FactorRavnUsl1} takes place
for any parameters, whereas for \eqref{1FactorRavnUsl2} is true only
for    $\beta<0$.

Thus from  \eqref{genLCME}, \eqref{genLCVar} we get the following
values for the conditional expectation and variance:
\begin{equation}
\label{1FactorRanvCME} \bar{f}(t,x) = -\frac{(\beta x+B)\alpha
e^{-\beta
t}}{\beta^{2}}-\frac{(B\alpha-A\beta)t}{\beta}+\frac{(\beta
x+B)\alpha}{\beta^{2}}+f_0,
\end{equation}

\begin{equation}
\label{1FactorRanvCVar}
\begin{array} {rcl}
\bar{v}(t,x) &=& \displaystyle -\frac{(-4 \Sigma_2 \alpha^2+4 \alpha
\Sigma_3 \beta)e^{-\beta t}}{2\beta^3} - \frac{\Sigma_2 \alpha^2
e^{-2 \beta t}}{2\beta^3}
-\\
&& \displaystyle -  \frac{(4 \alpha \beta^2 t \Sigma_3-2 \Sigma_1 t
\beta^3+3 \Sigma_2 \alpha^2-2 \alpha^2 \beta t \Sigma_2-4 \alpha
\Sigma_3 \beta)}{2\beta^3},
\end{array}
\end{equation}
where $A, \alpha, \Sigma_1, \Sigma_2, \Sigma_3$ are given in
\eqref{1FactorCoefs}, \eqref{1FactorSigma}.

For the case of two assets depending on one factor we get
$$Q_{\gamma}(t,x;h) = K_2 h^2 + K_1 h+K_0,$$
where  $h$ and  $(1-h)$ are the proportions of capital invested in
the first and second assets, respectively , where  $K_2, K_1, K_0$
are functions, which expressions are cumbersome, nevertheless, the
dependence on time is only exponential  or polynomial.
Since $K_2= - \frac{1}{2}\sum_{i=1}^3 (\sigma_{1i}^2+\sigma_{2i}^2)
- \gamma \bar{v}(t,x;h)<0$, then the optimal strategy in the sense
of Definition \ref{def}  is $H_{\gamma}=-\frac{K_1}{2K_2}$. In the
next section we find the explicit optimal strategy for the classical
example of portfolio containing of two assets one of which is a bank
account.

\begin{remark}
We performed a series of numerical experiments and showed that for a
real market parameters the difference in optimal strategies depends
very weakly on initial distribution of the factor. Namely, for the
Gaussian distribution  with $s>0.001$ the result is very similar to
the limiting case of uniform initial distribution.
\end{remark}


\subsection{Comparing with the Bielecki and Pliska
strategy}\label{comparing} T.Bielecki and S.Pliska in their works
considered  a classical example of portfolio consisting of two
assets, where one asset is a bank account and a factor is the
interest rate. The formula for the Bielecki and Pliska optimal
strategy
 $H_{\theta}$ and the maximal value of the functional
 $\rho(\theta)$ is written out in Sec.\ref{BPstrategy}. To compare our strategy with  the Bielecki-Pliska one
 we also consider this classical example.

Thus, let the assets of the portfolio obey  SDEs:
$$\frac{d S_{1}(t)}{S_{1}(t)} = (A_{1} + \alpha_{1} R(t))dt + \sigma_{1} d W_{1}(t), \quad S_{1}(0) = s > 0,$$
$$\frac{dS_{2}(t)}{S_{2}(t)} = R(t)dt, \quad S_{2}(0) = 1,$$
the dynamics of the interest rate  $R$ is
$$dR(t) = (B + \beta R(t))dt + \lambda dW_{2}(t), \quad R(0) = r $$
(the Vasicek model). Here $A_{1}, \alpha_{1}, B, \beta, \sigma_{1},
\lambda$ are given constants, moreover $B>0, \beta<0$, and $W_{1},
W_{2}$ are independent Brownian motions.

The equation for the capital of investor  is the following:
$$\frac{dV(t)}{V(t)} = \left[h_{1}A_{1}+(h_{1}\alpha_{1} +h_{2})R(t) \right]dt + h_{1}\sigma_{1}dW_{1}(t), \quad V(t) = v > 0.$$

Since we consider the portfolio consisting of two asset, the we
denote $h_{1}=h$ the proportion of capital invested in the risky
asset and $h_{2} = 1-h$ the share invested in the bank account.

If $\ln V(t) = F(t)$, then
$$dF(t) = \left[h A_{1} - \frac{h^{2}\sigma_{1}^{2}}{2} + (h\alpha_{1} + 1 - h)R(t) \right]dt + h\sigma_{1}dW_{1}(t).$$
We consider the case of uniform initial distribution of the interest
rate $R$. Here  $\eqref{1FactorRanvCME}$ $\eqref{1FactorRanvCVar}$
have the form
\begin{equation}
\label{linCME_CVar} \bar{f}(t,r)=M_2 h^{2}(t) + M_1 h(t) +
M_0,\qquad \bar{v}(t,r) = L_{2} h^{2}(t) + L_{1}h(t) + L_{0},
\end{equation}
where
$$
\begin{array}{rcl}
M_2 & = & \displaystyle -\frac{\sigma_{1}^{2}t}{2}, \quad M_1 =
\displaystyle \frac{(\alpha_{1}-1)(\beta r+B)(1-e^{-\beta
t})}{\beta^{2}} -\frac{(B(\alpha_{1}-1)-\beta
A_{1})t}{\beta}, \\
M_0 & = & \displaystyle \frac{(\beta r+B)(1-e^{-\beta
t})}{\beta^{2}} -\frac{Bt}{\beta} + f_{0}, \quad L_{2} =
\displaystyle -\frac{\lambda^{2}}{2\beta^{3}}(\alpha_{1}-1)^2
\phi(t)
+\sigma_{1}^{2}t, \\
L_{1} & = & \displaystyle
-\frac{\lambda^{2}}{\beta^{3}}(\alpha_{1}-1)\phi(t), \quad L_{0} =
\displaystyle -\frac{\lambda^{2}}{2\beta^{3}}\phi(t), \phi(t)=
(e^{-2\beta t}-4e^{-\beta t}-2\beta t +3).
\end {array}
$$
Then
\begin{equation}
\label{linQ} \bar{Q}_{\gamma}(t, r; h) = (M_2 - \gamma
L_{2})h^{2}(t) + (M_1 - \gamma L_{1})h(t) + M_0 - \gamma L_{0}.
\end{equation}
Since $L_2(0)=0$, $\frac{\partial L_2(t)}{\partial t} =
\displaystyle \frac{\lambda^2 (\alpha_1-1)^2 (e^{-\beta
t}-1)^2}{\beta^2} + \sigma_{1}^2 > 0,$ then the  coefficient of the
leading term of the quadratic with respect to $h$ function
$\bar{Q}_{\gamma}(t, r)$ is negative and the unique point of maximum
is
\begin{equation}
\label{linH} \bar H_{\gamma} = \frac{M_{1}-\gamma L_1}{\sigma_1^2 t
+  2 \gamma L_2}.
\end{equation}

Thus, at any moment of time the investor, knowing a current interest
rate, can maximize his income investing  a proportion  $\bar
H_{\gamma}$ of capital to the risky asset and the rest  $1-\bar
H_{\gamma}$ to the bank account.

It follows from $\eqref{linCME_CVar}$  that as  $t\rightarrow
\infty$ the conditional expectation $\bar{f}(t,r)$ and the
conditional variance $\bar{v}(t,r)$ increase as $\displaystyle
e^{-\beta t}$ and $\displaystyle e^{-2\beta t}$, respectively (we
recall that $\beta<0$).  Introducing the risk coefficient, we
describe the subjective influence of randomness to the expected mean
income of portfolio.

\begin{figure}[!htp]
  \begin{center}
  \includegraphics[width=0.5\columnwidth]{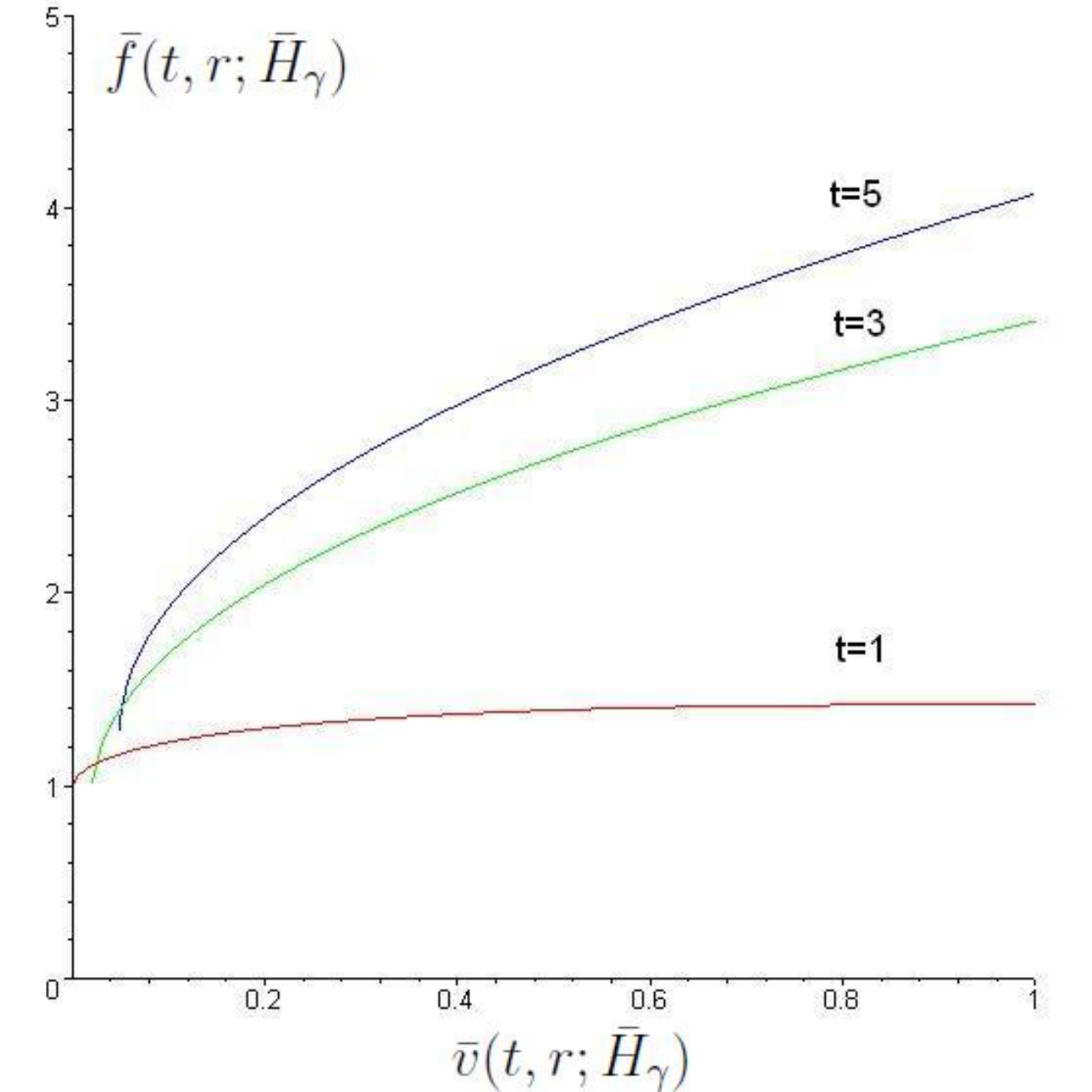}
  \caption{$\bar f(t,r;\bar{H}_{\gamma})$ as a function of $\bar v (t,r;\bar{H}_{\gamma})$
  at fixed moment of time  and different  $\gamma$.}
  \label{E(Var)}
  \end{center}
\end{figure}

Fig.\ref{E(Var)} shows the dependence  $\bar f(t,r;
\bar{H}_{\gamma})$ on $\bar v (t,r; \bar{H}_{\gamma})$ (the
effective frontier) at different moments of time for different
values of the parameter of risk $\gamma$ and given other parameters
of model $A_{1}=0.15, \alpha_{1}=-1,\sigma_{1}=0.2, B=0.05,
\beta=-1, \lambda=0.02, r=0.01, f_0=1$ (the values of parameters are
chosen as in example from \cite{biplsh}).

Let us compare the conditional expectation of the portfolio at a
fixed value of factor  for two strategies under consideration. We
substitute \eqref{linH} and \eqref{exBPH} in the formula
\eqref{linCME_CVar} for $\bar{f}(t,r)$ and after computations we get
the following proposition.
\begin{utv}
For $\gamma=0, \theta=0$,  $t>0$ the following inequality holds:
$$
(\bar{f}(t,r; \bar{H}_{\gamma})-\bar{f}(t,r;
H_{\theta}))|_{\gamma=0, \theta = 0} = \frac{(\alpha_1-1)^2(r +
\frac{B}{\beta})^2 t (\frac{e^{-\beta t}-1}{t\beta} +
1)^2}{2\sigma_1^2} >0.
$$
\end{utv}


\begin{utv}
There exists $ t_{*}>0$ such that for all $ t\in(0,t_{*})$ the
following statement hold:
\begin{enumerate}

\item
if $\alpha_1\neq 1$ and the factor satisfies the condition
\begin{equation}
\label{usl} (r+\frac{B}{\beta})(r+\frac{A_1}{(\alpha_1-1)})>0,
\end{equation}
then there exists  $ \gamma_{*}>0$ such that for all
$\gamma\in(0,\gamma_{*})$
$$\bar{f}(t,r;
\bar{H}_{\gamma})-\bar{f}(t,r; H_{\theta})>0;$$
\item
if $\alpha_1=1$ and $A_1>0$, then  $\bar{f}(t,r;
\bar{H}_{\gamma})-\bar{f}(t,r; H_{\theta})>0$ for all $\gamma>0.$
\end{enumerate}

\end{utv}

\textbf{Proof.} We consider the difference  $\bar{f}(t,r;
\bar{H}_{\gamma})-\bar{f}(t,r; H_{\theta})$ as a function of  $t$ at
other parameters fixed. We denote this difference as $q(t)$. The
computations show that  $q'(0)=0$ and
$$
\begin{array} {rcl}
q''(0) &=& (8 r^2 \gamma^2 \lambda (\alpha_1-1)^3 -2((1+2
\gamma)\gamma \beta \sigma_1 r^2+\\
&& +((1+2 \gamma)\gamma B \sigma_1-8 \gamma^2 \lambda A_1)
r) (\alpha_1-1)^2-\\
&& -2 \Big((-2(2 \gamma + 1)\gamma^2 \lambda \sigma_1^2+(1+2
\gamma)\gamma A_1 \beta \sigma_1) r +\\
&&+ (1+2 \gamma)\gamma A_1 B \sigma_1-4 \gamma^2 \lambda A_1^2 \Big)
(\alpha_1-1)+ \\
&& + 4(2 \gamma+ 1)\gamma^2 \lambda \sigma_1^2 A_1)(\sigma_1^3(1+6
\gamma+12 \gamma^2+8 \gamma^3))^{-1}.
\end{array}
$$
The denominator of this expression is positive. One can see that if
$\alpha_1=1$ and $A_1>0$, then $q''(0)>0$. This proves the second
part of the proposition.

If  $\alpha_1\neq1$, than we expand the numerator
$q''(0)=q''(0,\gamma)$  in series with respect to $\gamma$ about
zero:
$$q''(0,\gamma) = -2\beta\sigma_1(\alpha_1-1)^2(r+\frac{B}{\beta})(r+\frac{A_1}{(\alpha_1-1)})\gamma + O(\gamma^2).$$
This expression is positive if the condition \eqref{usl} holds.
$\Box$

\begin{figure}[!htp]
   \begin{center}
    \includegraphics[width=0.4\columnwidth]{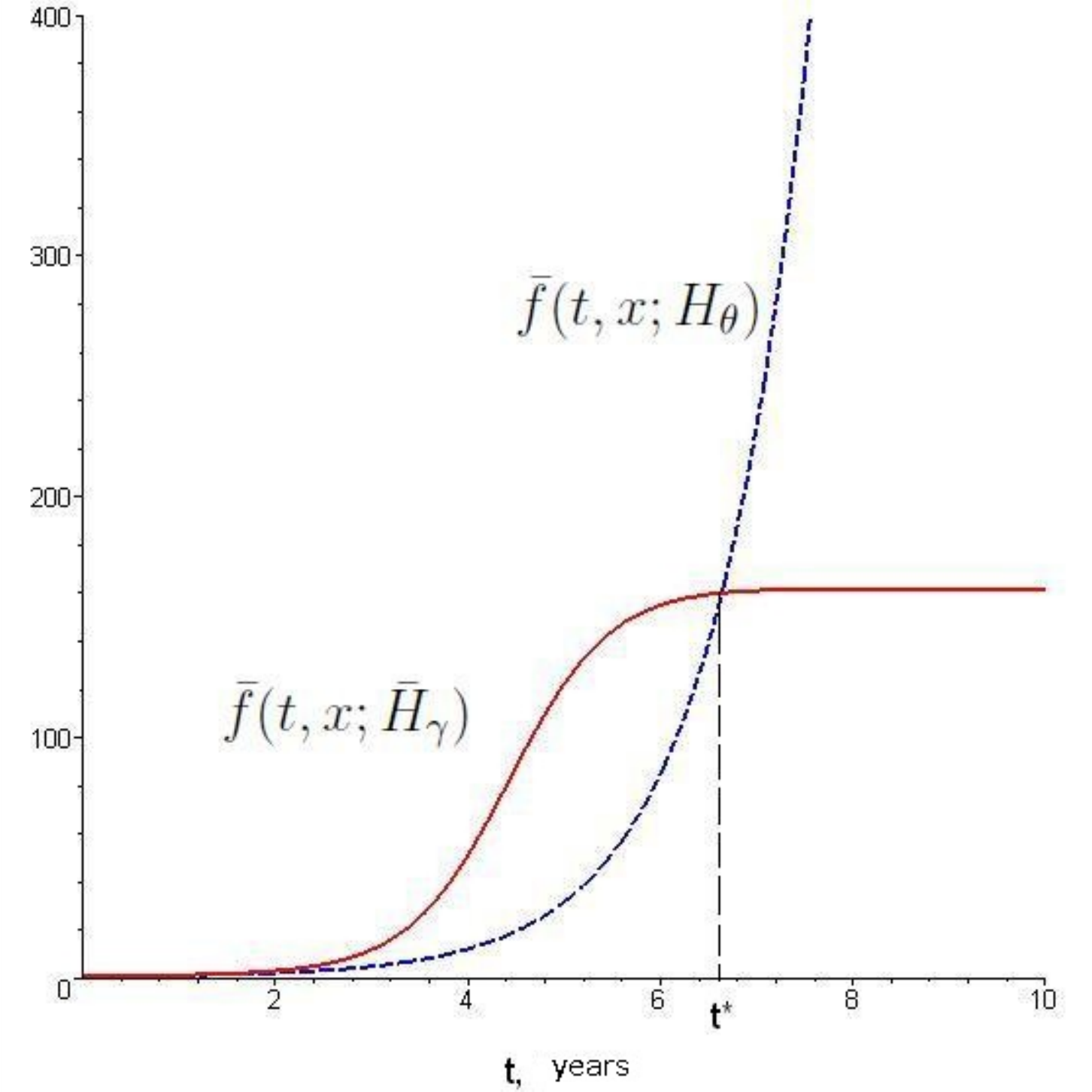}
  \caption{The conditional expectation of portfolio's long-run expected growth
rate  $\bar{f}(t, r; h)$ for our strategy  $\bar{H}_{\gamma}$ (solid
line) and in the case of the strategy of Bielecki and
Pliska~$H_{\theta}$ (dashed line).}
  \label{E_theta_E_gamma}
  \end{center}
\end{figure}

\vskip0.7cm Fig.\ref{E_theta_E_gamma} shows the graphs of
conditional expectations for two strategies for the same parameters
of model as in Fig.\ref{E(Var)}. Here $4\gamma=\theta=0.1$. This
example shows that our strategy for realistic values of parameter
gives greater expectation of portfolio's long-run expected growth
rate till the moment $t^{*}$ and this situation can hold within
several years.

\begin{figure}[!htp]
   \begin{center}
    \includegraphics[width=0.7\columnwidth]{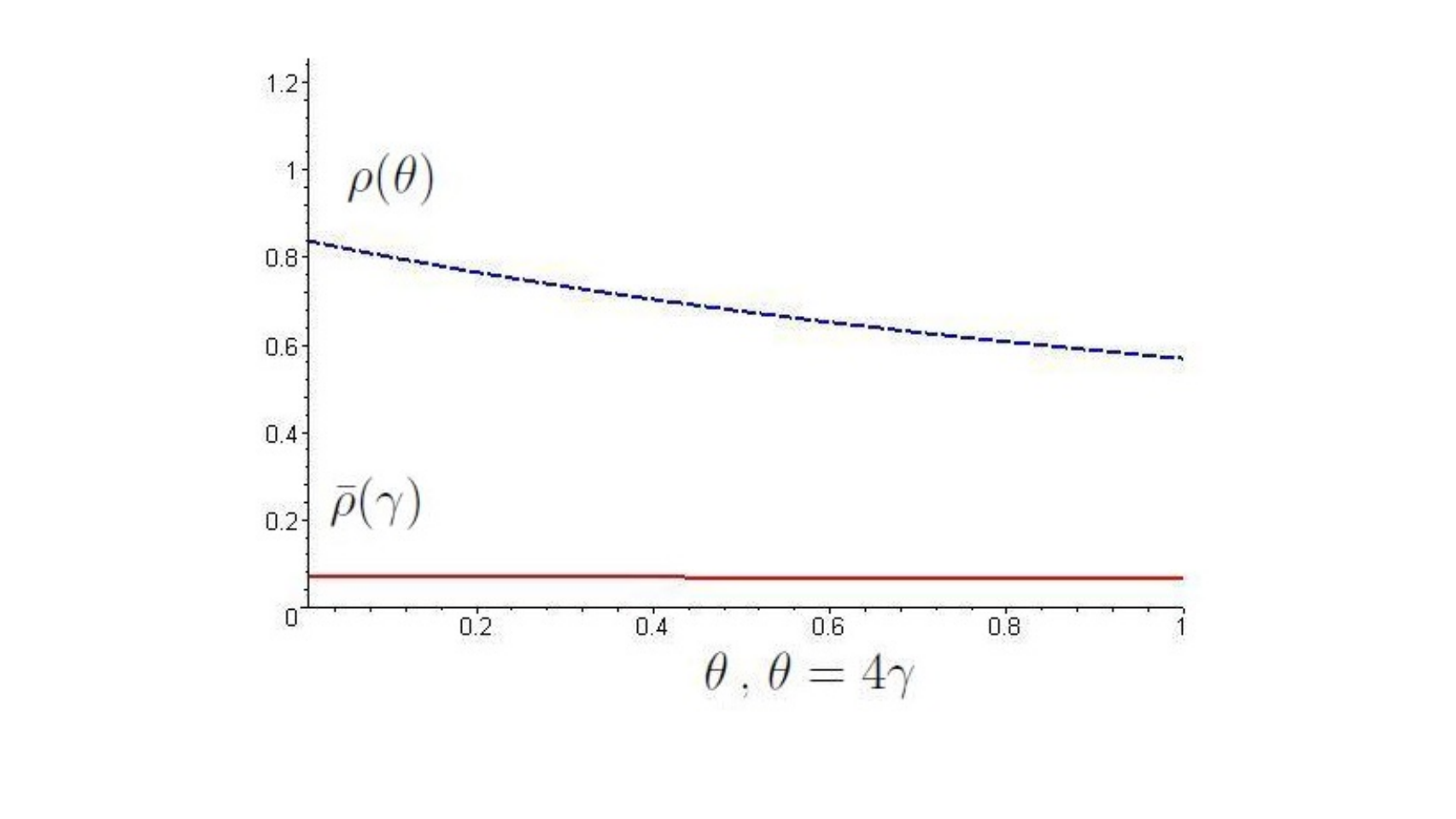}
  \caption{The expected rate of
growth of capital at infinity: $\bar \rho(\gamma)$ for our strategy,
$\rho(\theta)$ for the Bielecki and Pliska strategy, $\theta =
4\gamma$.}
  \label{rho_theta_rho_gamma}{}
  \end{center}
\end{figure}

\vskip0.7cm Taking into account the principle of constructing the
function $\bar Q_\gamma (t, r; h)$ it makes sense to compare two
strategies for $\gamma= \theta/4$ and small  $\theta$. First we
compare the results as $t\rightarrow \infty.$  At any fixed $x$ we
get  $\displaystyle \lim_{t\rightarrow \infty} \bar
H_{\gamma}(t)=-\frac{1}{\alpha_{1}-1} $ in the case $\alpha_1\ne 1$
and $\displaystyle \lim_{t\rightarrow \infty} \bar
H_{\gamma}(t)=\frac{A_1}{\sigma^2_1}$ in the case $\alpha_{1}=1.$
Thus, the limit $ \bar H_{\gamma}(t)$ is discontinuous as a function
of   $\alpha_1$. We introduce the following denotation:
$$\bar{\rho}(\gamma) := \lim_{t\rightarrow\infty} \frac{\bar{Q}_{\gamma}(t,r; \bar{H}_{\gamma})}{t}.$$
Computations show that
$$
\begin{array} {rcl}
\bar{\rho}(\gamma) &=&
\frac{A_{1}}{1-\alpha_{1}}-(\gamma+\frac{1}{2})\frac{\sigma_{1}^{2}}{(\alpha_{1}-1)^{2}},\quad
\alpha_1\ne 1,\\
\bar{\rho}(\gamma) &=& \frac{A_1^2}{2\sigma_1^2}-\frac{B}{\beta},
\quad \alpha_1=1.
\end{array}
$$
The value  $\bar{\rho}(\gamma)$ corresponds to the expected rate of
growth of capital at infinity, it is analogous to  $\rho(\theta)$ in
the model of Bielecki and Pliska, see\eqref{exBP}.
Fig.\ref{rho_theta_rho_gamma} shows this function at the same values
of parameter as at Fig.\ref{E(Var)}, $\theta$ varies from  $0$ to
$1$, $ \gamma=\frac{\theta}{4}.$

After the limit pass as $\theta\rightarrow 0 $ we get from
\eqref{exBP0}
\begin{equation}
\label{exBP00} \lim_{\theta\rightarrow 0} \rho_{\theta} =
-\frac{B}{\beta}+\frac{1}{2\sigma_1^2}(A_1-\frac{B}{\beta}(\alpha_1-1))^2-
\frac{\lambda^2(\alpha_1-1)^2}{4\sigma^2\beta}.
\end{equation}
If we set directly $\gamma=0$ in
$\frac{\bar{Q}_{\gamma}(t,x;\bar{H}_{\gamma})}{t} $, and then
perform the limit pass as $t\rightarrow\infty$, we get
\eqref{exBP00} without the last term, containing $\lambda$.

\subsection{Asymptotics of the proportions of capitals}
In this section we study the asymptotics of proportions of the
portfolio capital as times goes to infinity for the cases of two and
three assets depending on one factor with uniform initial
distribution. We recall that for the Gaussian initial distribution
basically there arises a restriction on application of strategy for
large  $t$, therefore the computation of asymptotics is impossible.

For the case of two assets depending on one factor we have the
system  $\eqref{1FactorFCapital},$ $\eqref{1Factor}$ for $m=2$. The
functions corresponding to the proportions of capital are found in
Sec.\ref{uniform}:
\begin{equation}
\label{H12} H_{1}(t) = -\frac{K_{1}}{2K_{2}},
\end{equation}
\begin{equation}
\label{H22} H_{2}(t) = 1 + \frac{K_{1}}{2K_{2}}.
\end{equation}

We denote the asymptotic limits of the proportions of capitals as
follows:
$$H_{1(2)}^{\infty} := \lim_{t\to\infty}(H_{1}(t)), \quad H_{2(2)}^{\infty} := \lim_{t\to\infty}(H_{2}(t)).$$

The computation shows that the following proposition holds:
\begin{utv}
Let $H_{1}(t)$ and $H_{2}(t)$ be  defined as  $\eqref{H12}$,
$\eqref{H22}$, then
\begin{enumerate}
\item
if $\alpha_{1} \neq \alpha_{2}$, then $ H_{1(2)}^{\infty} =
\displaystyle -\frac{\alpha_{2}}{\alpha_{1}-\alpha_{2}}, \quad
H_{2(2)}^{\infty} = 1 - H_{1(2)}^{\infty};$
\item
if $\alpha_{1} = \alpha_{2}$, then $ H_{1(2)}^{\infty} = \mathcal
K_{1} \cdot \infty,$

$\mathcal K_{1} = \displaystyle
\frac{\gamma\alpha_{1}((\sigma_{11}-\sigma_{12})\lambda_{1}+(\sigma_{21}-\sigma_{22})\lambda_{2}+(\sigma_{31}-\sigma_{32})\lambda_{3})}
{(\sigma_{11}^2+\sigma_{12}^2+\sigma_{13}^2+\sigma_{21}^2+\sigma_{22}^2+\sigma_{23}^2)(2\beta\gamma-1)};$
\item
if $\mathcal K_{1} = 0$, then $H_{1}(t) \equiv  H_{1(2)}^{\infty} =
\displaystyle
\frac{(2\beta\gamma-1)\sigma_2^2+A_2-A_1}{(\sigma_1^2+\sigma_2^2)(2\beta\gamma-1)}.$
\end{enumerate}
\end{utv}

 Thus, the limit values of proportions at infinity in the case of
two assets depends only on parameters  $\alpha_{1}$ and
$\alpha_{2}$, if $\alpha_{1} \ne \alpha_{2}$. In the case
$\alpha_{1} = \alpha_{2}$ the limit value depends on other
parameters, too. For large $t$ it worth to invest to an asset that
is less depending on the factor (the respective $\alpha_{i}$ is the
smallest my modulus), despite of trends and volatilies.

In the case of three assets depending on one factor ($m=3,$ $n=1$)
the proportions of the capital can be also found according to
Sec.\ref{algorithm}. Let us introduce the denotation of limits of
proportions of the capital $H_{i}(t), i=1, 2, 3$ as the time tends
to infinity:
$$H_{1(3)}^{\infty}:=\lim_{t\to\infty}(H_{1}(t)), H_{2(3)}^{\infty}:=\lim_{t\to\infty}(H_{2}(t)),
H_{3(3)}^{\infty}:=\lim_{t\to\infty}(H_{3}(t)).$$ The explicit
formulas for asymptotic limits are the following:
$$ H_{1(3)}^{\infty} = (2\beta(U_{3}(\alpha_{2}^2-\alpha_{1}\alpha_{2}) + U_{2}(\alpha_{3}^2-\alpha_{1}\alpha_{3}))\gamma + (-A_{1}+A_{3}-U_{3})\alpha_{2}^2 + $$
$$ + (-A_{1}+A_{2}-U_{2})\alpha_{3}^2 + (2A_{1}-A_{2}-A_{3})\alpha_{2}\alpha_{3} + (A_{2}-A_{3}+U_{3})\alpha_{1}\alpha_{2} + (-A_{2}+A_{3}+U_{2})\alpha_{1}\alpha_{3})/$$
$$ (2\beta((U_{2}+U_{3})\alpha_{1}^2+(U_{1}+U_{3})\alpha_{2}^2+(U_{1}+U_{2})\alpha_{3}^2) + 4\beta(-U_{3}\alpha_{1}\alpha_{2}-U_{2}\alpha_{1}\alpha_{3}-U_{1}\alpha_{2}\alpha_{3}))\gamma + $$
$$ + (-U_{2}-U_{3})\alpha_{1}^2 + (-U_{1}-U_{3})\alpha_{2}^2 + (-U_{1}-U_{2})\alpha_{3}^2 + 2(U_{3}\alpha_{1}\alpha_{2}+U_{2}\alpha_{1}\alpha_{3}+U_{1}\alpha_{2}\alpha_{3})),$$

$$ H_{2(3)}^{\infty} = (2\beta(U_{3}(\alpha_{1}^2-\alpha_{1}\alpha_{2}) + U_{1}(\alpha_{3}^2-\alpha_{2}\alpha_{3}))\gamma + (-A_{2}+A_{3}-U_{3})\alpha_{1}^2 + $$
$$ + (A_{1}-A_{2}-U_{1})\alpha_{3}^2 + (-A_{1}+2A_{2}-A_{3})\alpha_{1}\alpha_{3} + (A_{1}-A_{3}+U_{3})\alpha_{1}\alpha_{2} + (-A_{1}+A_{3}+U_{1})\alpha_{2}\alpha_{3})/$$
$$ (2\beta((U_{2}+U_{3})\alpha_{1}^2+(U_{1}+U_{2})\alpha_{2}^2+(U_{1}+U_{2})\alpha_{3}^2) + 4\beta(-U_{3}\alpha_{1}\alpha_{2}-U_{2}\alpha_{1}\alpha_{3}-U_{1}\alpha_{2}\alpha_{3}))\gamma + $$
$$ + (-U_{2}-U_{3})\alpha_{1}^2 + (-U_{1}-U_{3})\alpha_{2}^2 + (-U_{1}-U_{2})\alpha_{3}^2 + 2(U_{3}\alpha_{1}\alpha_{2}+U_{2}\alpha_{1}\alpha_{3}+U_{1}\alpha_{2}\alpha_{3})),$$

$$ H_{3(3)}^{\infty} = (2\beta(U_{2}(\alpha_{1}^2-\alpha_{1}\alpha_{3}) + U_{1}(\alpha_{2}^2-\alpha_{2}\alpha_{3}))\gamma + (A_{2}-A_{3}-U_{2})\alpha_{1}^2 + $$
$$ + (A_{1}-A_{3}-U_{1})\alpha_{2}^2 + (-A_{1}-A_{2}+2A_{3})\alpha_{1}\alpha_{2} + (A_{1}-A_{2}+U_{2})\alpha_{1}\alpha_{3} + (-A_{1}+A_{2}+U_{1})\alpha_{2}\alpha_{3})/ $$
$$ (2\beta((U_{2}+U_{3})\alpha_{1}^2+(U_{1}+U_{2})\alpha_{2}^2+(U_{1}+U_{2})\alpha_{3}^2) + 4\beta(-U_{3}\alpha_{1}\alpha_{2}-U_{2}\alpha_{1}\alpha_{3}-U_{1}\alpha_{2}\alpha_{3}))\gamma + $$
$$ + (-U_{2}-U_{3})\alpha_{1}^2 + (-U_{1}-U_{3})\alpha_{2}^2 + (-U_{1}-U_{2})\alpha_{3}^2 + 2(U_{3}\alpha_{1}\alpha_{2}+U_{2}\alpha_{1}\alpha_{3}+U_{1}\alpha_{2}\alpha_{3})),$$
where
$$U_{1}=\sigma_{11}^2+\sigma_{12}^2+\sigma_{13}^2+\sigma_{14}^2,$$
$$U_{2}=\sigma_{21}^2+\sigma_{22}^2+\sigma_{23}^2+\sigma_{24}^2,$$
$$U_{3}=\sigma_{31}^2+\sigma_{32}^2+\sigma_{33}^2+\sigma_{34}^2.$$

Let us note that the limit behavior depends  in general case on all
parameters of the model and this difference from the case of two
assets seems strange. Nevertheless, if parameters  $\alpha_i$ for a
pair of assets coincide, then the situation is analogous to the case
of two assets. For example, if $\alpha_2=\alpha_3$, then for any
values of other parameters  $H_{1(3)}^{\infty} = \displaystyle
\frac{\alpha_2}{\alpha_2-\alpha_1}$. Moreover, $H_{2(3)}^{\infty} $
and  $H_{3(3)}^{\infty}$ depend on other parameters of the model,
too. The case where all $\alpha_i$ are equal, is degenerate, as
above:
$$H_{1(3)}^{\infty} = \mathcal K_{1} \cdot \infty, \quad \mathcal K_{1} =
\displaystyle \frac{\gamma\alpha_{1}(\sum_{l=1}^4 \lambda_l \sum
_{i,j,k} (-1)^{i+j-1} \sigma_k^2(\sigma_{il}-\sigma_{jl})}
{(2\beta\gamma-1)\sum _{i\ne j,i,j=1 }^3\sigma^2_i \sigma_{j}^2)},$$
where $\displaystyle \sigma_i^2=\sum_{k=1}^4
\sigma^2_{ik},\,i=1,2,3,$ where $i,j,k$ are all even combinations of
indices $(1,2,3)$. If $\mathcal K_1=0$, then
$$H_{1}(t)\equiv H_{1(3)}^{\infty} = \displaystyle
\frac{(2\beta\gamma-1) \sigma_2^2
\sigma_3^2+(A_2-A_1)\sigma_3^2+(A_3-A_1)\sigma_2^2}{(2\beta\gamma-1)\sum
_{i\ne j,i,j=1 }^3\sigma^2_i \sigma_{j}^2}.$$

\subsection{Influence of different parameters of model
on the optimal strategy of investment for small time}

It is interesting to note that for small time the strategy of
investment depends on all parameters and differs significantly on
the limit behavior as $t\to\infty$.  We show the results of
computations for the case of two and three assets (the values of
parameters are given in the tables \ref{3.1} and  \ref{3.2},
respectively).

\begin{figure}[!htp]
  \begin{center}
    \includegraphics[width=0.4\columnwidth]{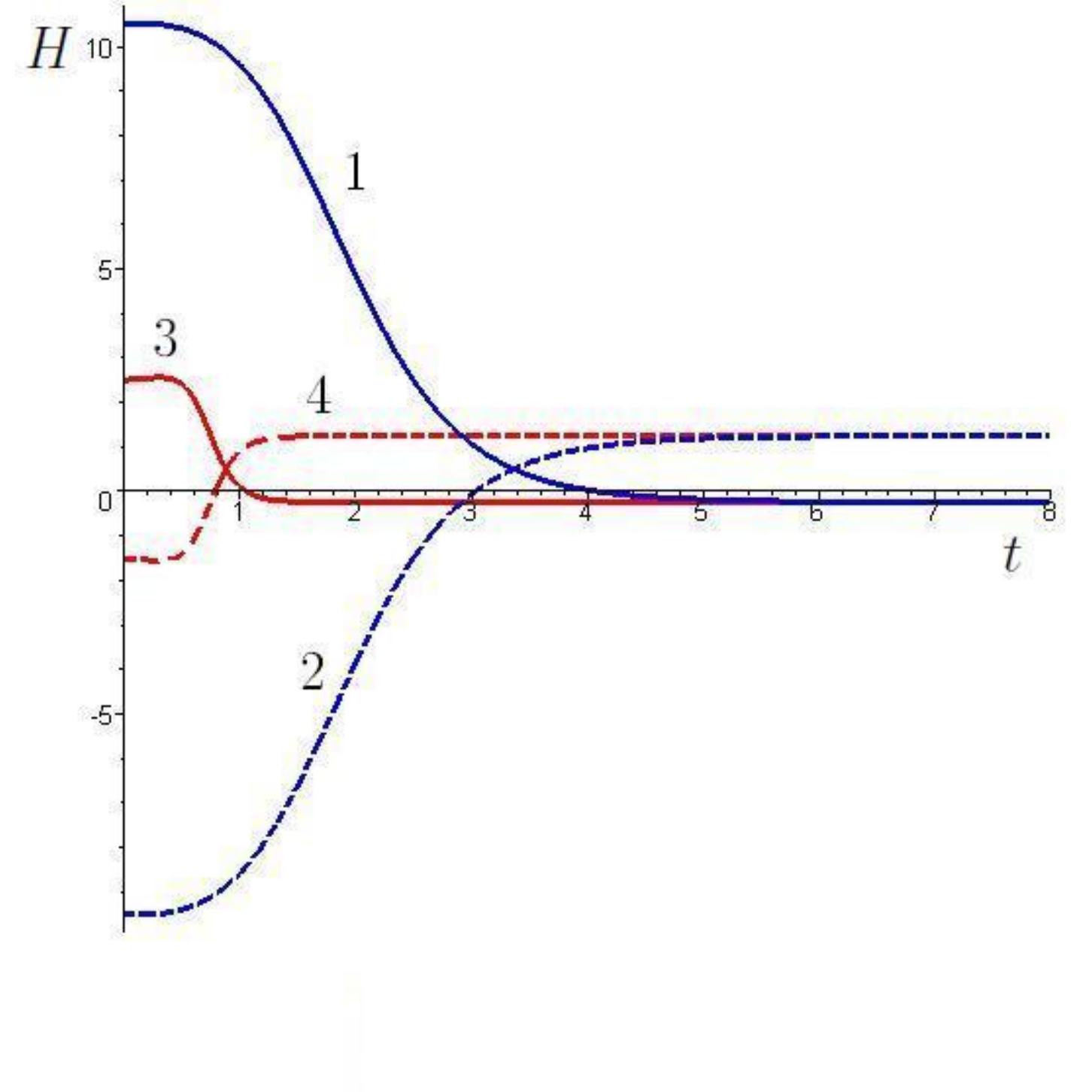}
  \caption{Influence of parameter $\beta$ on the strategy $H=(H_1,H_2)$: \textbf{1.} $H_1,$ for $\beta=-1$; \,
  \textbf{2.} $H_2,$ for
$\beta=-1$; \, \textbf{3.} $H_1,$ for $\beta=-5$; \, \textbf{4.}
$H_2,$ for $\beta=-5$.}
  \label{2assets_beta}
  \end{center}
\end{figure}

\begin{figure}[!htp]
  \begin{center}
    \includegraphics[width=0.4\columnwidth]{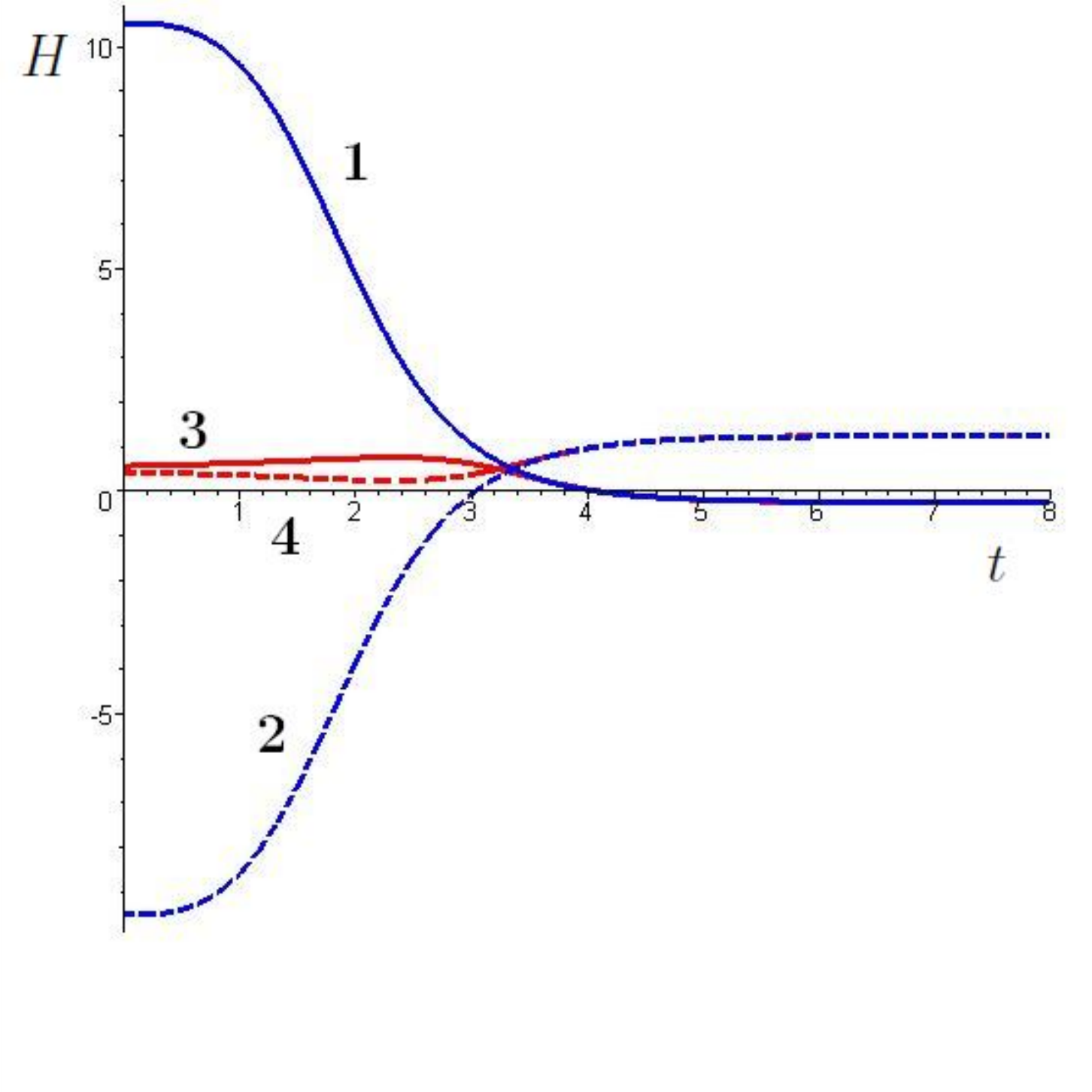}
  \caption{Influence of parameter  $\sigma_{11}$ on the strategy $H=(H_1,H_2)$: \textbf{1.} $H_1,$ for $\sigma_{11}=0.1$; \,
  \textbf{2.} $H_2,$ for $\sigma_{11}=0.1$; \, \textbf{3.} $H_1,$ for $\sigma_{11}=0.5$; \,
  \textbf{4.} $H_2,$ for $\sigma_{11}=0.5$.}
  \label{2assets_sigma11}
  \end{center}
\end{figure}

\begin{figure}[!htp]
  \begin{center}
    \includegraphics[width=0.4\columnwidth]{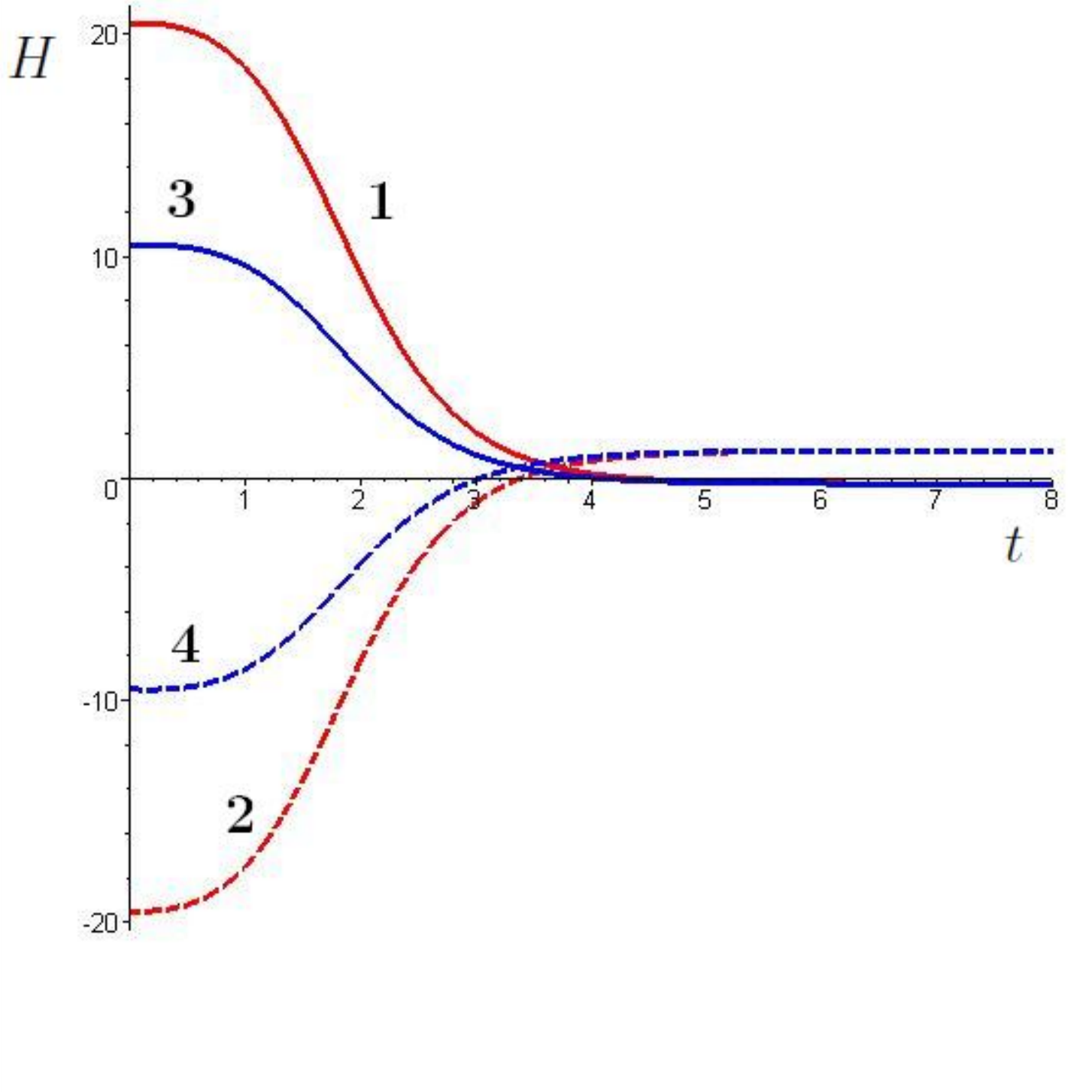}
  \caption{Influence of parameter  $A_1$ on the strategy  $H=(H_1,H_2)$: \textbf{1.} $H_1,$ for $A_1=0.1$; \,
  \textbf{2.} $H_2,$ for $A_1=0.1$; \, \textbf{3.} $H_1,$ for $A_1=0.5$; \,
  \textbf{4.} $H_2,$ for $A_1=0.5$.}
  \label{2assets_A1}
  \end{center}
\end{figure}

As we have seen for the case of two assets ($n=2$) the limit
behavior of the proportions depends only on parameters
$\alpha_{1},\alpha_{2}$. For small times the strategy is different:
\begin{itemize}
\item
 Fig. \ref{2assets_beta} presents graphs for the proportions of capital
for different  $\beta$. For greater $\beta$  the function reaches
its asymptotical value quicker;
\item
Fig. \ref{2assets_sigma11} presents graphs for the proportions of
capital in dependence on parameter $\sigma_{11}$. For small $t$ an
increasing of $\sigma_{11}$ results to a decreasing of the
proportion of corresponding asset in the portfolio;

\item
Parameter $A_{1}$ influences very weakly at large times,
nevertheless, for small $t$ its influence is significant (see Fig.
\ref{2assets_A1}).
\end{itemize}

\begin{figure}[!htp]
  \begin{minipage}{0.3\columnwidth}
  \centerline{\includegraphics[width=0.9\columnwidth]{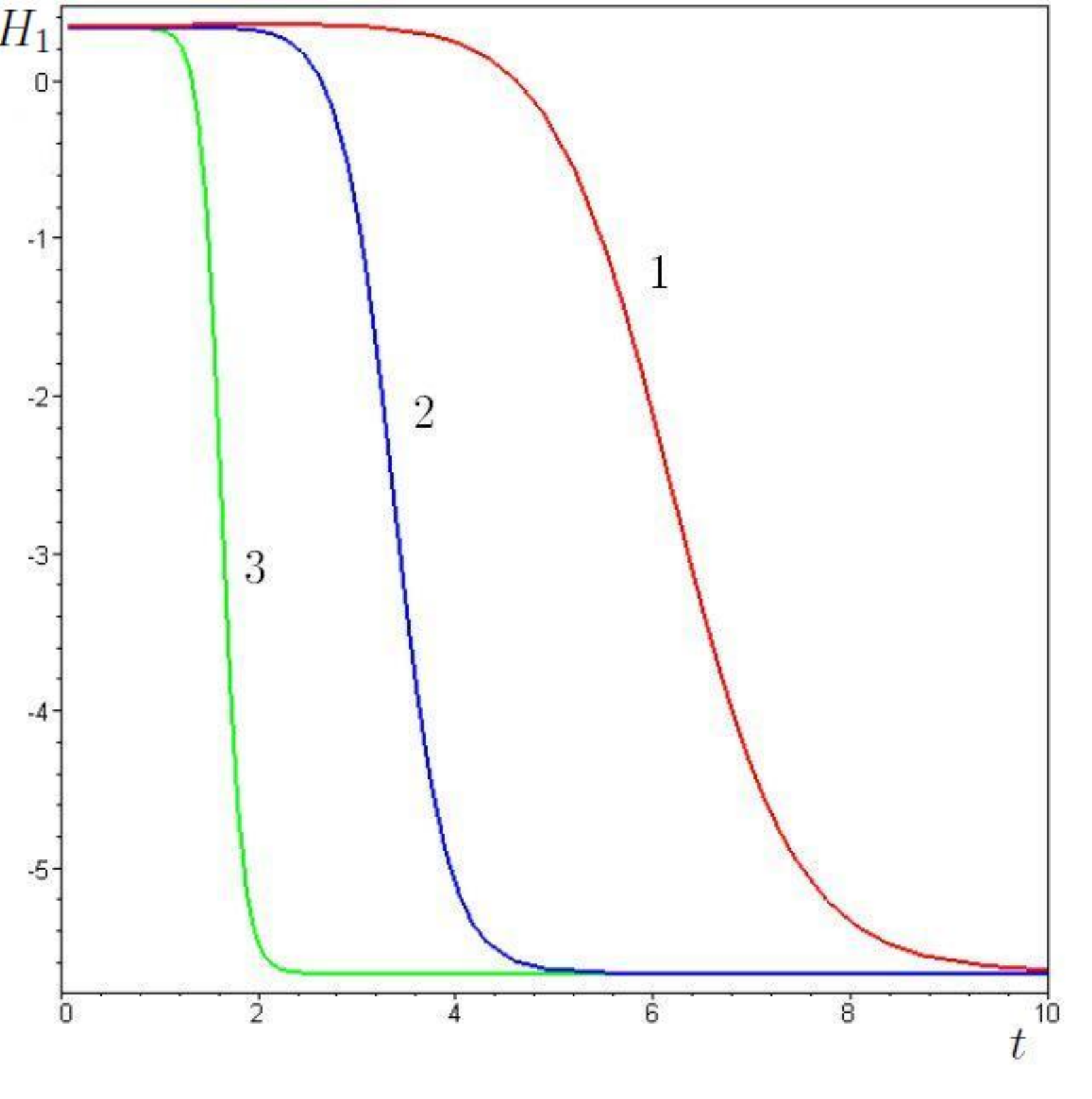}}
  \end{minipage}
\begin{minipage}{0.3\columnwidth}
  \centerline{\includegraphics[width=0.9\columnwidth]{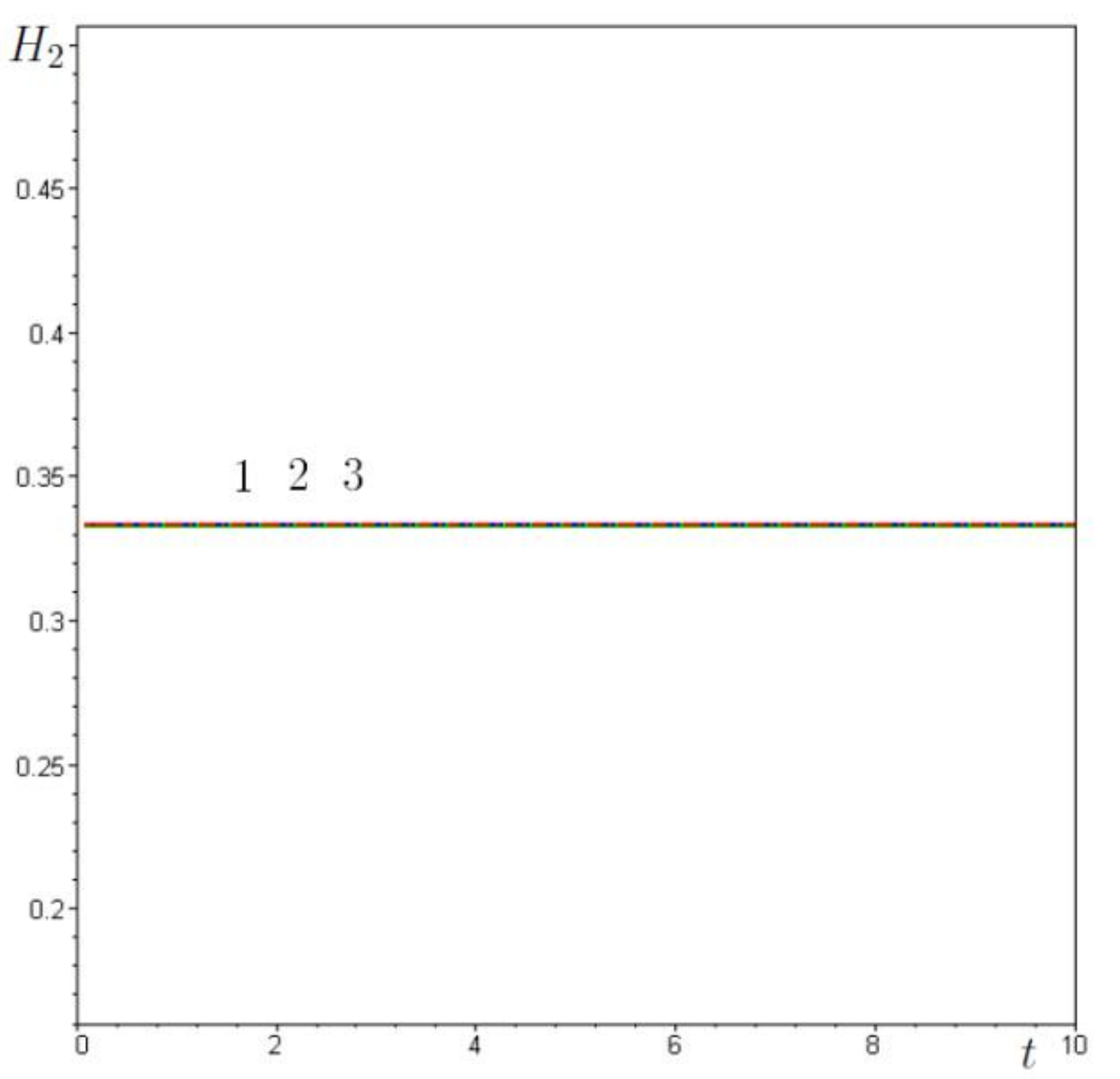}}
  \end{minipage}
\begin{minipage}{0.3\columnwidth}
  \centerline{\includegraphics[width=0.9\columnwidth]{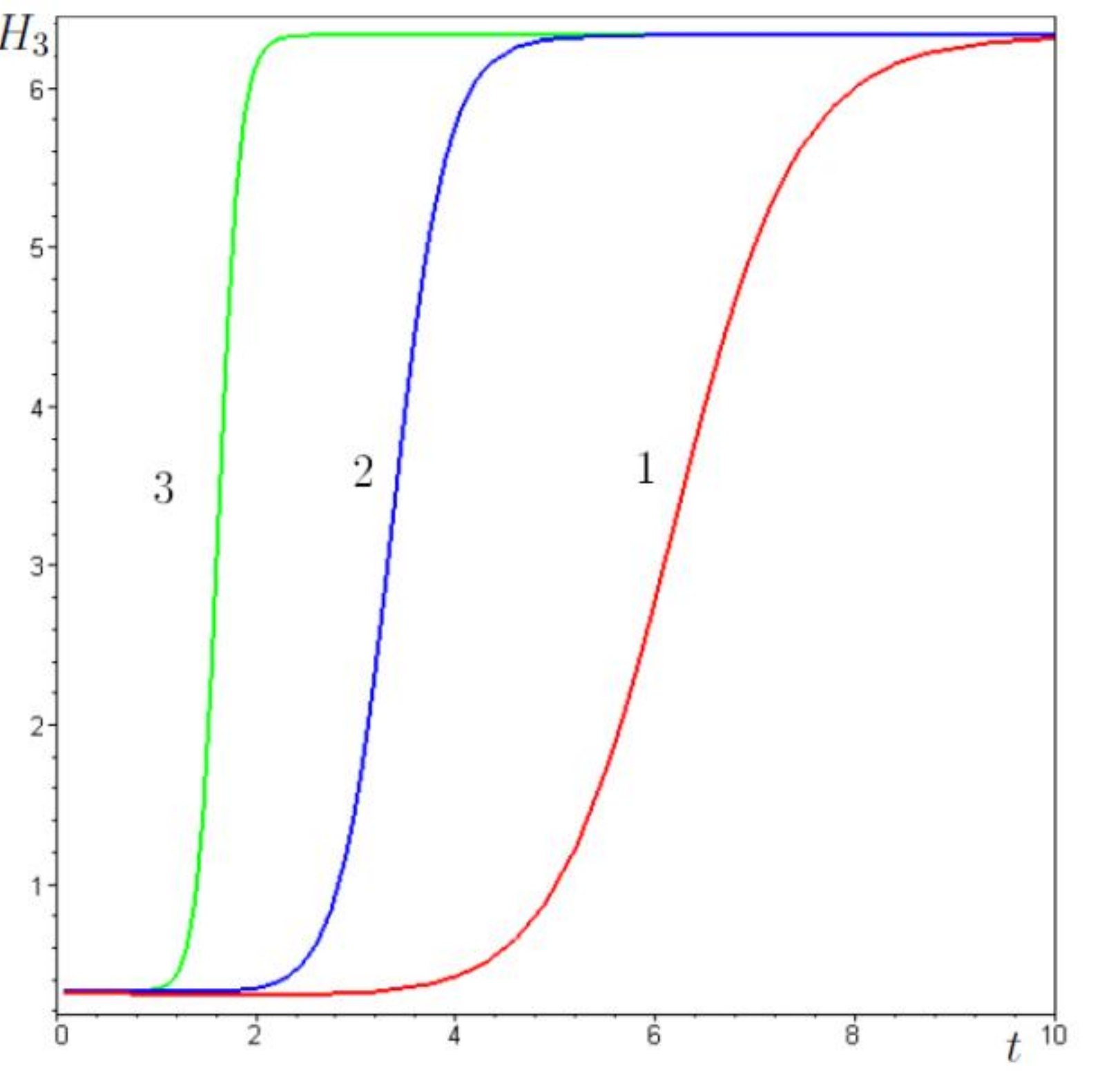}}
 \end{minipage}
   \caption{Influence of parameter  $\beta$ on the strategy  $\qquad(H_1, H_2,
  H_3)$: \textbf{1.}  $\beta=-0.9$; \, \textbf{2.}  $\beta=-2$; \, \textbf{3.}
   $\beta=-5$.}\label{3assets_beta}
\end{figure}

\begin{figure}[!htp]
  \begin{minipage}{0.3\columnwidth}
  \centerline{\includegraphics[width=0.9\columnwidth]{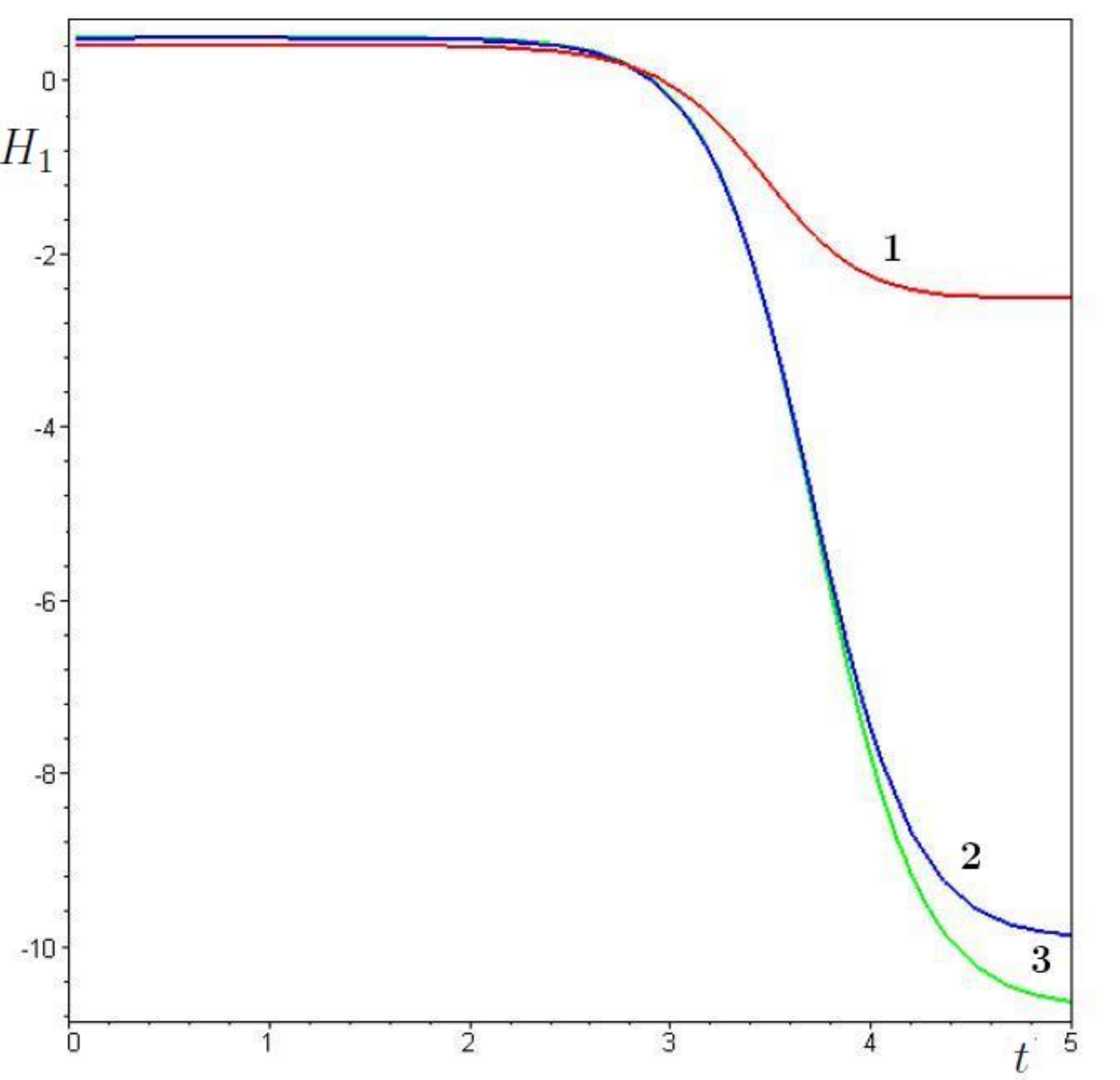}}
  \end{minipage}
\begin{minipage}{0.3\columnwidth}
  \centerline{\includegraphics[width=0.9\columnwidth]{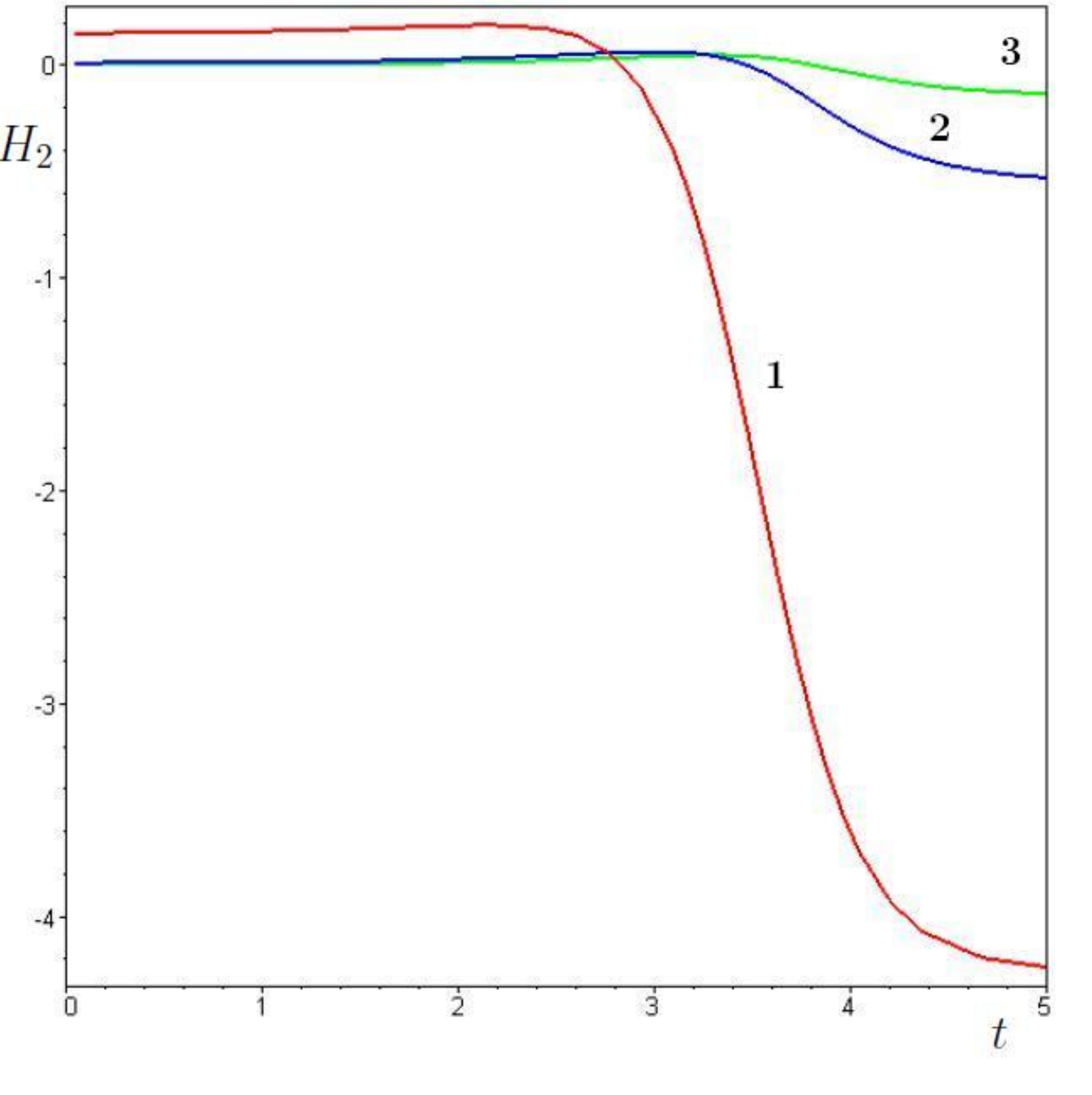}}
  \end{minipage}
\begin{minipage}{0.3\columnwidth}
  \centerline{\includegraphics[width=0.9\columnwidth]{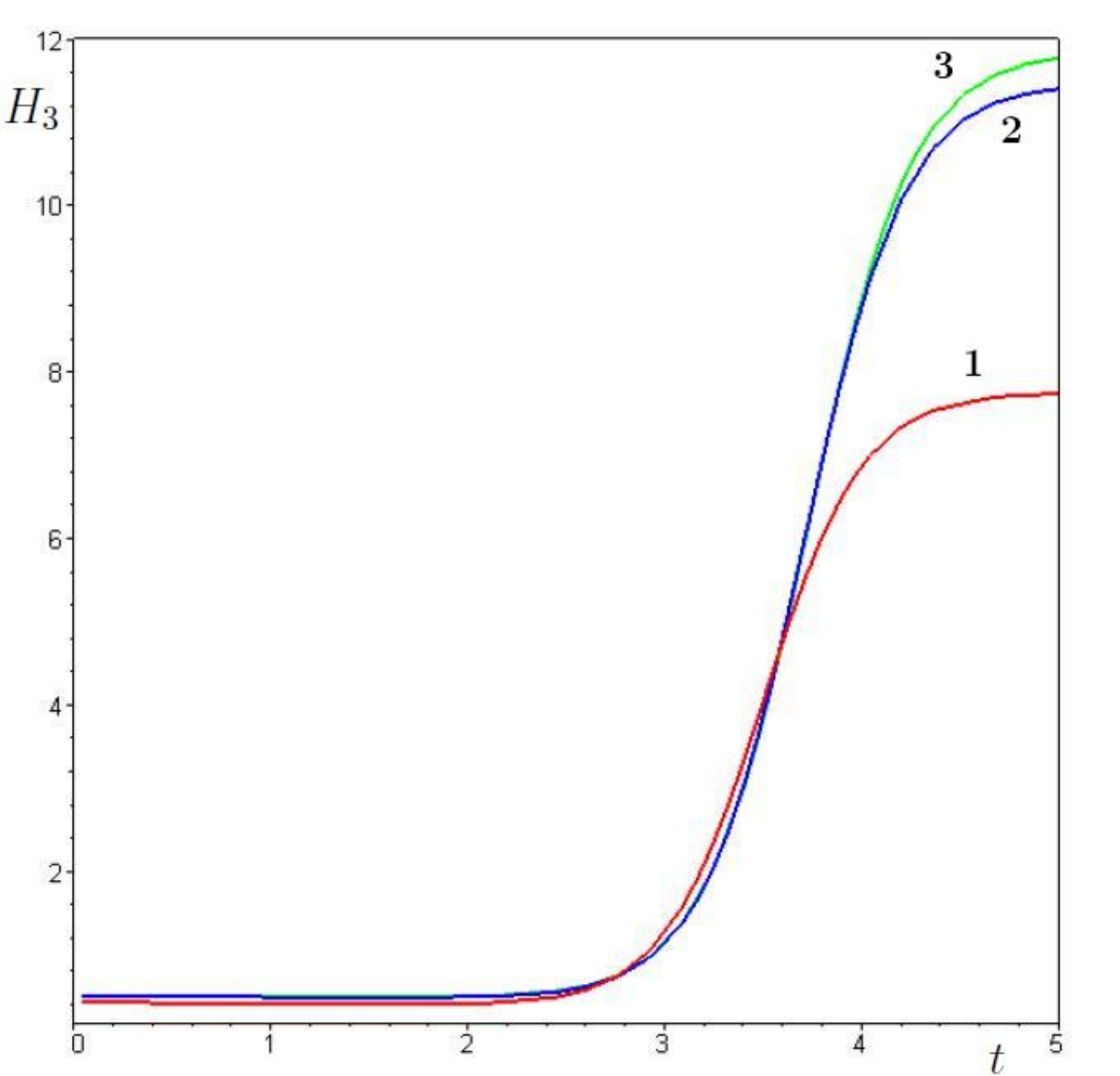}}
 \end{minipage}
 \caption{Influence of parameter  $\sigma_{11}$ on the strategy  $(H_1, H_2,
 H_3)$:
  \textbf{1.}  $\sigma_{11}=0.3$; \, \textbf{2.}  $\sigma_{11}=1.5$; \, \textbf{3.}
   $\sigma_{11}=3$.}
   \label{3assets_sigma}
\end{figure}

\begin{figure}[!htp]
  \begin{minipage}{0.3\columnwidth}
  \centerline{\includegraphics[width=0.9\columnwidth]{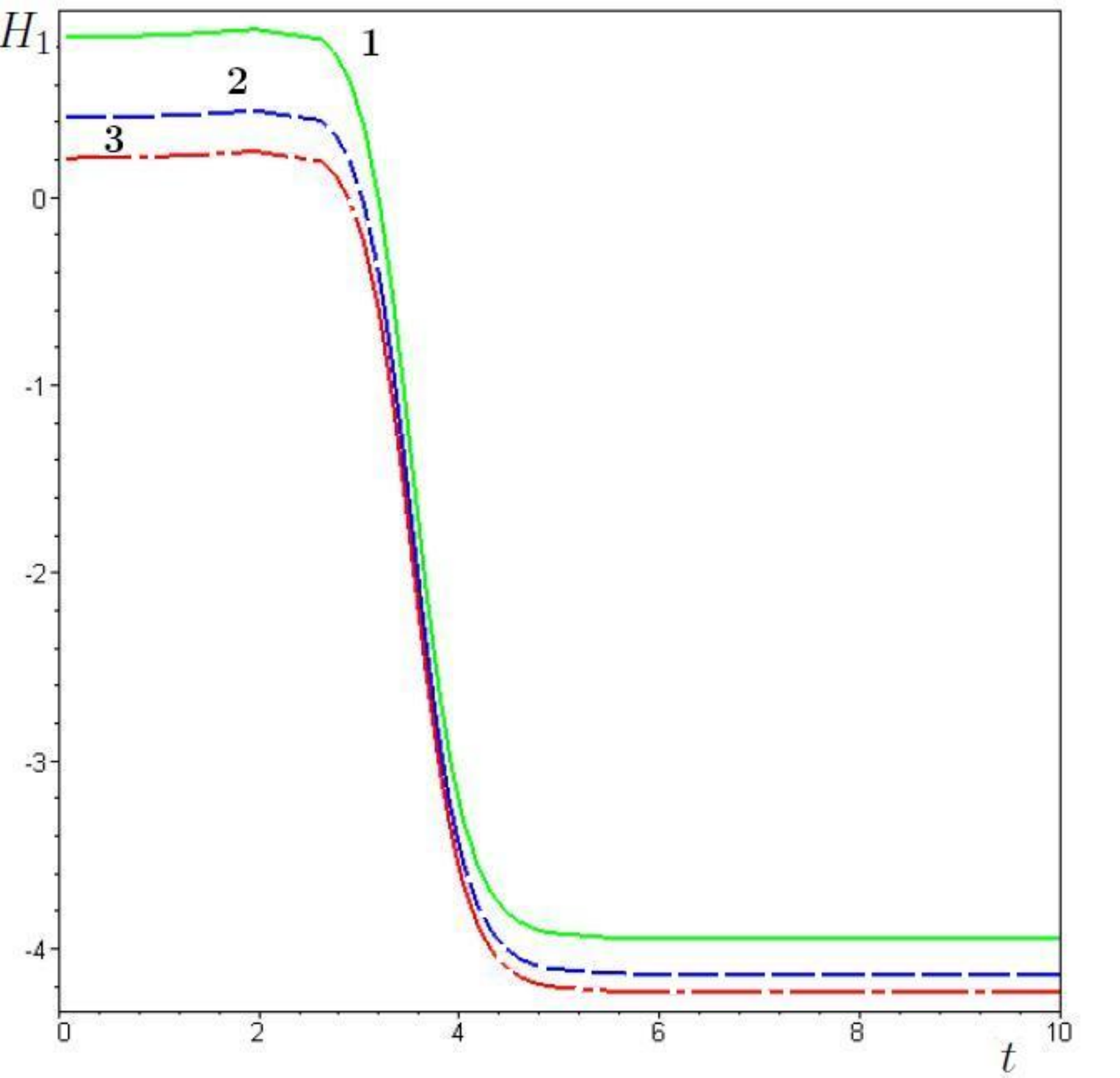}}
  \end{minipage}
\begin{minipage}{0.3\columnwidth}
  \centerline{\includegraphics[width=0.9\columnwidth]{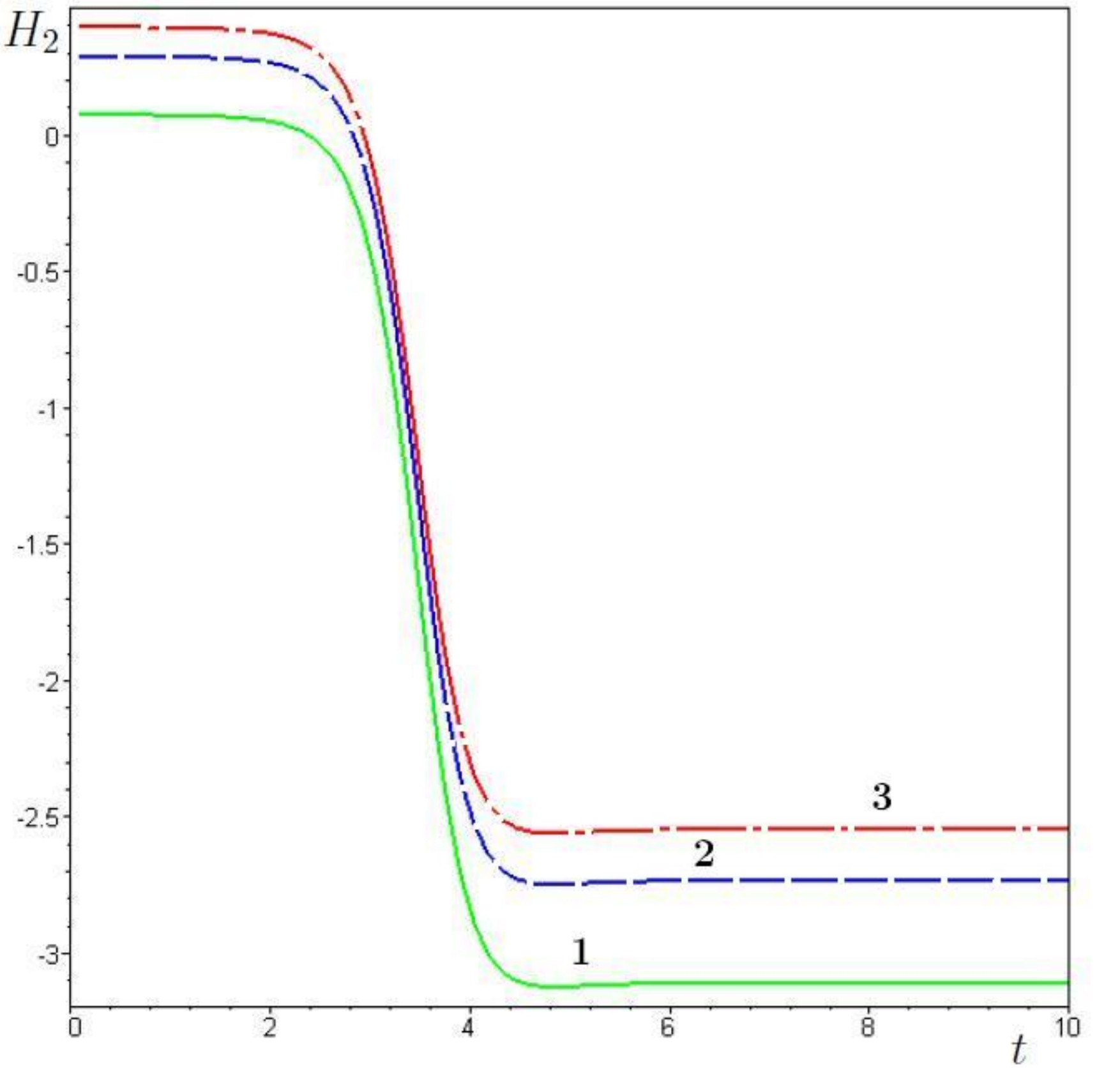}}
  \end{minipage}
\begin{minipage}{0.3\columnwidth}
  \centerline{\includegraphics[width=0.9\columnwidth]{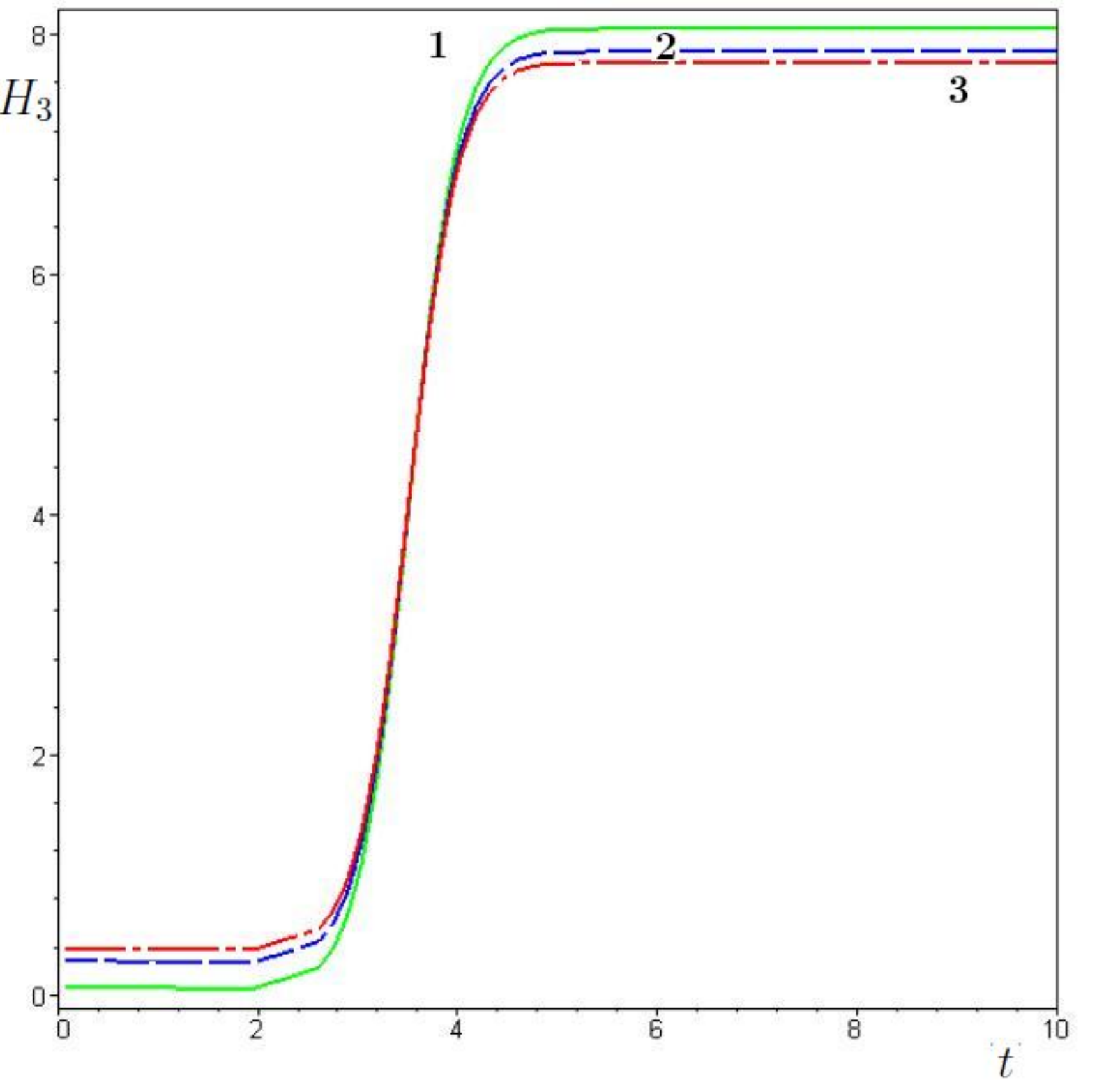}}
 \end{minipage}
 \caption{Influence of parameter $A_1$ on the strategy $(H_1, H_2,
 H_3)$:
  \textbf{1.}  $A_1=5$; \, \textbf{2.}  $A_1=2$; \, \textbf{3.}
   $A_1=0.5$.}
   \label{3assets_A}
\end{figure}

To study the strategy of optimal investment in the case of three
assets ($n=3$), we analyze functions  $H_{i}(t), i=1,2,3,$ changing
the values of parameters $\beta,\sigma_{ij},A_{i},$
$i=1,2,3;j=1,2,3,4$:
\begin{itemize}
\item
First we set the parameters $\alpha_{i}$ very close and study the
influence of $\beta$, $\beta<0$, other parameters are fixed.
Fig.\ref{3assets_beta} illustrates the dependence of $H_{i}(t)$ юn
$\beta$.
\item
Then we fix the parameter  $\beta$ ($\beta=-2$) and change
$\sigma_{11}$, the volatility of the first asset, other parameters
are as in Fig.\ref{3assets_beta}. Fig.\ref{3assets_sigma} the
dependence of $H_{i}(t)$ on  $\sigma_{11}$. Since
$$H_{1(3)}^{\infty} = \lim_{t\to\infty} H_{1}(t)=\frac{\Psi_{1}}{\Psi_{2}\sigma_{11}^2+\Psi_{3}},$$
where $\Psi_{1},\Psi_{2},\Psi_{3}$ do not depend of  $\sigma_{11}$,
if  $\sigma_{11}$ increases, then the limit value  $H_{1}(t)$
becomes smaller. Thus, the asset with a small volotility is
preferable.

\item
Then we fixe  $\beta$ and $\sigma_{11}$ and study the influence of
$A_{1}$. First of all we group the expression $H_{1(3)}^{\infty}$
with respect to  $A_{1}$ and get
$$H_{1(3)}^{\infty}=\frac{(\alpha_{2}-\alpha_{3})^2 A_{1}}{\Phi_{1}}+\Phi_{2},$$
where $\Phi_{1},\Phi_{2}$ do not depend of  $A_{1}$. Thus, the
parameter $A_{i}$ influences on the strategy for small $t$, whereas
for large $t$ this dependence is very weak provided
 $\alpha_{i}$ are close
(see Fig.\ref{3assets_A}).

\item The influence of an increasing  of the risk parameter $\gamma$
analogous to an increasing of  $\beta$ by modulus.
\end{itemize}

Let us summarize the influence of parameters on the character of the
optimal strategy, analogous in the case of two and three assets:
\begin{enumerate}
\item an increasing of parameters $\beta$ and $\gamma$ (by modulus) results a quicker
attainment of the limit value as $t\to\infty$;

\item for small time a decreasing of volatility of  $i$-th asset
(the values of $\sigma_{ik}$) results an increasing of the
proportion of this asset in the portfolio;

\item despite the fact that the trend  $A_{i}$ does not influence on the limit behavior as $t\to
\infty$, for small time the influence of this parameter if
significant (increasing of $A_{i}$ results the increasing of
proportion of the corresponding asset).
\end{enumerate}

\begin{remark}
Let us note that for our strategy any moment of time can be taken as
the initial one. Therefore, we can reasonably find the moment of
time $T$ for actualization of parameters of the model, i.e. for
setting the time to zero. For every set of parameters of the model
there exists its own "infinity", that is the time of achievement of
the asymptotical value. For real data this time has an order of
several years. It is natural to take this time as a time $\,T$ for
actualization. As it follows from our considerations, if the risk
parameter  $\gamma$ increases, then  $T$ becomes smaller, therefore,
we have to actualize the model more frequently.
\end{remark}

\begin{table}[!htp]\label{3.1}
\caption{Values of parameters for the case of two assets}
\begin{center}
\begin{tabular}{|c|c|c|c|}
\hline
& Fig.\ref{2assets_beta} & Fig.\ref{2assets_sigma11} & Fig.\ref{2assets_A1} \\
\hline
$\beta$ & -1;-5& -1 & -1 \\
\hline
$\alpha_{1}$ & 0.5 & 0.5 & 0.5 \\
\hline
$\alpha_{2}$ & 0.1 & 0.1 & 0.1 \\
\hline
$\gamma$ & 1 & 1 & 1 \\
\hline
$A_{1}$ & 0.1 & 0.1 & 0.1;0.5 \\
\hline
$A_{2}$ & 0.1 & 0.1 & 0.1 \\
\hline
$B$ & 1 & 1 & 1 \\
\hline
$\sigma_{11}$ & 0.1 & 0.1;0.5 & 0.1 \\
\hline
$\sigma_{21}$ & 0.1 & 0.1 & 0.1 \\
\hline
$\sigma_{12}$ & 0.1 & 0.1 & 0.1 \\
\hline
$\sigma_{22}$ & 0.1 & 0.1 & 0.1 \\
\hline
$\sigma_{13}$ & 0.1 & 0.1 & 0.1 \\
\hline
$\sigma_{23}$ & 0.1 & 0.1 & 0.1 \\
\hline
$\lambda_{1}$ & 0.1 & 0.1 & 0.1 \\
\hline
$\lambda_{2}$ & 0.1 & 0.1 & 0.1 \\
\hline
$\lambda_{3}$ & 0.1 & 0.1 & 0.1 \\
\hline
\end{tabular}
\end{center}
\end{table}

\begin{table}[!htp]\label{3.2}
\caption{Values of parameters for the case of three assets}
\begin{center}
\begin{tabular}{|c|c|c|c|}
\hline
& Fig.\ref{3assets_beta} & Fig.\ref{3assets_sigma} &Fig.\ref{3assets_A} \\
\hline
$\beta$ & -0.9;-2;-5 & -2 & -2 \\
\hline
$\alpha_{1}$ & 0.13 & 0.13 & 0.13 \\
\hline
$\alpha_{2}$ & 0.12 & 0.12 & 0.12 \\
\hline
$\alpha_{3}$ & 0.11 & 0.11 & 0.11 \\
\hline
$\gamma$ & 1 & 1 & 1 \\
\hline
$A_{1}$ & 0.1 & 0.1 & 0.5;2;5 \\
\hline
$A_{2}$ & 0.1 & 0.1 & 0.1 \\
\hline
$A_{3}$ & 0.1 & 0.1 & 0.1 \\
\hline
$B$ & 1 & 1 & 1 \\
\hline
$\sigma_{11}$ & 0.1 & 0.3;1.5;3 & 0.3 \\
\hline
$\sigma_{21}$ & 0.1 & 0.1 & 0.1 \\
\hline
$\sigma_{31}$ & 0.1 & 0.1 & 0.1 \\
\hline
$\sigma_{12}, \sigma_{22}, \sigma_{32}$ & 0.1 & 0.1 & 0.1 \\
\hline
$\sigma_{13}, \sigma_{23}, \sigma_{33}$ & 0.1 & 0.1 & 0.1 \\
\hline
$\lambda_{1}, \lambda_{2}, \lambda_{3}, \lambda_{4}$ & 0.1 & 0.1 & 0.1 \\
\hline
\end{tabular}
\end{center}
\end{table}

\section{A nonlinear interest rate (the Cox-Ingersoll-Ross
model)}\label{CIR}

\subsection{An auxiliary  problem: the conditional expectation and variance}

The strict theory by Bielecki and Pliska is restricted to  the case
of the factor with a constant volatility. The reason is that for
this model the Hamilton-Jacobi-Bellman is reduced to  a  very
special parabolic second order PDE  where a sum of the order of
derivatives and the order of polynomial in the  coefficients at
these derivatives is equal to two. In this section we consider
another model of the interest rate where the volatility is
proportional to the square root of the rate itself.  The solution to
problem can be found in this case for a special initial distribution
of the interest rate.

 We consider a particular case of system
\eqref{genSDE}:
\begin{equation}
\label{nonlinFCapital} dF = (A + \alpha R)dt + \sigma dW_{1},
\end{equation}
\begin{equation}
\label{nonlinFactor}   dR = (B + \beta R)dt + \lambda \sqrt{R}
dW_{2}.
\end{equation}
Here $B>0, \beta<0,\sigma>0,\lambda>0, A,\alpha$ are constants.

The first equation describes the return $F$ of asset with the trend
that linearly depends on the interest rate  $R,$ that obeys  the
Cox-Ingersoll-Ross model \cite{cir1985}.  The inequality  $- 2\beta
B>\lambda^2$ implies  a positivity of the random process describing
the interest rate  \cite{feller}. The interest rate of the form
\eqref{nonlinFactor} as a factor was considered, in particular, in
the work \cite{biplsheu}, a step toward to the finding the optimal
strategy in the sense of Bielecki-Pliska \cite{bielplis}.
Nevertheless the authors obtain only partial results.

Let us assume that initially the interest rate $R$ is distributed
uniformly on the interval  $(0,L), L>0.$

\begin{remark}
For the Vasicek model we obtain explicit formulae for initial
Gaussian distribution including limiting cases. Nevertheless, for
the Cox-Ingersoll-Ross we can get an explicit formula only for
uniform initial distribution.
\end{remark}

The Fokker-Planck equation for the join distribution of random
values $F$ and $R$ given by system \eqref{nonlinFCapital},
\eqref{nonlinFactor}, is
\begin{equation}
\label{nonlinFPK}
\begin{array}{rcl}
&& \displaystyle \frac{\partial P(t,f,r)}{\partial t}+(A+\alpha r)
\frac{\partial P(t,f,r)}{\partial f}+\beta P(t,f,r)+\\
&& \displaystyle +(B+\beta r-\lambda^2)\frac{\partial
P(t,f,r)}{\partial r}- \displaystyle
\frac1{2}\sigma^2\frac{\partial^2 P(t,f,r)}{\partial f^2}-\frac1{2}
\lambda^2 r \frac{\partial^2 P(t,f,r)}{\partial r^2}=0,
\end{array}
\end{equation}
subject to initial conditions
\begin{equation}
\label{nonlinFPKID} P \left( 0,f,r \right) =\delta \left(
f-f_{0}\right) {\chi_{(0,L)}(r)}.
\end{equation}
Rigourously speaking, to define a probability density function $P
\left( 0,f,r \right)$ we have to divide  the expression
\eqref{nonlinFPKID} by $L$. Nevertheless, as follows from the
linearity of equation \eqref{nonlinFPK} and definitions
\eqref{FurCME}, \eqref{FurCVar} this multiplier does no influence on
the result of computations.

The Fourier transform with respect to $(f,r),$ the function $\hat P
\left( t,\mu,\xi \right),$ obeys the equation
\begin{equation}
\begin{array}{rcl}
\label{nonlinFurFPK} \displaystyle {\frac {\partial }{\partial
t}}\hat P \left( t,\mu,\xi \right)-
 \left( \alpha\,\mu+\beta\,\xi -i\frac{1}{2}{\lambda}^{2}{\xi }^{2}
 \right) {\frac {\partial }{\partial {\xi }}}\hat P \left( t,\mu,\xi
 \right)+
 \\+ \displaystyle \left(\frac1{2}\,
 {\sigma}^{2}{\mu}^{2}+A\mu i+B\xi i\right) \hat P \left( t,\mu,\xi
\right)=0,
\end{array}
\end{equation}
with initial conditions
\begin{equation}
\label{nonlinFurFPKID} \hat P(0, \mu,
\xi)=\frac{1}{2\pi}\,e^{-i\mu\,f_{0}}\,\frac{e^{-i\xi
L}-1}{i\xi}\rightarrow \frac{1}{2\pi}\,e^{-i\mu\,f_{0}}\,\delta(\xi)
\,\mbox{\quad as \quad}\,L\to \infty.
\end{equation}

Equation \eqref{nonlinFurFPK} has the first order and can be
integrated. The solution to the problem
\eqref{nonlinFurFPK},\eqref{nonlinFurFPKID} in the limit case
$L\to\infty$ can be found by a standard way:
$$\hat P(t,\mu,\xi)={e^{-{\frac {2\,i f_{0}\mu\,\lambda^{2}+2\,t\,A\mu\,{\lambda}^{2}\,i+2\,t\,B\,{\beta}+t\,{\sigma}^{2}{\mu}^{2}
{ {\lambda} }^{2}}{2\,{{\lambda}}^{2}}}}}{\delta}
 \left(s(t,\mu,\xi) \right)\times$$
$$\times\left(
\frac{{{\lambda}}^{2} ( 2\,i{
\alpha}\,\mu+2\,i{\beta}\,{\xi}+{{\lambda}}^{2}{{\xi} }^{2})
\cosh\left(\frac {t\sqrt {2\,i{\alpha}\,\mu\,{
{\lambda}}^{2}+{{\beta}}^{2}}}{2}+i\arctan ( {\frac
{{{\lambda}}^{2}{\xi}+i{\beta}}{\sqrt
{2\,i{\alpha}\,\mu\,{{\lambda}}^{2}+{{\beta}}^{2}}}}) \right)^2}{
2\,i{\alpha}\,\mu\,{{\lambda}}^{2}+{{\beta}}^{2}
 }\right) ^{-{\frac {B}{{{\lambda}}^{2}}}},$$
where
$$
\begin{array}{rcl}
s(t,\mu,\xi)=  -{\frac { \left(  \left( {\beta}-\sqrt
{2\,i{\alpha}\, \mu\,{{\lambda}}^{2}+{{\beta}}^{2}} \right)
{\xi}+2\,{\alpha}\,\mu \right) {e^{-t\sqrt
{2\,i{\alpha}\,\mu\,{{\lambda}}^{ 2}+{{\beta}}^{2}}}}+ \left(
-{\beta}-\sqrt {2\,i{\alpha}\, \mu\,{{\lambda}}^{2}+{{\beta}}^{2}}
\right) {\xi}-2\,{\alpha}\,\mu}{\sqrt
{2\,i{\alpha}\,\mu\,{{\lambda}}^{2}+{{\beta}}^{2}}+i{{\lambda}}^{2}{\xi}-{\beta}+
\left( \sqrt {2\,
i{\alpha}\,\mu\,{{\lambda}}^{2}+{{\beta}}^{2}}-i{{\lambda}
}^{2}{\xi}+{\beta} \right) {e^{-t\sqrt {2\,i{\alpha}\,\mu\,{{
\lambda}}^{2}+{{\beta}}^{2}}}}}}.
\end{array}
$$

We apply  \eqref{FurCME} and \eqref{FurCVar}:
\begin{equation}
\label{FurP} \hat P(t,0,\xi)=\theta(t,\xi) {\delta}
 \left( s(t,0,\xi)\right),
\end{equation}
\begin{equation}
\label{FurdP}
\partial_\mu \hat P
(t,0,\xi)=\phi (t,\xi)\,\delta(s(t,0,\xi))+ \psi(t,\xi)\,
\delta'_{\mu}(s(t,0,\xi)),
\end{equation}
\begin{equation}
\label{FurddP}
\partial^2_\mu \hat P
(t,0,\xi)=q_{1}(t,\xi)\,\delta(s(t,0,\xi))+ q_{2}(t,\xi)\,
\delta'_{\mu}(s(t,0,\xi))+q_{3}(t,\xi)\delta''_{\mu}(s(t,0,\xi)),
\end{equation}
where
$$s(t,0,\xi)=
 \frac{2\beta\xi}{i\xi\lambda^2+(2\beta-i\xi\lambda^2)e^{-\beta t}},$$

$$\theta(t,\xi) = \left( \frac{-i \lambda^2 (i\xi^2\lambda^2  -2\beta\xi)\cosh\left( \frac{\beta t}{2}+
i\arctan(\frac{\lambda^2\xi+i\beta}{\beta}) \right)^2 e^{\beta
t}}{\beta^2} \right) ^{\frac{-B}{\lambda^2}},$$

$$\phi (t,\xi)= \theta(t,\xi) \left( L_1 - i(At+f_{0}) \right), \quad \psi(t,\xi)=\theta (t,\xi)L_{2},$$

$$L_{1}=\frac{B\alpha \left( (4\beta \lambda^4\xi^2-i\lambda^6\xi^3+4i\beta^2\lambda^2\xi)t+(4\beta^2-2\lambda^4\xi^2-6i\beta\lambda^2\xi) \right)
} {\beta \lambda^2\xi (\lambda^2\xi+2i\beta)^2} \times $$
$$ \times \frac{\sinh\left( \frac{\beta
t}{2}+i\arctan(\frac{\lambda^2\xi+i\beta}{\beta})\right)}{\cosh\left(\frac{\beta
t}{2}+i\arctan(\frac{\lambda^2\xi+i\beta}{\beta})\right)} +
\frac{2\alpha B (i\lambda^2\xi(\lambda^4\xi^2 - 5\beta^2) -
4\beta\lambda^4\xi^2)+ 2\beta^3
}{\beta^2\lambda^2\xi(\lambda^2\xi+2i\beta)^2},$$

$$L_{2} = \frac{-\alpha\xi^2\lambda^4 e^{2\beta t} + 2((\lambda^2\xi+2i\beta)\lambda^2\xi t - 2i\lambda^2\xi +2\beta)\alpha \beta e^{\beta t}
+4i\alpha\beta\lambda^2\xi+\alpha\lambda^2\xi-4\alpha\beta^2}{i\beta
\lambda^2\xi(e^{\beta t}-1)+2\beta^2}.$$

We do not write here the explicit values of $q_{i}(t,\xi), i=1, 2,
3$, since they are very long.

We substitute \eqref{FurP},\eqref{FurdP},\eqref{FurddP} in
\eqref{FurCME} and \eqref{FurCVar} and after cumbersome but standard
computations get
\begin{equation}
\begin{array}{rcl}
\label{nonlinFurCME} && \bar f(t,r)= \displaystyle f_{0} +
\left(A-\frac{\alpha}{\beta}\right)t + \frac{(1-e^{\beta t})\alpha
r}{\beta} -{\frac { \left( 2+{\lambda}^{2} \right) \alpha\,
e^{2\beta\,t}
}{{\beta}^{2}}}+\\
&&+\left( {\frac { \left( 1+{\lambda}^{2}\right) \alpha\,t}{\beta}}
+{\frac { \left( 2+{\lambda}^{2} \right) \alpha}{{\beta}^{2}}}
\right) {e^{\beta\,t}},
\end{array}
\end{equation}

\begin{equation}
\begin{array}{rcl}
\label{nonlinFurCVar} && \bar{v}(t,r) = \displaystyle t\sigma^2 +
\left( -\frac{2e^{3\beta t}}{\beta^3} + \frac{(2\beta t+3)e^{2\beta
t}}{\beta^3} -\frac{2e^{\beta t}}{\beta^3} +\frac{1}{\beta^3}
\right)\alpha^2\lambda^2 r + \\
&& \displaystyle +\left( {4 e^{4\beta t}} - \frac{3(4\beta t
+5)e^{3\beta t}}{2} + (\beta^2 t^2+4\beta t+6)e^{2\beta t} +
\frac{(2\beta^2 t^2-2\beta
t-5)e^{\beta t}}{2}\right)\frac{\alpha^2\lambda^4}{\beta^4}+\\
&& \displaystyle +\left( \frac{5Be^{4\beta t}}{2} -{2B(\beta
t+3)e^{3\beta t}} +\frac{15Be^{2\beta t}}{2} -{4Be^{\beta
t}}-{B{\beta}t} \right)\frac{\alpha^2\lambda^2}{\beta^4}.
\end{array}
\end{equation}

\subsection{Example of portfolio consisting of two assets}

We consider
 a simple asset
allocation example, featuring an interest rate which affects a stock
index and also serves as a second investment opportunity,
illustrates how factors which are commonly used for forecasting
returns can be explicitly incorporated in a portfolio optimization
model. This example was systematically considered in the works by
T.Bielecki and S.Pliska
 (see \cite{bielplis},
\cite{biplsh}, \cite{biplsheu}).
The dynamics of the security prices
is
$$\frac{dS_{1}(t)}{S_{1}(t)} = (A_{1} + \alpha_{1} R(t))dt + \sigma_{1} dW_{1}(t),$$
$$\frac{dS_{2}(t)}{S_{2}(t)} = R(t)dt,$$
where $R(t)$ is the Cox-Ingersoll-Ross interest rate
\eqref{nonlinFactor}.

The capital of portfolio obeys the equation $\displaystyle \frac{d
V}{V}=h\frac{d S_{1}}{S_{1}}+(1-h)\frac{d S_{2}}{S_{2}}$, where the
scalar valued function $h$ is interpreted as the proportion of
capital invested in the risky asset, leaving the proportion $1-h$
invested in the bank account.
\begin{remark}
We can consider a portfolio consisting of any number of assets, as
it was done for the case of the Vasicek-type interest rate.
\end{remark}

Let us denote $\ln V = F$. Then
\begin{equation}
\label{nonlin2Capital}{dF}=\left( A_{1}h-\frac{1}{2}\sigma_{1}^2
h^2+\Big((h-1)\, \alpha_{1}+1\Big)\,R \right)dt+\sigma_{1}hdW_{1}.
\end{equation}

For the system  \eqref{nonlin2Capital}, \eqref{nonlinFactor} we
apply the formulae for conditional mathematical expectation and
variance  \eqref{nonlinFurCME}, \eqref{nonlinFurCVar} after
substitution
\begin{equation}
\label{nonlin2Param} A=A_{1} h-\frac{1}{2}\sigma_{1}^2h^2, \quad
\alpha=(\alpha_{1}-1)h+1, \quad \sigma=\sigma_{1}h.
\end{equation}
We consider again the functional \eqref{FTfunc}:
$$\bar {Q}_{\gamma}(t,r; h)=\bar {f}(t,r; h)-\gamma\bar {v}(t,r; h),$$
where $\gamma$ is the risk aversion coefficient, and find the
optimal strategy in the sense of Definition  \ref{def}.

According to \eqref{nonlinFurCME}, \eqref{nonlinFurCVar} and
\eqref{nonlin2Param} we get
$$\bar {Q}_{\gamma}(t,r; h)=K_{2}h^2+K_{1}h+K_{0},$$
where $K_{i}$ and  smooth functions of $t, r$ and coefficients
$A_{1}, \alpha_{1}, \sigma_{1}, B, \beta, \lambda, \gamma$. These
functions  can be expressed through elementary functions,
nevertheless, these expressions are cumbersome and we do not write
them.
Since   $\bar {Q}_{\gamma}(t,r; h)$ is quadratic with respect  $h$,
and
$$K_{2}=-\frac{t\sigma_{1}^{2}h^{2}}{2} -\gamma \bar{v}(t,r; h)|_{(\alpha=\alpha_{1}-1,\, \sigma=\sigma_{1})}<0,$$
then $\bar {Q}_{\gamma}(t,r;h)$ has a unique point of maximum
 (analogous to the linear case, see Sec.\ref{Vasicek}), the respective
 optimal strategy in the sense of definition
 \ref{def} is the following:
$$\bar{H}_{\gamma} = \frac{-K_{1}}{2 K_{2}} =$$
$$= \frac{(1-\alpha_{1})(M_{4}e^{4\beta t} + M_{3}e^{3\beta t} + M_{2}e^{2\beta t} + M_{1}e^{\beta t}+M_{0}) + A_{1}\beta^{4}t}
{(1-\alpha_{1})^{2}(M_{4}e^{4\beta t} + M_{3}e^{3\beta
t}+N_{2}e^{2\beta t}+N_{1}e^{\beta t}+N_{0}) +
(2\gamma+1)\sigma_{1}^{2}\beta^{4}t},$$ where
$$M_{4}=\gamma\lambda^{2}(8\lambda^{2}+5B), \quad M_{3}=-\gamma\lambda^{2}\left((3\lambda^{2} + B)4\beta t +
15\lambda^{2} + 4r\beta + 12B \right),$$
$$M_{2}=2\gamma\lambda^{4}\beta^{2}t^{2} + (2\lambda^{2} + \beta r)4\gamma \beta \lambda^{2} t + 12\gamma\lambda^{4} +
(5B+2\beta r)3\gamma\lambda^{2} -(\lambda^{2}+B)\beta^{2},
$$
$$M_{1}=2\gamma\lambda^{4}\beta^{2}t^{2} + (\beta^{2} - 2\gamma\lambda^{2})\beta\lambda^{2}t - 5\gamma\lambda^{4} +
(\beta^{2}-4\gamma\beta r - 8\gamma B)\lambda^{2} + \beta^{3}r +
\beta^{2}B,$$
$$M_{0}=(\beta^{2}-2\gamma\lambda^{2})\beta B t +2\gamma\lambda^{2}\beta r - \beta^{3}r,
\quad N_{1}=\gamma\lambda^{2}(2\beta^{2}\lambda^{2}t^{2} - 2\beta
\lambda^{2}t - 5\lambda^{2} - 4\beta r - 8B),$$
$$N_{2}=\gamma\lambda^{2}(2\lambda^{2}\beta^{2}t^{2}+(2\lambda^{2}+\beta r)4 \beta t+12\lambda^{2}+6\beta
r+15B),\quad N_{0}=2\gamma\beta \lambda^{2}(-Bt+r).$$

We get
\begin{equation}
\label{nonlin2Lim} \lim_{t\rightarrow\infty} \bar{H}_{\gamma} =
\lim_{t\rightarrow\infty} \bar{H}_{\gamma}(t) =
\frac{(\alpha_{1}-1)(\beta^{2} B - 2 \gamma\lambda^{2}
B)-A_{1}\beta^{3}}{(\alpha_{1}-1)^{2} 2 \gamma\lambda^{2}
B+(2\gamma+1)\sigma_{1}^{2}\beta^{3}}.
\end{equation}

As was shown in Sec.\ref{Vasicek}, in the case of a linear interest
rate model
$$\displaystyle \lim_{t\rightarrow\infty}
\bar{H}_{\gamma}(t)=\frac{-1}{\alpha_{1}-1},\quad \alpha_1\neq1.$$
This expression can be obtained by a limit pass in
\eqref{nonlin2Lim} as $\gamma\rightarrow\infty$ and
$\sigma_{1}\rightarrow0$.

\section{Comparing the optimal strategies for linear and nonlinear models of the interest
rate}\label{comparing_V_CIR}

We set  the inferest rate $r=0.05,$ the initial capital of portfolio
$f_{0}=0.08,$ the risk aversion coefficient $\gamma=0.1,$ the
parameters  $B=0.05,$ $\beta=-1,$ $\lambda=0.04,$ $A_{1}=0.15,$
$\alpha_{1}=-1,$ $\sigma_{1}=0.2$ are taken from \cite{biplsh}.

\begin{figure}[!htp]
  \begin{center}
    \includegraphics[width=0.5\columnwidth]{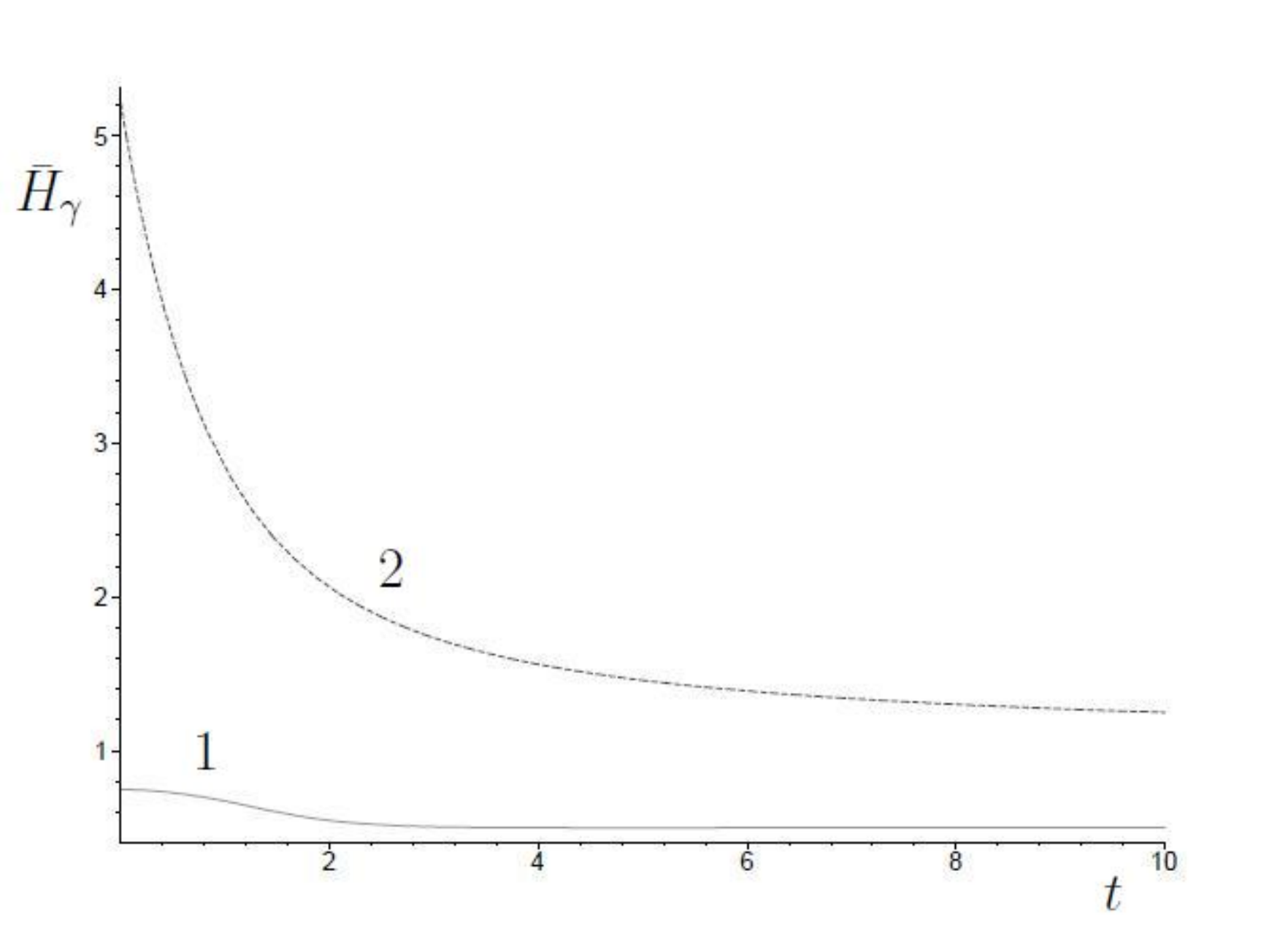}
  \caption{Comparing the optimal strategies for different models of
  the interest rate: \textbf{1.} Vasicek-type interest rate \,
  \textbf{2.}   Cox-Ingersoll-Ross interest rate.}
  \label{lin_nonlin}
  \end{center}
\end{figure}

Fig. \ref{lin_nonlin} illustrates the corresponding optimal
strategies of investment for the case of linear (solid line) and
nonlinear  (dashed line) interest rate for their uniform initial
distribution. We see that these strategies are very different. To
approach the optimal strategy for the nonlinear case to the strategy
for the linear case we should choose A larger risk sensitive
parameter $\gamma$.

As we have seen, in the case of the Vasicek-type interest rate the
asset less dependent on the factor is preferable for the investment
 for a large time. As follows from different combination of parameters, for
the Cox-Ingersoll-Ross interest rate the properties of the factor
are taken into account  more effectively.

\begin{remark}
If we assume additionally that the variance of the return of the
risky asset satisfying \eqref{nonlinFCapital} is proportional to the
interest rate, we fall in the situation of the Heston model
\cite{heston}, one of the most popular models of stochastic
volatility.  In \cite{martroz} we analyze the value of mean
dispersion, and formula  \eqref{FurCME} turns useful there, too.
\end{remark}

\begin{remark}
Papers \cite{hatasek2006} and \cite{hata2011} extend the work
 \cite{biplsheu}, they deal with the problem of long-run optimal investment in the frame if the Cox-Ingersoll-Ross
 model.
\end{remark}
\begin{remark}
As follows from  \cite{alroz2}, the behavior of strategy of
investment should depend on the speed of  decay  at infinity the
initial distribution of the interest rate.
\end{remark}
\begin{remark}
Equations \eqref{nonlinFCapital}, \eqref{nonlinFactor} refer to the
so called  ``affine'' model  \cite{duffilsch2003}, therefore the
respective Fokker-Planck equation can be solved explicitly.
\end{remark}

\section*{Acknowledgements}

 The work was partially supported by RFBR Project Nr. 12-01-00308 (OR).

\bibliographystyle{amsalpha}

\end{document}